\documentclass[superscriptaddress,aps,prd,twocolumn,showpacs,nofootinbib,longbibliography, superscriptaddress, showkeys]{revtex4-2}

\DeclareUnicodeCharacter{02BC}{\textendash}
\usepackage[utf8]{inputenc}
\usepackage[T1]{fontenc}
\usepackage{amsmath,amssymb,amsthm}
\usepackage{easybmat}
\usepackage[colorlinks=true,citecolor=blue,urlcolor=blue, linkcolor= magenta]{hyperref}
\usepackage[pdftex]{graphicx}
\usepackage{times, txfonts}
\usepackage{braket}
\usepackage{color}
\usepackage{ragged2e}
\usepackage{booktabs}
\usepackage{epstopdf}
\usepackage{gensymb}
\usepackage{natbib}
\usepackage{subcaption}
\usepackage{float}
\newcommand{\be}{\begin{equation}}
	\newcommand{\ee}{\end{equation}}
\newcommand{\ba}{\begin{eqnarray}}
	\newcommand{\ea}{\end{eqnarray}}

\usepackage{tikz,xcolor,hyperref}

\definecolor{lime}{HTML}{A6CE39}
\DeclareRobustCommand{\orcidicon}{\hspace{-4pt}
	\begin{tikzpicture}
		\draw[lime, fill=lime] (0,0) 
		circle [radius=0.16] 
		node[white] {\hspace{0.1mm}{\fontfamily{qag}\selectfont \tiny ID}};
		\draw[white, fill=white] (-0.07,0.1) 
		circle [radius=0.01];
	\end{tikzpicture}
	\hspace{-3.2mm}
}

\foreach \x in {A, ..., Z}{\expandafter\xdef\csname 
	orcid\x\endcsname{\noexpand\href{https://orcid.org/\csname orcidauthor\x\endcsname}
		{\noexpand\orcidicon}}
}

\begin{document}
	
 \title{Probing $CP$ violation and mass ordering in neutrino oscillations in matter through quantum speed limits}
	\author{Subhadip Bouri\orcidA{}}\email{subhadipb@iisc.ac.in}
\author{Abhishek Kumar Jha\orcidB{}}\email{kjabhishek@iisc.ac.in}
\affiliation{Department of Physics, Indian Institute of Science, Bangalore 560012, India}
\author{Subhashish Banerjee\orcidC{}}\email{subhashish@iitj.ac.in}
\affiliation{Indian Institute of Technology Jodhpur, Jodhpur 342011, India}

	


\date{\today}

\begin{abstract}
{The quantum speed limits (QSLs) set fundamental lower bounds on the time required for a quantum system to evolve from a given initial state to a final state. In this work, we investigate $CP$ violation and the mass ordering problem of neutrino oscillations in matter using the QSL time as a key analytical tool. We examine the QSL time for the unitary evolution of two- and three-flavor neutrino states, both in vacuum and in the presence of matter. Two-flavor neutrino oscillations are used as a precursor to their three-flavor counterparts. We further compute the QSL time for neutrino state evolution and entanglement in terms of neutrino survival and oscillation probabilities, which are experimentally measurable quantities in neutrino experiments. A difference in the QSL time between the normal and inverted mass ordering scenarios, for neutrino state evolution as well as for entanglement, under the effect of a $CP$ violation phase is observed.
Our results are illustrated using the length scales and energies of ongoing long-baseline accelerator neutrino experiments such as T2K, NOvA, and the upcoming DUNE experiment. Notably, three-flavor neutrino oscillations in constant matter density exhibit faster state evolution across all these neutrino experiments in the normal mass ordering scenario. Additionally, we observe fast entanglement suppression in DUNE assuming a normal mass ordering.}
\end{abstract}

\maketitle

\section{Introduction}
\label{sect1}
The uncertainty relations are widely regarded as a fundamental aspect of modern physics, representing the uncertainty relationship between conjugate variables in a quantum mechanical system \cite{PhysRev.34.163,PhysRevLett.113.260401}. Mandelstam and Tamm (MT) demonstrated that the energy, time uncertainty relation can be understood as imposing limits on the evolution rate of a quantum system \cite{Mandelstam1991}. Another bound on the quantum evolution time was subsequently obtained by Margolus and Levitin (ML) in terms of the initial mean energy of the quantum system \cite{MARGOLUS1998188}. Both MT and ML were combined to provide a tight bound on the minimum time required for a quantum system to evolve from an initial state to the final state under a given dynamical process \cite{PhysRevLett.103.160502}. The limit arises due to the intrinsic dynamical timescales of a quantum system. This limitation on a quantum system’s evolution time is known as the quantum speed limit (QSL) time. Over the last decade, by employing the geometric approach, the topic of QSL time has been extensively researched \cite{PhysRevLett.65.1697,PATI1991105,ANANDAN199729,PATI199540,PhysRevA.67.052109,Deffner_2013,Deffner:2017cxz}. In Ref.\,\cite{Deffner_2013}, the QSL time was shown to provide the fundamental limit on the rate of evolution of closed quantum systems, whose dynamics is unitary. The Bures angle can be considered a measure of the “closeness” between the initial and final states of the evolved quantum system. The QSL time for the evolution of a quantum system through the state space can be expressed as a function of the Bures angle and is inversely proportional to the variance of the driving Hamiltonian. Moreover, the generalization of QSL time has been explored for driven closed \cite{Thakuria_2024} and open quantum systems \cite{PhysRevLett.111.010402,Baruah:2022zqu}. The QSL time has broad applications in various quantum tasks, such as quantum information \cite{PhysRevA.85.052327,Aggarwal:2021xha,PhysRevLett.46.623,Paulson:2021jmi,Paulson:2022lih}, quantum correlations \cite{Tiwari:2022kxh}, quantum computation \cite{Lloyd_2000}, quantum metrology \cite{PhysRevLett.105.180402,Giovannetti_2011}, nonequilibrium quantum entropy production \cite{PhysRevLett.111.010402}, quantum thermodynamics \cite{Campo_2014,Funo_2019}, quantum optimal control algorithms \cite{PhysRevLett.103.240501,PhysRevA.82.022318}, quantum batteries \cite{Campaioli2018,PhysRevResearch.2.023113,PhysRevA.104.042209,PhysRevA.106.042436}, and quantum gravity \cite{Liegener:2021zhs,Wei:2023jrx,Wang:2022qgj,Maleki:2019cqn}. In recent times, in the realm of particle physics, the impact of open quantum system and information theoretic ideas, such as, the QSL time, has been made on the neutral meson \cite{Banerjee:2014vga,Naikoo:2018vug,Banerjee:2022ckm} and the neutrino system \cite{Khan:2021kai}. 

Neutrinos are weakly interacting particles with nonzero mass \cite{10.1093/acprof:oso/9780198508717.001.0001}, as demonstrated by experiments with atmospheric \cite{Super-Kamiokande:1998kpq,kajita2016nobel} and solar neutrinos \cite{mcdonald2016nobel}. It is known that a neutrino of a specific flavor $\nu_\alpha$ (where $\alpha \in{e, \mu, \tau}$) undergoes time evolution and may be detected in a distinct flavor state \cite{Pontecorvo:1957cp,Pontecorvo:1957qd,Pontecorvo:1967fh,Bilenky:1978nj,Bilenky:2004xm,Giganti:2017fhf}. This phenomenon constitutes the essence of neutrino oscillations. Flavor oscillations occur because neutrinos exist as a mixture of different masses. A flavor state can be written as a linear superposition of three nondegenerate mass eigenstates. Both flavor and mass eigenstates can be related by the unitary transformation matrix known as the Pontecorvo-Maki-Nakagawa-Sakata (PMNS) mixing matrix  \cite{maki1962remarks}. 
In the standard three-flavor neutrino oscillations in vacuum, the PMNS matrix of Dirac neutrinos is fully described by three mixing angles $\theta_{12}$, $\theta_{23}$, and $\theta_{13}$ and a complex phase $\delta_{\rm CP}$ related to charge-conjugation and parity-reversal ($CP$) symmetry violations~\cite{Giganti:2017fhf}. The survival and oscillation probabilities of initial neutrino flavor states can be related to the three mixing angles, two squared-mass differences and the ratio of length to energy ($L/E$), where $ L$ is the distance between the neutrino source and detector, and $ E$ is the neutrino source energy. Parameters linked to these probabilities are measurable in neutrino experiments. The quest to investigate precise measurements of PMNS mixing parameters \cite{Giganti:2017fhf}, $CP$-violation phases~\cite{PhysRevD.23.1666,T2K:2019bcf,PhysRevD.57.4403}, and other issues such as neutrino mass ordering (whether the neutrino mass ordering is normal or inverted in nature) \cite{Kobzarev:1981ra} and sterile neutrinos, are the primary focus of the current neutrino experiments including solar \cite{KamLAND:2013rgu}, atmosphere \cite{Denton:2022een}, reactor (short and long baseline) \cite{DayaBay:2013yxg,RENO:2018dro,DoubleChooz:2019qbj}, and accelerator (long baseline) \cite{DiLodovico:2015nfq,Kudryavtsev_2016}. 

Neutrinos interact very weakly with matter, with interactions typically dominated by coherent forward elastic scattering rather than incoherent scattering \cite{10.1093/acprof:oso/9780198508717.001.0001}. We can easily estimate the order of magnitude of a mean free path due to incoherent scatterings from dimensional arguments. In the laboratory frame (when the target particle is at rest), the cross section for neutrino weak interactions with a charged lepton or hadron is approximately $\sigma_{\text{lab}} \sim G_{\mathrm{F}} E M \sim 10^{-38} \, \text{cm}^2 \left( E M / \text{GeV}^2\right)$, where $E$ is the neutrino energy, $G_F$ is the Fermi constant, and $M$ is the mass of the target particle (assuming the neutrino mass is negligible). The mean free path of a neutrino in a medium with a number density \( N \) of target particles is $\ell \sim \left(1/{N \sigma}\right) \sim \left(10^{38} \text{cm} / N \,\text{cm}^{-3} \left(\frac{EM}{\text{GeV}^2} \right) \right)$. In ordinary matter, the primary targets are nucleons, with a mass \( M \sim 1 \, \text{GeV} \) and a number density \( N \sim \left(N_{\mathrm{A}} / \text{cm}^3\right) \sim 10^{24} \, \text{cm}^{-3} \), where \( N_{\mathrm{A}} \) is Avogadro's number. This gives a mean free path as $\ell_{\text{matter}} \sim \left(10^{14} \, \text{cm}/(E / \text{GeV})\right).$ Thus, the accelerator long baseline neutrino source experiments on Earth, which we shall consider in this work, are typically of the order of hundreds of kilometers and are largely transparent to neutrinos with energies in the range of a few GeV. As a result, it is not necessary to consider neutrino incoherent scattering, allowing the oscillating neutrino system to be treated as a close quantum system. Moreover, there is a scenario when neutrinos encounter extremely high matter densities (for example, in neutron stars and supernova cores where $N\approx 10^{12}N_A/\text{cm}^3$ and $E\approx1\text{MeV}$), where it is necessary to take into account neutrino incoherent scattering, allowing the oscillating neutrino system to be treated as an open quantum system.

In a close quantum system, the dynamical evolution of the neutrino system is governed by unitary transformation. The driving Hamiltonian for neutrino flavor oscillations in the presence of Earth's matter background is the effective Hamiltonian \cite{PhysRevD.17.2369,RevModPhys.61.937,Ohlsson:1999xb,Ohlsson:1999um,Denton_2016, Barenboim_2019,Nguyen:2022snr}, which can be described by two regions of neutrino state propagation: in vacuum and in the presence of constant matter. Thus, a natural question arises: how quickly does the neutrino flavor state change with time in these two regions of propagation? In this paper, we aim to answer this question by examining in detail the QSL time for the unitary dynamics of two- and three-flavor neutrino oscillations in matter. In this context, we quantify the Bures angle in relation to neutrino survival probability, providing insight into the proximity between the initial and final states of time-evolved flavor neutrino states in vacuum and in a constant matter background. Furthermore, we compute the variance in the driving Hamiltonian for various regions of neutrino flavor state propagation with time. The study of QSL time in two-flavor neutrino oscillations in matter serves as a precursor to understanding three-flavor neutrino oscillations actually occurring in nature. In three-flavor neutrino oscillations, considering matter effects and $CP$ violation, we discuss the behavior of the QSL time for ongoing accelerator long-baseline neutrino source experiments like NOvA and T2K, as well as for the upcoming experiment DUNE. Further, we analyze the QSL time for the initial muon neutrino and muon anti-neutrino flavor states. Our analysis reveals that the QSL time demonstrates sensitivity to the normal and inverted mass-ordering scenarios and $CP$ violation in neutrino physics.

Furthermore, quantum entanglement and coherence are two fundamental features arising from the principle of quantum superposition \cite{Nielsen_Chuang_2010}. Since the time-evolved neutrino state is a coherent superposition state, it is natural to study quantum entanglement in the neutrino system. In the plane-wave framework of two- and three-flavor neutrino oscillations, the linear superposition of neutrino states have been mapped to two- and three-mode quantum states, representing two- and three-qubit Bell’s type states, respectively \cite{Blasone_2009,BLASONE2013320,PhysRevD.77.096002,PhysRevA.77.062304,Banerjee:2015mha,Alok:2014gya,Dixit:2017ron,Dixit:2018kev,Dixit:2019swl,Dixit:2020ize,KumarJha:2020pke,Yadav:2022grk,Li:2022mus,wang2023monogamy,Bittencourt:2023asd,Blasone:2023qqf,Quinta:2022sgq,Jha:2022yik,Dixit:2023fke,Konwar:2024nrd}. It is known that the concept of entanglement plays a crucial role in quantifying and characterizing entanglement in the time-evolved neutrino mode (flavor) state. Various multimode entanglement measures, such as concurrence, linear entropy, teleportation fidelity, geometric discord, von Neumann entropy, tangle, negativity, three-tangle, and three-$\pi$, have been quantified in terms of neutrino flavor transition probabilities. Also, other aspects like Bell’s inequality and Bell-Clauser-Horn-Shimoy-Holt inequality violations have been studied in the context of neutrino oscillations \cite{Blasone_2009,BLASONE2013320,PhysRevD.77.096002,PhysRevA.77.062304,Banerjee:2015mha,Alok:2014gya,Dixit:2017ron,Dixit:2018kev,Dixit:2019swl,Dixit:2020ize,KumarJha:2020pke,Yadav:2022grk,Li:2022mus,wang2023monogamy,Bittencourt:2023asd,Blasone:2023qqf,Quinta:2022sgq,Jha:2022yik,Dixit:2023fke,Konwar:2024nrd}. As a three-qubit system, the three-flavor neutrino states resemble the generalized W states \cite{KumarJha:2020pke}. Additionally, the study of the mode entanglement using the wave-packet approach in neutrino oscillations has been done \cite{Blasone_2015,Blasone:2021cau,Blasone:2022ete,Ravari:2022yfd,Ettefaghi:2023zsh}. These  suggest that genuine multipartite entanglement persists in the neutrino flavor-changing states. Moreover, the nonclassical nature of neutrino oscillations was investigated using temporal analogues of Bell inequalities, such as Leggett-Garg inequalities \cite{PhysRevLett.117.050402,PhysRevD.99.095001,NAIKOO2020114872,Blasone:2022iwf,Chattopadhyay:2023xwr,Soni:2023njf,Shafaq:2020sqo,Sarkar:2020vob}. Quantum coherence in neutrino oscillations can be maintained over a long distance and observed in different experiments \cite{PhysRevA.98.050302,Dixit:2018gjc,Li:2022zic}.
Furthermore, by encoding neutrinos in a multiqubit system, potential use of neutrino oscillations has been made for quantum computation as well \cite{Arguelles:2019phs,Molewski:2021ogs,Jha:2021itm,Nguyen:2022snr}. 

The study of the QSL time for quantum evolution has unveiled fundamental constraints on the rate of change of various entanglement measures \cite{Mohan:2021vfu,PhysRevA.107.052419,Das:2017nty}. Specifically, QSL time for entanglement entropy has been investigated for the bipartite pure quantum system \cite{PhysRevA.106.042419}. The simplest case of a multipartite quantum system is a bipartite quantum system comprising two subsystems. Entanglement entropy serves as an important entanglement measure to assess the degree of quantum entanglement between the subsystems of a bipartite quantum system. The entanglement entropy determines the von Neumann entropy on the reduced density matrix of one of the subsystems \cite{PhysRevLett.101.010504}. It quantifies the uncertainty resulting from the lack of a well-defined quantum state in a subsystem due to potential entanglement with another subsystem. The variance in entanglement entropy is defined as the capacity of entanglement of the reduced state of any of the subsystems \cite{DeBoer:2018kvc,PhysRevLett.105.080501}. The QSL time for entanglement entropy shown in Ref. \cite{PhysRevA.106.042419} estimates how fast entanglement may be formed or degraded in a bipartite quantum system.

Recently, to probe the astrophysical neutrinos, bipartite entanglement entropy has been proposed to quantify entanglement in quantum many-body collective neutrino flavor oscillations \cite{PhysRevC.101.065805,Roggero:2021asb,PhysRevD.107.023019,Balantekin:2023qvm}. These studies are indeed useful for gaining a comprehensive understanding of quantum entanglement in neutrino systems. However, in this work, we consider a single-particle neutrino system as a bipartite quantum system. A single particle dynamics of a flavor neutrino state in the relativistic regime can be mapped to a two-dimensional Hilbert space. It has been shown that the Hilbert spaces for flavor and massive neutrinos are orthogonal and that a condensate structure of neutrino-antineutrino pairs is present in the vacuum for the flavor neutrinos \cite{Blasone:1995zc}. These effects become significant in the nonrelativistic regime, where they lead to modifications in the neutrino oscillation formulas. In the particular scenarios in the relativistic regime where vacuum corrections are negligible, the flavor neutrinos “live" in a Hilbert space, which is the tensor product of the Hilbert spaces for massive neutrinos (two or three). An alternative perspective, as discussed in Ref.\,\cite{Blasone:2022quj}, is that particle states represent excitations of quantum fields. It is a well-known result that two free fields with different masses generate unitarily inequivalent Hilbert spaces \cite{Haag:1955ev}. Thus, we cannot accommodate two states with different masses in the same Hilbert space. Therefore, it is necessary to form a tensor product of these states, and a generic combination of these will be entangled. Subsequently, defining occupation numbers as a multiqubit space, a single-particle neutrino flavor state can be interpreted as multpartite mode-entangled states \cite{PhysRevD.77.096002}. This is further reinforced by a strong activity in the field of exploring quantum correlations for oscillating neutrinos \cite{Blasone_2009,BLASONE2013320,PhysRevD.77.096002,PhysRevA.77.062304,Banerjee:2015mha,Alok:2014gya,Dixit:2017ron,Dixit:2018kev,Dixit:2019swl,Dixit:2020ize,KumarJha:2020pke,Yadav:2022grk,Li:2022mus,wang2023monogamy,Bittencourt:2023asd,Blasone:2023qqf,Quinta:2022sgq,Jha:2022yik,Dixit:2023fke,Konwar:2024nrd}, recently reviewed in Ref.\,\cite{Blasone:2023qqf}. These investigations support a single-particle neutrino system to be treated as a bipartite quantum system. By mapping the two-flavor neutrino state to the two-qubit state, we quantify the entanglement entropy and the capacity of entanglement in two-flavor neutrino oscillations in vacuum and in the presence of constant Earth-matter background. We use these concepts to explore the QSL time for entanglement entropy in two-flavor neutrino oscillations. Furthermore, we extend our investigation to study the fundamental limit on the time required for entanglement entropy in three-flavor neutrino oscillations with the $CP$-violation phase, considering normal and inverted mass ordering scenarios. We use the length scales and energies of ongoing long-baseline accelerator neutrino experiments such as T2K, NOvA, and the upcoming DUNE experiment to characterize our results.

The study of QSL in neutrino oscillations in matter with the $CP$ violation phase in the context of open quantum systems was discussed in Ref.\,\cite{Khan:2021kai}. However, mass-ordering scenarios were not considered there. Here, the study of the QSL time for neutrino oscillations in matter (both for muon neutrino and muon antineutrino) is conducted, within the ambit of closed quantum systems, taking into account the $CP$ violation phase and mass-ordering scenarios, incorporating the length scales and energies of long-baseline accelerator neutrino experiments such as T2K, NOvA, and DUNE as well as the fundamental vacuum parameters from NuFIT data \cite{Esteban:2024eli,NuFIT}. Furthermore, this work utilizes the concepts of multimode entanglement to study neutrino oscillations and examine their relation with the QSL time for entanglement entropy using the expected baseline length and energy from different long-baseline neutrino source experiments.

The organization of the paper is as follows: In Sec.\,\ref{Sect2}, we briefly review the QSL time for quantum evolution and for entanglement. In Sec.\,\ref{Sect4}, the QSL time is computed in two-flavor scenarios, especially for the time evolution of  muon-flavor neutrino states propagating in vacuum and in presence of a constant Earth-matter potential. Moreover, we map neutrinos to two-qubit states and explore various bipartite entanglement measures such as entanglement entropy and capacity of entanglement. In three-flavor scenarios, Sec.\,\ref{Sect5} explores the QSL time for the initial muon neutrino and muon antineutrino flavor states evolving in matter in the presence of the $CP$-violation phase, and under the assumptions of normal and inverted mass ordering. Furthermore, in Sec.\,\ref{Sect6}, we map the initial muon flavor neutrino state to a three-qubit state and investigate various bipartite entanglement measures such as entanglement entropy and capacity of entanglement as the initial muon flavor neutrino propagates in space with a constant Earth-matter background. We estimate the QSL time for entanglement entropy in the presence of matter and the $CP$-violation phase, considering normal and inverted mass ordering. Discussions and conclusions are made in Sec.\,\ref{Sect7}.

\section{Quantum speed limit time for quantum evolution and for entanglement}
\label{Sect2}
 In quantum mechanics, the state of a quantum system is described by a wave function $\ket{\psi}$, and the time evolution of this wave function is given by the Schrödinger equation \cite{Nielsen_Chuang_2010}
 \begin{equation}
 i\hbar\frac{d\ket{\psi}}{dt}=\mathcal{H}\ket{\psi}\Longrightarrow \ket{\psi_\tau}=e^{-i\mathcal{H}t/\hbar}\ket{\psi_0}\equiv U_t\ket{\psi_0},
 \label{1}
 \end{equation}
  where $U_t$ is the unitary time evolution operator ($U_t U_t^{*}=I$). An initial pure state remains pure due to the unitary dynamics. In this case, the initial and final states can be represented in the density matrix form as $\rho_{0}=\left|\psi_{0}\right\rangle\left\langle\psi_{0}\right|$ and $\rho_{\tau}=\left|\psi_{\tau}\right\rangle\left\langle\psi_{\tau}\right|$, respectively. The driving Hamiltonian $\mathcal{H}$ for the initial pure state $\rho_0$ can be either time independent or time dependent, and is Hermitian $\mathcal{H}^\dagger=\mathcal{H}$.

  MT and ML-type bounds on speed limit time are estimated by using geometrical approaches to quantify the closeness between the initial and final states \cite{Deffner:2017cxz}. Here, the Bures angle is employed to measure the distance between two quantum states
\begin{equation}
    \mathcal{B}=\cos^{-1}\sqrt{\mathcal{F}(\psi_0,\psi_\tau)}.
    \label{2}
\end{equation} 
In Eq.\,(\ref{2}), ${\mathcal{F}}=\rm Tr\,(\rho_0\rho_\tau)=|\langle\psi_0|\psi_\tau\rangle|^2$ represent the survival probability of the initial pure state $\rho_0$ and is known as fidelity. $\mathcal{F}=1$ implies that $\rho_\tau$ is same as $\rho_0$. For the initial pure state $\rho_0$, if the given driving Hamiltonian ($\mathcal{H}$) is time independent, the QSL time ($\tau_{\rm QSL}$) for such a system is defined as \cite{Thakuria_2024}
\begin{equation}
\tau \geq \tau_{\rm QSL}=\frac{\hbar \mathcal{B}}{\Delta{\mathcal{H}}},
\label{3}
\end{equation}
 where $\tau$ is the propagation time of the initial state, the ratio $\tau_{\rm QSL}/\tau\leq 1$ is the QSL time-bound condition, and
 \begin{equation}
 \Delta{\mathcal{H}}= \sqrt{\langle   \mathcal{H}^2\rangle - \langle 
\mathcal{H}\rangle ^2}
\label{4}
\end{equation}
is the variance (or fluctuation) in the driving Hamiltonian ($\mathcal{H}$).
Here, Eq.\,({\ref{3}}) represents the MT bound \cite{Mandelstam1991}, which corresponds to Eq.\,(12) from Ref.\,\cite{Thakuria_2024}, where the Bures angle is denoted by $\mathcal{B}$.

Furthermore, in a bipartite pure quantum system, a state of the quantum system can be represented as the tensor product of minimum two-qubit Hilbert spaces \cite{Nielsen_Chuang_2010}: $\ket{\psi}\in H_A\otimes H_B$. For example, a two-qubit Bell-type superposition state can be considered as a bipartite quantum state: $\ket{\psi}=\alpha{\ket{10}+\beta{\ket{01}}}$, $|\alpha|^2+|\beta|^2=1$, where $\ket{01}=\ket{0}\otimes\ket{1}$ and $\ket{10}=\ket{1}\otimes\ket{0}$ are two-qubit basis states of $\ket{\psi}$. In general, Bell's state of two-qubit or more than two-qubit representation can be regarded as a bipartite pure entangled state. Any entangled bipartite pure quantum system $\rho_{AB}=\ket{\psi}\bra{\psi}$ consists of two subsystems $\rho_A=\rm Tr_B(\rho_{AB})$ and $\rho_B=\rm Tr_A(\rho_{AB})$ such that $\rho\neq\rho_A\otimes\rho_B$, where $\rho_A$ and $\rho_B$ are two reduced density matrices. The entanglement entropy $S_{EE}$ for $\rho_A$ (or $\rho_B$) is defined as \cite{PhysRevLett.101.010504}
\begin{equation}
    S_{EE}=S(\rho_A)=-\rm Tr\,(\rho_A log_2\rho_A).
    \label{5}
\end{equation}
Moreover, if the dynamics of the system is unitary and the driving Hamiltonian remains time independent, the quantum speed limit for the entanglement entropy ($S_{EE}$) adheres to the following bound \cite{PhysRevA.106.042419}
\begin{equation}
    \tau\geq \tau^E_{\rm QSL}=\frac{\hbar|S_{EE}(\tau)-S_{EE}(0)|}{2\Delta \mathcal{H} \frac{1}{\tau}\int ^\tau_0 \sqrt{C_E(t)}dt}.
    \label{6}
\end{equation}
The variance in the entanglement entropy $S_{EE}$ is defined as capacity of entanglement $C_{E}$ \cite{DeBoer:2018kvc,PhysRevLett.105.080501}, and is 
\begin{equation}
    C_E=\sum_i\lambda_i log_2^2\lambda_i-S^2_{EE}.
    \label{7}
\end{equation}
Here, $\lambda_i's$ are the eigenvalues of the reduced density matrix $\rho_A$ of one of the subsystems of a given pure bipartite quantum system $\rho_{AB}$, where $\rm Tr\,(\rho_{AB}^2)=1$ and $\rm Tr\,(\rho^2_A)<1$. Equation\,(\ref{6}) establishes a quantum speed limit on the production and decay of entanglement in the case of pure bipartite states when the driving Hamiltonian is time independent. 

\section{Quantum Speed Limit time of two-flavor neutrino oscillations}
\label{Sect4} 
In this section, we first explore the QSL time for the evolution of two-flavor neutrino states in the presence of matter backgrounds. Later, we investigate the QSL time for entanglement entropy in the two-flavor neutrino oscillations in matter.

\subsection{QSL time for evolution of two-flavor neutrino states in matter}
\label{Sect4A}
 In the two-flavor neutrino oscillations in vacuum, at time $t=0$, the neutrino flavor state can be written as \cite{10.1093/acprof:oso/9780198508717.001.0001}
\begin{equation}
    \begin{pmatrix}
        \ket{\nu_\alpha}\\
        \ket{\nu_\beta}\\
\end{pmatrix}=U(\theta)\begin{pmatrix}
        \ket{\nu_j}\\
        \ket{\nu_k}\\
    \end{pmatrix},
    \label{25}
\end{equation}
where $\alpha,\beta=e,\mu,\tau$; $j,k=1,2,3$ and $U U^\dagger=I$,
\begin{equation}
 U(\theta)=  \begin{pmatrix}
        U^*_{e1} & U^*_{e2}\\
        U^*_{\mu 1} & U^*_{\mu 2}\\
    \end{pmatrix}\equiv \begin{pmatrix}
        \cos\theta & \sin\theta\\
        -\sin\theta & \cos\theta\\
    \end{pmatrix}.
    \label{26}
\end{equation}
 When the neutrino starts propagating in matter, its effective Hamiltonian in the two-flavor scenario can be expressed as~\cite{10.1093/acprof:oso/9780198508717.001.0001,PhysRevD.17.2369,Blennow:2013rca}
\begin{eqnarray}
{\mathcal{H}_\mathrm{M}
=\frac{1}{4E}\Bigg[\begin{pmatrix}
     -\Delta m^2\cos2\theta & \Delta m^2\sin2\theta\\
     \Delta m^2\sin2\theta &  \Delta m^2\cos2\theta\\
     \end{pmatrix} +\begin{pmatrix}
     A_{\rm CC} & 0\\
     0 & -A_{\rm CC}\\
     \end{pmatrix}\Bigg]},\hspace{0.48cm}
     \label{27}
     \end{eqnarray}
where $\Delta m^{2}$ and $\theta$ are the squared-mass difference and mixing angle in vacuum, respectively, and $E$ is the energy of the neutrino, which is different for different neutrino experiments. Here,
\begin{equation}
    A_{\rm CC}=2EV_{\rm CC},
    \label{27a}
     \end{equation}
   where $V_{\rm CC}$ is the Wolfenstein matter potential  \cite{PhysRevD.17.2369,Blennow:2013rca}. This charged-current potential\footnote{The neutral current (NC) contribution has been separated from the effective Hamiltonian in matter ($\mathcal{H}_M$), given by eqn.\,(\ref{27}), because it introduces a phase common to all flavors, which can be eliminated by a phase shift and does not affect neutrino flavor transition probabilities \cite{10.1093/acprof:oso/9780198508717.001.0001}.} $V_{\rm CC}$, which originates from the neutrino’s interaction with electrons in matter through coherent forward elastic scattering (assuming negligible noncoherent effects), can be expressed as \cite{Majhi:2022fed,DUNE:2020jqi} 
\begin{equation}  
V_{\text{CC}} = \sqrt{2} G_F N_e \simeq 7.5\, X_e \frac{\rho}{10^{14} \, (\text{g/cm}^3)} \, \text{eV}, 
\end{equation} where $G_F$ is the Fermi coupling constant, $N_e$ is the electron number density, and $X_e$ is the relative electron number density, which is 0.5 for the neutral medium. The approximate matter density in the Earth's crust along the long-baseline experiment is
 $\rho \approx 2.848 \, \text{g/cm}^3$\,\cite{DUNE:2020jqi}. Consequently, a constant matter potential is obtained as 
\begin{equation}  
V_{\text{CC}} \approx 1.01 \times 10^{-13} \,\text{eV}. 
\label{A14}
\end{equation}   
Substituting Eq.\,(\ref{A14}) into Eq.\,(\ref{27a}), we can determine the value of matter parameter $A_{CC}$ for various neutrino energies.

The evolution equation for the two-flavor neutrino in matter is
\begin{equation}
     i\frac{d}{dt}\ket{\nu_\alpha}
    =H_\mathrm{M}\ket{\nu_\alpha}.
     \label{28}
     \end{equation}
     \begin{figure}[!ht]
    \centering
    \includegraphics[scale=0.5]{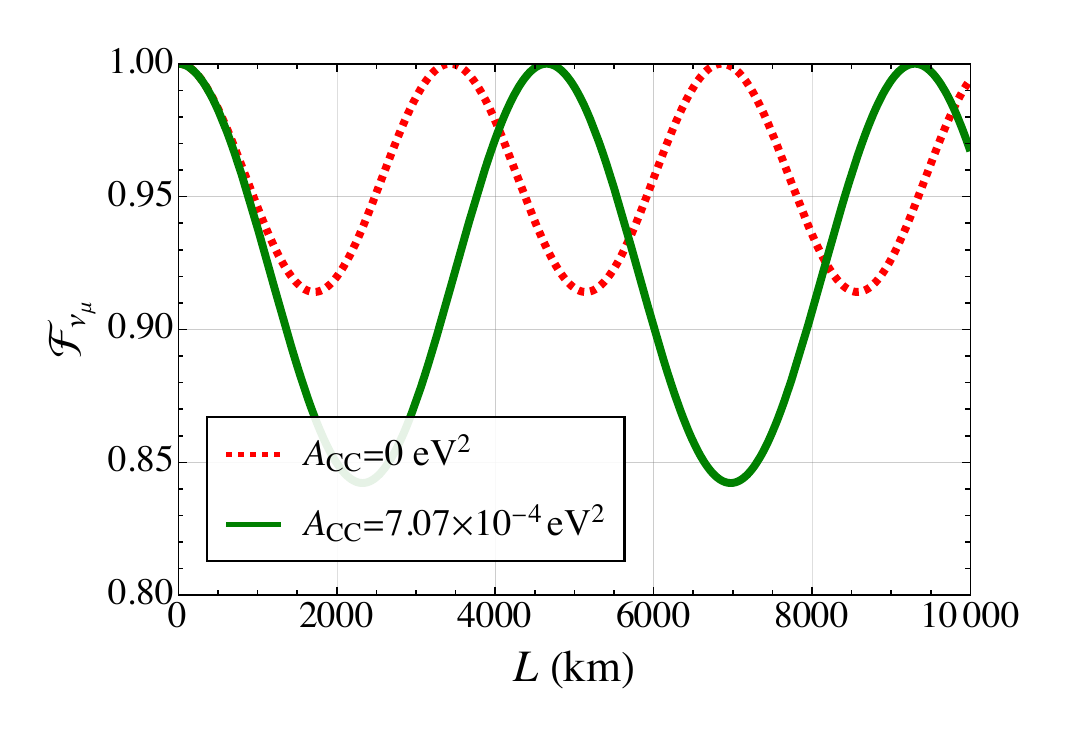}
    \caption{\justifying{In the two-flavor neutrino oscillation, the fidelity $\mathcal{F}_{\nu_\mu}$ of the initial muon neutrino state $\ket{\nu_\mu}$ is depicted as a function of propagation length $L\,(\text{km})$. This representation includes vacuum ($A_{\rm CC}=0$), indicated by a red dashed line and in a constant matter potential ($A_{\rm CC}=7.07\times10^{-4} \text{eV}^2$), represented by a green solid line. The various neutrino mixing parameters used are taken from Table\,\ref{tab1}.}}
    \label{Fig1}
\end{figure}
The effective Hamiltonian $\mathcal{H}_{M}$ in matter is Hermitian $\mathcal{H}_{M}^\dagger=\mathcal{H}_{M}$ and can be diagonalized by the orthogonal transformation
\begin{equation}
\mathcal{H}_{\mathrm{M}}=\frac{1}{4 E} (U_{\mathrm{M}} (\mathbb{M}_\mathrm{M})^{2} U_{\mathrm{M}}^{\dagger}),
\label{29}
\end{equation}
where $(\mathbb{M}_\mathrm{M})^{2}=\frac{1}{4E} \operatorname{diag}\left(-\Delta m_{\mathrm{M}}^{2}, \Delta m_{\mathrm{M}}^{2}\right)
$. The unitary matrix
\begin{equation}
U_{\mathrm{M}}(\theta_{\mathrm{M}})=\left(\begin{array}{cc}
\cos \theta_{\mathrm{M}} & \sin \theta_{\mathrm{M}} \\
-\sin \theta_{\mathrm{M}} & \cos \theta_{\mathrm{M}}
\end{array}\right),
\label{30}
\end{equation}
is the effective mixing matrix in matter. The neutrino squared-mass difference and mixing angle in vacuum are related to matter parameters as \cite{10.1093/acprof:oso/9780198508717.001.0001,Agarwalla:2013tza}
\begin{equation}
\Delta m_{\mathrm{M}}^{2}=\sqrt{\left(\Delta m^{2} \cos 2 \theta-A_{\mathrm{CC}}\right)^{2}+\left(\Delta m^{2} \sin 2 \theta\right)^{2}},
\label{31}
\end{equation}
\begin{table}[htbp]
    \centering
    \begin{tabular}{|c|c|c|c|c|}
        \hline
        $\Delta m^2_{31}$ & $\theta^\circ_{13}$ & Energy\,(E)  \\
        \hline
        $2.534^{+0.025}_{-0.023} \times 10^{-3} \,\text{eV}^2$ & $8.52^{+0.11}_{-0.11} $ & $3.5\,\rm GeV$  \\
        \hline
    \end{tabular}
    \caption{\justifying{The list of neutrino mixing parameters with the best fit$\pm 1\sigma$ errors \cite{Esteban:2024eli,NuFIT}.}}
    \label{tab1}
\end{table}
and 
\begin{equation}
\tan 2 \theta_{\mathrm{M}}=\frac{\tan 2 \theta}{1-\frac{A_{\mathrm{CC}}}{\Delta m^{2} \cos 2 \theta}},
\label{32}
\end{equation}
respectively. Using Eqs.\,(\ref{30})-(\ref{32}) in Eq.\,(\ref{29}), the time-evolved neutrino flavor state $\ket{\nu_\alpha(t)}$ in the presence of matter background can be written as
\begin{align}
&\ket{\nu_\alpha(t)}= e^{-i\mathcal{H}_{\mathrm{M}}t}\ket{\nu_\alpha(0)} \nonumber \\
&= U_M\Lambda_M(t) U^\dagger_M \ket{\nu_\alpha(0)} \nonumber \\
&= U_M \begin{pmatrix}
        e^{i\Delta m^2_M\frac{t}{4E}} & 0 \\
        0 & e^{-i\Delta m^2_M\frac{t}{4E}} \\
    \end{pmatrix} U_{M}^{\dagger} \ket{\nu_\alpha(0)} \nonumber \\
&= \begin{pmatrix}
 \tilde{\mathcal{A}}_{\alpha\alpha}(t) & \tilde{\mathcal{A}}_{\alpha\beta}(t) \\
        \tilde{\mathcal{A}}_{\beta \alpha}(t) & \tilde{\mathcal{A}}_{\beta\beta}(t) \\
\end{pmatrix} \ket{\nu_{\alpha}(0)}.
\label{33}
\end{align}

\begin{figure}[ht]
    \centering
    \includegraphics[scale=0.51]{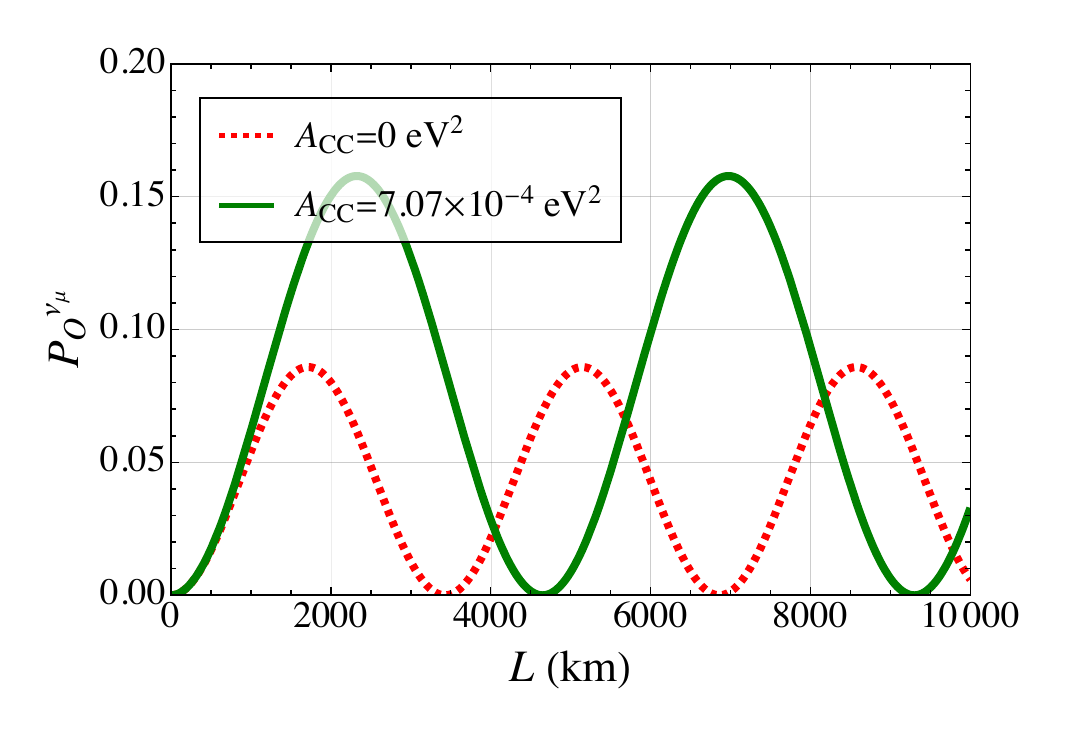}
    \caption{\justifying{In the two-flavor neutrino oscillation, the oscillation probability ($P^{\nu_\mu}_O=1-\mathcal{F}_{\nu_\mu}$) of the initial muon neutrino state $\ket{\nu_\mu}$ is depicted as a function of propagation length $L\,(\rm km)$. This representation includes vacuum ($A_{\rm CC}=0$) indicated by a red dashed line and in a constant matter potential ($A_{\rm CC}=7.07\times10^{-4} \text{eV}^2$) represented by a green solid line. The various neutrino mixing parameters used are taken from Table\,\ref{tab1}.}}
    \label{Fig2a}
\end{figure}

Thus, from Eq.\,(\ref{33}), the time-evolved flavor neutrino state $\ket{\nu_\alpha(t)}$ can be written in a coherent superposition of flavor basis in matter as
\begin{equation}
    \ket{\nu_\alpha (t)}=\tilde{\mathcal{A}}_{\alpha \alpha}(t)\ket{\nu_\alpha}+\tilde{\mathcal{A}}_{\alpha \beta}(t)\ket{\nu_\beta},
    \label{34}
\end{equation}
 where $|\tilde{\mathcal{A}}_{\alpha\alpha}(t)|^2+|\tilde{\mathcal{A}}_{\alpha \beta}(t)|^2=1$. In the ultrarelativistic approximation ($t\approx L$), the survival probability of the time-evolved flavor neutrino states $\ket{\nu_\alpha(t)}$ in the Earth-matter background (when $A_{\rm CC}\neq0$) is given as $P^{\rm Matter}_s=|\langle \nu_\alpha(L)|\nu_\alpha\rangle|^2=|\tilde{\mathcal{A}}_{\alpha\alpha}(L)|^2$. For any initial neutrino state in matter
\begin{equation}
   P_S^\text{Matter}=1-\sin^2(2\theta_{M})\sin^2(\phi_{M}),
    \label{35}
\end{equation}
where $\phi_M=\frac{\Delta m^2_{M\rm }  L}{4E}$. Since $\theta_{M}$ and $\Delta m^2_M$ are dependent on $A_{\rm CC}$, $P_S^{\rm Matter}$ may be analyzed as a function of the matter parameter $A_{\rm CC}$. When $A_{\rm CC}=0$, the survival probability of the initial state $\ket{\nu_\alpha}$ in matter, Eq.\,(\ref{35}), reduces to its vacuum counterpart $P^{\rm Vacuum}_s=|\langle \nu_\alpha(L)|\nu_\alpha \rangle|^2$ which is 
\begin{equation}
    P_S^\text{Vacuum}=1-\sin^2(2\theta)\sin^2(\phi),
    \label{36}
\end{equation}
 where $\phi=\frac{\Delta m^2 L}{4E}$. Therefore, the oscillation probability of the initial flavor neutrino state $\ket{\nu_\alpha}$ changes to $\ket{\nu_{\beta}}$ in vacuum as $P^{\rm Vacuum}_O=|\langle \nu_\beta(L)|\nu_\alpha \rangle|^2=1-P^{\rm Vacuum}_S$, where $P^{\rm Vacuum}_S+P^{\rm Vacuum}_O=1$. In the analysis of neutrino oscillation experimental data, the oscillatory term in Eq.\,(\ref{36}), $\sin^2(\phi)$, can be simplified to the following convenient form: $\sin^2(\frac{\Delta m^2 L}{4E})\equiv\sin^2(\frac{\Delta m^2 Lc^3}{4\hbar E}) $ $\rightarrow \sin^2(1.27\Delta m^2 (\text{eV}^2)\frac{L(\text{km})}{E(\text{GeV})})$.
\begin{figure}[ht]
    \centering
    \includegraphics[scale=0.51]{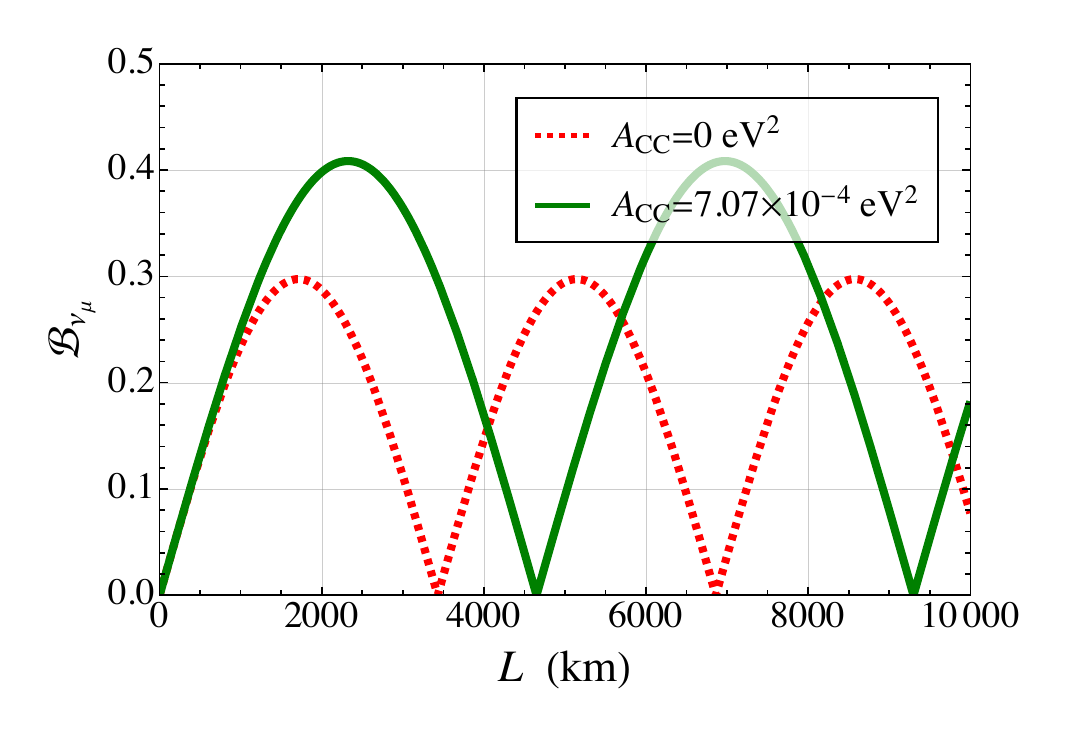}
    \caption{\justifying{In the two-flavor neutrino oscillation, the Bures angle $\mathcal{B}_{\nu_\mu}$ of the initial muon neutrino state $\ket{\nu_\mu}$ is plotted as a function of propagation length $L\,(\text{km})$. This representation includes vacuum ($A_{\rm CC}=0$) indicated by a red dashed line and in a constant matter potential ($A_{\rm CC}=7.07\times10^{-4} \text{eV}^2$) represented by a green solid line. The various neutrino mixing parameters used are taken from Table\,\ref{tab1}.}}
    \label{Fig2}
\end{figure}

Typically, for the three-flavor neutrino oscillation in vacuum, neutrino experiments are sensitive to the three mixing angles $\theta_{12}$, $\theta_{13}$, and $\theta_{23}$; the two squared-mass differences $\Delta m^2_{21}$ and $\Delta m^2_{31}$ (or $\Delta m^2_{32}$), where $\Delta m^2_{31}=\Delta m^2_{32}+\Delta m^2_{21}$; as well as the $CP$-violation phase $\delta_{CP}$ \cite{Giganti:2017fhf}. In experiments with baseline $L$ and beam energy $E$ such that $\Delta m^2_{21} L/E<<1$, a valid approximation is to set $\Delta m^2_{21}=0$ and $\delta_{CP}=0$ in the three-flavor neutrino oscillation formalism in vacuum, yielding the two-flavor neutrino oscillation approximation \cite{Giganti:2017fhf}. In this regime, two-flavor neutrino oscillations are driven by $\Delta m^2_{31}\approx\Delta m^2_{32}$, applicable to the long-baseline accelerator experiment.

Moreover, in the three-flavor neutrino oscillation, long-baseline accelerator experiments probe $\theta_{13}$ by measuring the oscillation probability of electron neutrinos in a muon neutrino beam ($\nu_\mu\rightarrow\nu_e$) \cite{T2K:2013ppw}. At leading order, under the condition $\Delta m^2_{21}L/E<<1$, the three-flavor neutrino oscillations reduce to two-flavor oscillations in vacuum, giving the oscillation probability approximately as \cite{Giganti:2017fhf,K2K:2006wtf,MINOS:2011amj} 
\begin{equation}
P(\nu_\mu\rightarrow\nu_e)\approx \sin^2(2\theta_{13})\sin^2(\Delta m^2_{31}L/4E).
\label{36a}
\end{equation}Here, we employ a two-flavor approximation, Eq.\,(\ref{36a}), that is particularly suitable for long-baseline accelerator experiments by following established treatments in Refs.\,\cite{Giganti:2017fhf,K2K:2006wtf,MINOS:2011amj}. We identify $\theta_{13}$ with the mixing angle $\theta$ and $\Delta m^2_{31}=m^2_3-m^2_1$ with $\Delta m^2$. This approximation provides a realistic description for long-baseline experiments while capturing the essential physics in vacuum. In the presence of matter, Eq.\,(\ref{36a}) can be modified by replacing the fundamental vacuum parameters $\theta_{13}$ and $\Delta m^2_{31}$ with the corresponding effective matter parameters, as defined in Eqs.\,(\ref{31}) and (\ref{32}). The list of neutrino mixing parameters used for our analysis is detailed in Table\,\ref{tab1}. For a constant matter potential [Eq.\,(\ref{A14})] and neutrino energy, $E\approx3.5\,\rm GeV$\,\cite{PhysRevD.99.095001}, we obtained
$A_{\mathrm{CC}} = 2EV_{\mathrm{CC}} \approx 7.07 \times 10^{-4}\,\mathrm{eV}^2$.

In general, comparing two quantum states and assessing their closeness is pertinent, especially in experimental settings where the observed quantum state needs to be compared with theoretical predictions. A valuable measure for this comparison is fidelity, denoted by $\mathcal{F}$. In the context of two-flavor neutrino oscillations, undergoing unitary evolution, the fidelity can be equated to the survival probability of the initial flavor neutrino states $\ket{\nu_\alpha}$ in the matter background as 
 \begin{equation}
    \mathcal{F}_{\nu_\alpha}(\ket{\nu_\alpha(0)},\ket{\nu_\alpha(L)})=P^{\rm Matter}_S.
    \label{37}
 \end{equation} 
Similarly, the oscillation probability of the initial flavor neutrino state $\ket{\nu_\alpha}$ related to the fidelity as 
 \begin{equation}
     P^{\rm {Matter}}_O=1- \mathcal{F}_{\nu_\alpha}(\ket{\nu_\alpha(0)},\ket{\nu_\alpha(L)})=1-P^{\rm Matter}_S,
 \end{equation}
 where $P^{\rm Matter}_O+P^{\rm Matter}_S=1$.
  We assume the initial state to be muon neutrino, i.e, $\alpha=\mu$. Figure\,\ref{Fig1} illustrates the fidelity $\mathcal{F}_{\nu_\mu}$ of the initial muon flavor neutrino state $\ket{\nu_\mu}$ as a function of propagation length $L\,(\text{km})$ in vacuum (red dashed line) and in a constant Earth-matter background (green solid line).  We observe that at $L=0$, $\mathcal{F}_{\nu_\mu}=1$. As $L$ increases, $\mathcal{F}_{\nu_\mu}$ varies periodically for the two different matter backgrounds. Similarly, in Fig.\,{\ref{Fig2a}}, the variation of the oscillation probability ($P^{\nu_\mu}_O$) for the initial state $\ket{\nu_\mu}$ as a function of $L\,(\rm km)$ is shown for both vacuum (red dashed line) and a constant Earth-matter potential (green solid line).  
  
  Substituting Eq.\,(\ref{37}) in Eq.\,(\ref{2}), the Bures angle of the state $\ket{\nu_\mu(t)}$ can be calculated as
\begin{equation}
\mathcal{B}_{\nu_\mu}=\cos^{-1}\sqrt{\mathcal{F}_{\nu_\mu}(\ket{\nu_\mu(0)},\ket{\nu_\mu(t)})}.
    \label{38}
\end{equation}
In Fig.\,\ref{Fig2}, the Bures angle $\mathcal{B}_{\nu_\mu}$ is plotted as a function of propagation length $L$ (in kilometers) for the initial state $\ket{\nu_\mu}$ in vacuum (red dashed line) and in the presence of a constant Earth-matter potential (green solid line). This graph illustrates the proximity between the initial and final states. We observe that the periodic behavior of $\mathcal{B}_{\nu_\mu}$ persists as the propagation length increases. The peak of {$\mathcal{B}_{\nu_\mu}$} in both vacuum and matter reaches its maximum when $\mathcal{F}_{\nu_\mu}$ (see Fig.\,\ref{Fig1}) is at its minimum and $P^{\nu_\mu}_{O}$ (see Fig.\,\ref{Fig2a}) is at its maximum.

\begin{figure}
    \centering
    \includegraphics[scale=0.5]{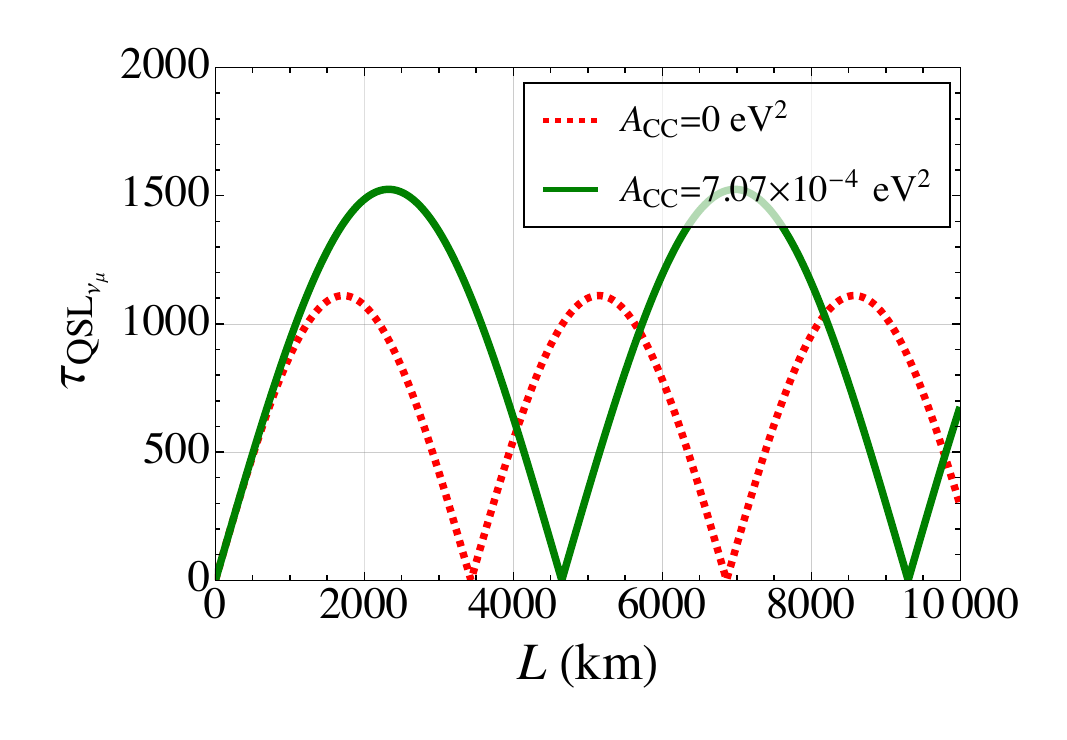}
    \caption{\justifying{In the two-flavor neutrino oscillation, we depict the QSL time $\tau_{\rm{QSL}_{\nu_\mu}}$ for the initial muon neutrino state $\ket{\nu_\mu}$ as a function of propagation length $ L\,(\text{km})$. This is shown in vacuum ($A_{\rm CC}=0$) with a red dashed line and in a constant matter potential ($A_{\rm CC}=7.07\times10^{-4} \,\text{eV}^2$) with a green solid line. The various neutrino mixing parameters used are taken from Table\,\ref{tab1}.}}
    \label{Fig3}
\end{figure}
\begin{figure}
    \centering
    \includegraphics[scale=0.5]{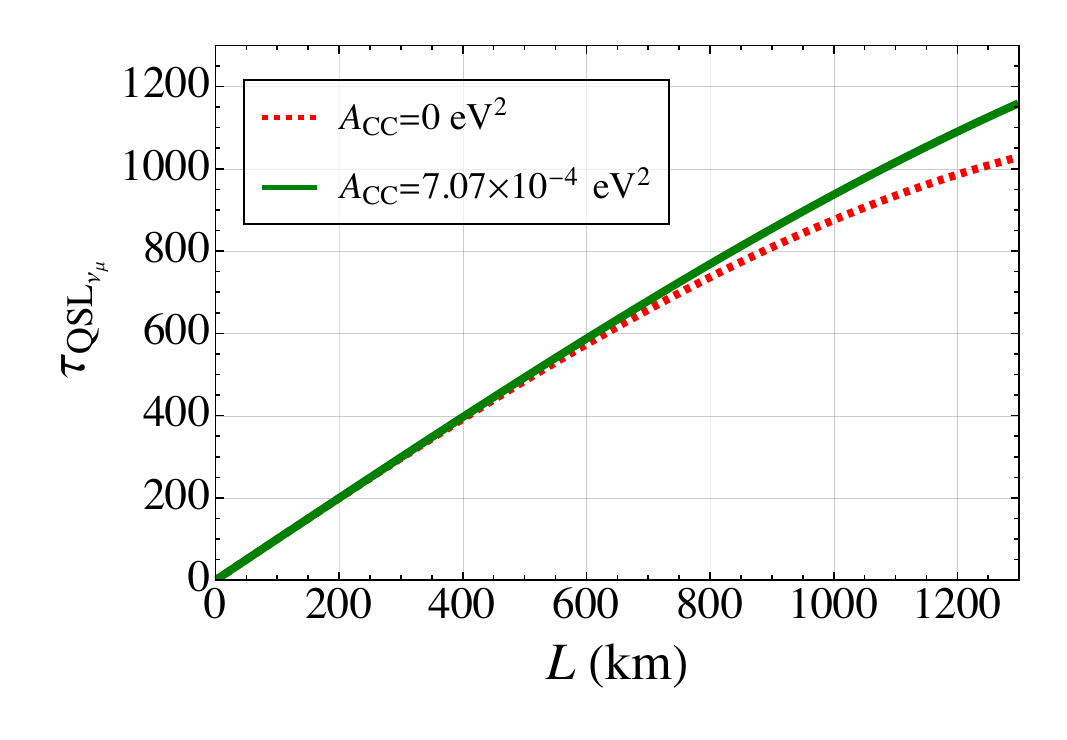}
    \caption{\justifying{In the two-flavor neutrino oscillation, we depict the QSL time $\tau_{\rm{QSL}_{\nu_\mu}}$ for the initial muon neutrino state $\ket{\nu_\mu}$ as a function of propagation length $ L\,(\text{km})$ up to a length of $L\approx 1300\, \text{km}$. This is shown in vacuum ($A_{\rm CC}=0$) with a red dashed line and in a constant matter potential ($A_{\rm CC}=7.07\times10^{-4} \,\text{eV}^2$) with a green solid line. The various neutrino mixing parameters used are taken from Table\,\ref{tab1}.}}
    \label{Fig3a}
\end{figure}

Furthermore, we utilize Eqs.\,(\ref{29}) and (\ref{34}) in Eq.\,(\ref{4}) to calculate the variance in the time-independent driving Hamiltonian ($\Delta\mathcal{H}_M$). Subsequently, employing the variance ($\Delta\mathcal{H}_M$) value, and the Bures angle, Eq.\,(\ref{38}), in Eq.\,(\ref{3}), we numerically compute the quantum speed limit time, $\tau_{\text{QSL}}$, for the initial muon neutrino flavor state $\ket{\nu_\mu}$ in vacuum and in a constant Earth-matter background. The effect of matter on $\tau_{{\rm QSL}_{\nu_\mu}}$, as a function of propagation length $ L~(\,\text{km})$, is illustrated in Fig.\,\ref{Fig3}. It is observed that $\tau_{{\rm QSL}_{\nu_\mu}}$ exhibits periodicity in vacuum  (red dashed line) and in the constant Earth-matter potential (green solid line) as the propagation length increases. We find that in both vacuum and matter, when the peak of $\mathcal{F}_{\nu_\mu}$ (see Fig.\,\ref{Fig1}) is at its minimum and $P^{\nu_\mu}_{O}$ (see Fig.\,\ref{Fig2a}) is at its maximum, the peak of $\tau_{{\rm QSL}_{\nu_\mu}}$ in Fig.\,\ref{Fig3} is also at its maximum.

Moreover, to demonstrate a realistic scenario of $\tau_{\rm QSL}$, we adopt the DUNE's baseline length $L \approx 1300\,\mathrm{km}$ and neutrino energy $E \approx 3.5\,\mathrm{GeV}$\,\cite{PhysRevD.99.095001}. Further discussion of the futuristic DUNE experiment is provided in Sec.~\ref{Sect5}.
It is worth mentioning that when the QSL time-bound condition, as defined in Eq.\,(\ref{3}), satisfies $\tau_{\rm QSL}/L=1$, the dynamic evolution of the neutrino state remains unchanged. In other words, the evolution speed has already reached its maximum. Conversely, $\tau_{\rm QSL}/L<1$ indicates the potential for the dynamic evolution of the neutrino state to speed up. The smaller the value of $\tau_{\rm QSL}/L$, the greater the potential for a quantum speedup. From Fig.\,\ref{Fig3a}, we observe that for the state $\ket{\nu_\mu(t)}$, the ratio of $\tau_{{\rm QSL}_{\nu_\mu}}/L<1$, lesser for evolution in vacuum (red dashed line) as compared to that in matter (green solid line) near $L\approx 1300\,\rm{km}$. Thus, a rapid evolution of the neutrino state is observed in vacuum (red dashed line) compared to matter (green solid line) in the DUNE setup.

\subsection{QSL time for entanglement in two-flavor neutrino oscillations in matter}
\label{Sect4B}
In this subsection, we initially employ the principles of quantum information theory to map neutrino flavor eigenstates at time $t=0$ onto a two-qubit-mode state as follows \cite{Blasone_2009,BLASONE2013320,PhysRevD.77.096002,PhysRevA.77.062304,Banerjee:2015mha,Alok:2014gya,Dixit:2017ron,Dixit:2018kev,Dixit:2019swl,Dixit:2020ize,KumarJha:2020pke,Yadav:2022grk,Li:2022mus,wang2023monogamy,Bittencourt:2023asd,Blasone:2023qqf,Quinta:2022sgq,Jha:2022yik,Dixit:2023fke,Konwar:2024nrd}: 
\begin{equation}
\ket{\nu_\alpha}\equiv\ket{1}_{\alpha}\otimes\ket{0}_{\beta}\equiv\ket{10};\hspace{0.5em} \ket{\nu_\beta}\equiv\ket{0}_{\alpha}\otimes\ket{1}_{\beta}\equiv\ket{\,01}.
\label{39}
\end{equation}
Under the ultrarelativistic approximation, neutrino flavor eigenstates can be considered as distinct modes. Entanglement represents a connection between the quantum states of the two modes. A specific form of entanglement that arises among various modes is termed ``neutrino mode (flavor) entanglement." Consequently, the time-evolved flavor state can be understood as an entangled superposition of flavor modes. This concept can be investigated within the framework of both two- and three-flavor neutrino oscillations.
 In the context of two-flavor neutrino oscillations, using Eq.\,(\ref{39}) in Eq.\,(\ref{34}), the time-evolved flavor neutrino state in the presence of a constant Earth-matter background can be represented as a two-qubit system 
\begin{equation}
    \ket{\nu_\alpha (t)}=\tilde{\mathcal{A}}_{\alpha \alpha}(t)\ket{10}+\tilde{\mathcal{A}}_{\alpha \beta}(t)\ket{01} .
    \label{40}
\end{equation}
The corresponding density matrix of the state $\ket{\nu_\alpha(t)}$ in the presence of a constant Earth-matter background is 
\begin{eqnarray} 
 &{\rho (t)=\ket{\nu_\alpha(t)}\bra{\nu_\alpha(t)}}\nonumber & \\
 &{=\begin{pmatrix}
 0 & 0 & 0 & 0\\
 0 & \vert{\tilde{\mathcal{A}}_{\alpha\alpha}(t)}\vert^2 & \tilde{\mathcal{A}}_{\alpha\alpha}(t) \tilde{\mathcal{A}}_{\alpha\beta}^*(t) & 0\\
 0 &  \tilde{\mathcal{A}}_{\alpha\beta}(t)\tilde{\mathcal{A}}_{\alpha\alpha}^*(t) & \vert{\tilde{\mathcal{A}}_{\alpha\beta}(t)}\vert^2 & 0 \\
 0 & 0 & 0 & 0\\
  \end{pmatrix}},& 
  \label{41}
 \end{eqnarray} 

\begin{figure}[ht]
    \centering
    \includegraphics[scale=0.5]{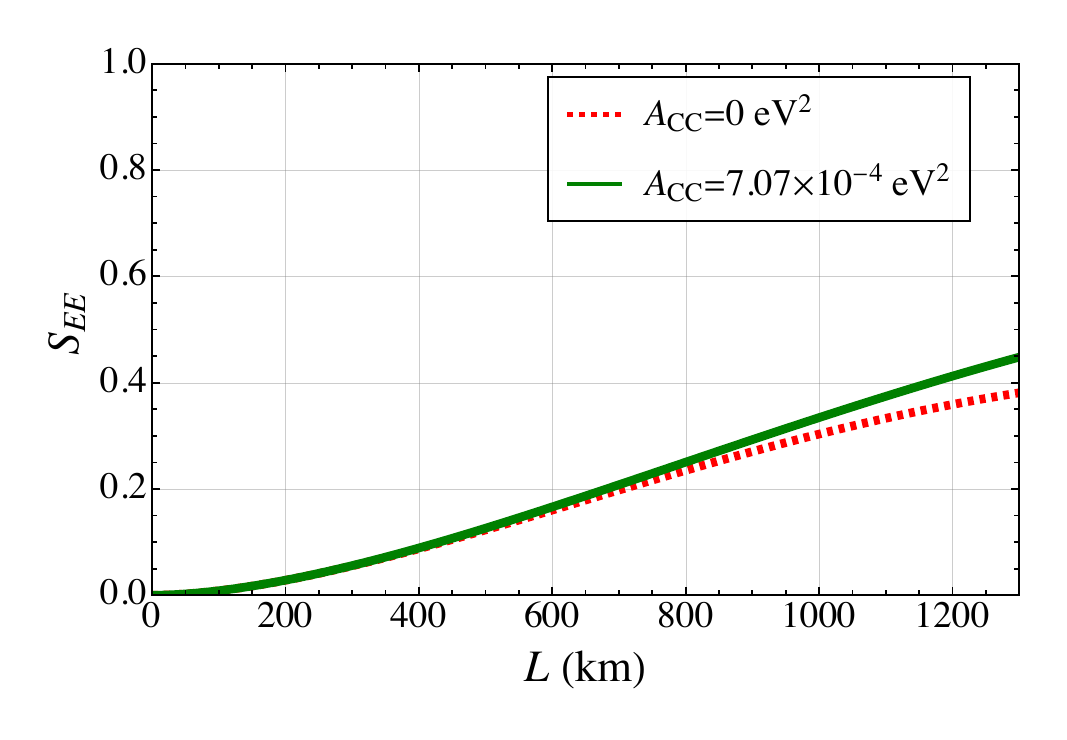}
    \caption{\justifying{In the two-flavor neutrino oscillation, the entanglement entropy ($S_{EE}(\rho_\mu(t))$), for the initial state $\ket{\nu_\mu}$, is shown as a function of propagation length $L\,(\text{km})$ up to a length of $L\approx 1300\,(\rm km)$. This is represented in vacuum ($A_{\rm CC}=0$) by a red dashed line and in a constant matter potential ($A_{\rm CC}=7.07\times10^{-4} \text{eV}^2$) by a green solid line. The various neutrino mixing parameters used are taken from Table\,\ref{tab1}.}}
    \label{Fig5}
\end{figure}
\begin{figure}
    \centering
    \includegraphics[scale=0.5]{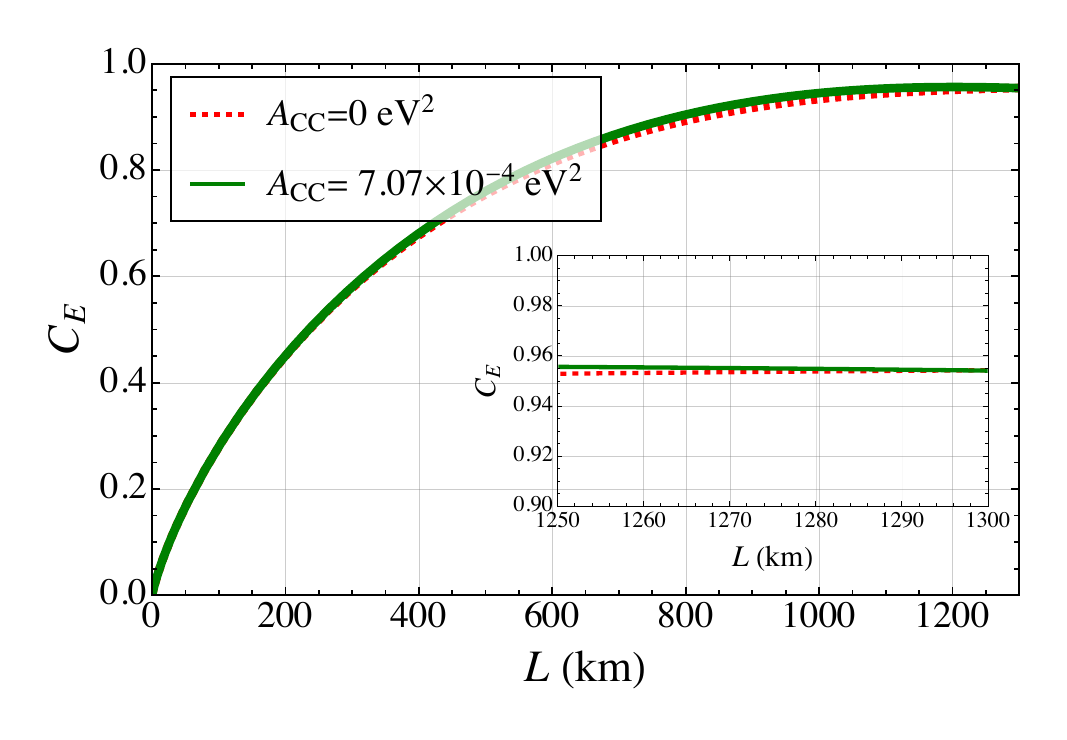}
    \caption{\justifying{In the two-flavor neutrino oscillation, the capacity of entanglement [$C_{E}(\rho_\mu(t))$], for the initial muon state $\ket{\nu_\mu}$, is depicted as a function of propagation length $L\,(\text{km})$ up to a length of $L\approx 1300\,(\rm km)$. This is represented in vacuum ($A_{\rm CC}=0$) by a red dashed line and in a constant matter potential ($A_{\rm CC}=7.07\times10^{-4} \text{eV}^2$) by a green solid line. The various neutrino mixing parameters used are taken from Table\,\ref{tab1}.}}
    \label{Fig6}
\end{figure}

where $\rho(t)=\rho^2(t)$ is an idempotent matrix and $\rm Tr\,(\rho^2(t))=1$. This implies that the state $\rho(t)$ is a pure state. However, the reduced state (by tracing over the other qubit) $\rho_{\alpha}(t)=\rm Tr_\beta(\rho(t))$ and $\rho_\beta(t)=\rm Tr_\alpha(\rho(t))$ follows 
$\rm Tr\,(\rho_\alpha^2(t))<1$ and $\rm Tr\,(\rho_\beta^2(t))<1$, which are in the mixed states. This shows that the state $\ket{\nu_\alpha(t)}$ is a bipartite quantum system of a two-qubit pure state. This is also true for the state $\ket{\nu_\beta(t)}$. The eigenvalues of the reduced state are $\rho_\alpha(t)$ are $\lambda_1,\lambda_2$. These two eigenvalues are the survival ($\lambda_1=P_S=|\tilde{\mathcal{A}}_{\alpha\alpha}(t)|^2$) and the oscillation ($\lambda_2=P_O=|\tilde{\mathcal{A}}_{\alpha\beta}(t)|^2=1-P_S$) probabilities of the initial state $\ket{\nu_\alpha}$ in matter background. Using Eq.\,(\ref{5}) and the eigenvalues of the reduced state $\rho_\alpha(t)$, the bipartite entanglement entropy $S_{EE}(t)$ of the reduced state $\rho_{\alpha}(t)$ can be calculated as 
\begin{align}
    S_{EE}(\rho_\alpha(t))=& -|\tilde{\mathcal{A}}_{\alpha\alpha}(t)|^2 \text{log}_2|\tilde{\mathcal{A}}_{\alpha\alpha}(t)|^2\nonumber\\&
    -(1-|\tilde{\mathcal{A}}_{\alpha\alpha}(t)|^2)\text{log}_2(1-|\tilde{\mathcal{A}}_{\alpha\alpha}(t)|^2).
    \label{42}
\end{align}
Consequently, using Eq.\,(\ref{7}), the capacity of entanglement $C_{E}$ of the reduced state $\rho_{\alpha}(t)$ is obtained as
\begin{align}
C_E(\rho_\alpha(t))=& |\tilde{\mathcal{A}}_{\alpha\alpha}(t)|^2\text{log}^2_2|\tilde{\mathcal{A}}_{\alpha\alpha}(t)|^2+(1-|\tilde{\mathcal{A}}_{\alpha\alpha}(t)|^2)\text{log}^2_2\nonumber\\&
    (1-|\tilde{\mathcal{A}}_{\alpha\alpha}(t)|^2)
    -(-|\tilde{\mathcal{A}}_{\alpha\alpha}(t)|^2 \text{log}_2|\tilde{\mathcal{A}}_{\alpha\alpha}(t)|^2\nonumber\\
    &-(1-|\tilde{\mathcal{A}}_{\alpha\alpha}(t)|^2)\text{log}_2(1-|\tilde{\mathcal{A}}_{\alpha\alpha}(t)|^2))^2.
    \label{43}
\end{align}
The bipartite entanglement measures quantified in Eqs.\,(\ref{42}) and (\ref{43}) are directly associated with two measurable quantities, survival ($P_S$) and oscillation ($P_O$) probabilities, enabling straightforward connections with experimentally determined values of neutrino mixing parameters. One of the prominent bipartite entanglement measures, concurrence, can be related to $P_S$ and $P_O$ of two-flavor neutrino oscillations as $C(\rho_\alpha(t))=2\sqrt{P_S P_O}$ \cite{Alok:2014gya}. This implies $P_S=1-\frac{1}{2}\sqrt{1-C^2(\rho_\alpha(t))}$ and $P_O=\frac{1}{2}\sqrt{1-C^2(\rho_\alpha(t))}$. Hence, the entanglement entropy $S_{EE}(\rho_\alpha(t))$, Eq.\,(\ref{42}), can be reexpressed in terms of concurrence as 
\begin{align}
    S_{EE}(C(\rho_\alpha(t)))&=(1-\frac{1}{2}\sqrt{1-C^2(\rho_\alpha(t))})\nonumber\\
    &\text{log}_2(1-\frac{1}{2}\sqrt{1-C^2(\rho_\alpha(t))})
    -\frac{1}{2}\sqrt{1-C^2(\rho_\alpha(t))}\nonumber\\
    &\text{log}_2(\frac{1}{2}\sqrt{1-C^2(\rho_\alpha(t))}).
    \label{44}
\end{align}
Since $P_S>0$ implies $P_O<1$, this in term implies  $C(\rho_\alpha(t))\neq0$. Consequently, $S_{EE}(C(\rho_\alpha(t)))\neq0$. Similarly, one can also relate capacity of entanglement, Eq.\,(\ref{43}), as a function of concurrence. In this vein, various other bipartite entanglement measures, such as Bell's inequality violation, teleportation fidelity, geometric discord, tangle, linear entropy, negativity, among others, have previously been investigated and simplified in terms of two-flavor neutrino survival and oscillation probabilities \cite{Blasone_2009,Alok:2014gya,KumarJha:2020pke}.
\begin{figure}
    \centering
    \includegraphics[scale=0.5]{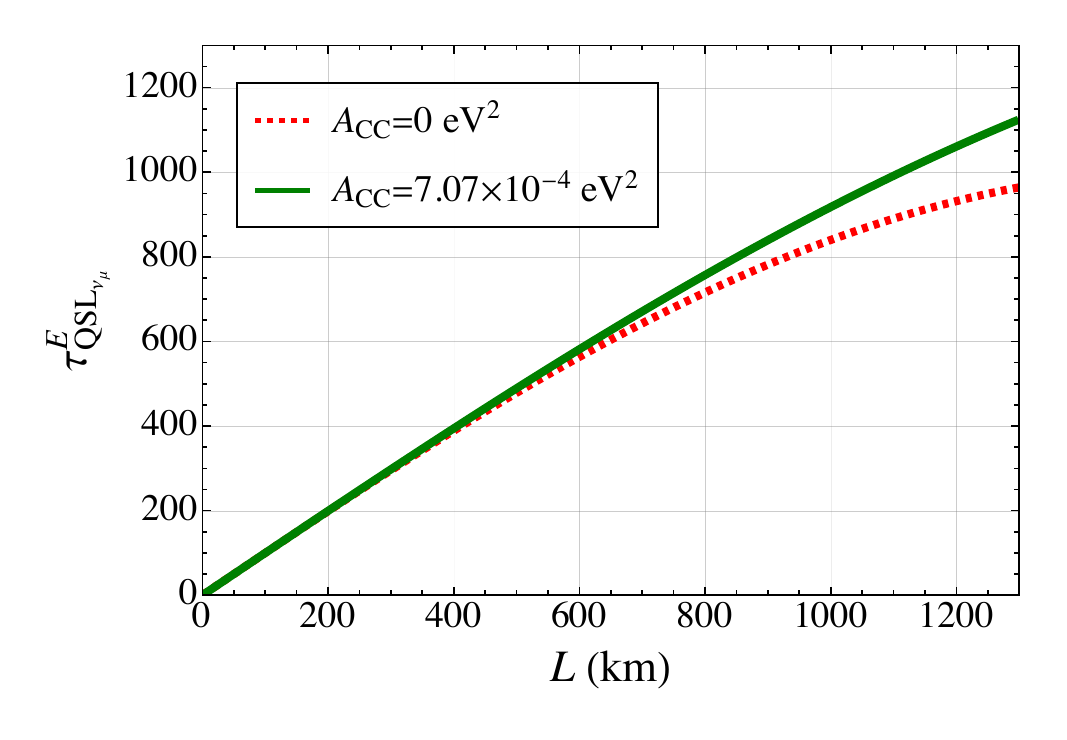}
    \caption{\justifying{
    QSL time for entanglement entropy  $\tau^E_{{QSL}_{\nu_\mu}}$ of the initial muon state $\ket{\nu_\mu}$ is shown as a function of propagation length $L\,(\text{km})$ up to a length of $L\approx 1300\,(\rm km)$ for two-flavor neutrino oscillations. This is represented in vacuum ($A_{\rm CC}=0$) by a red dashed line and in a constant matter potential ($A_{\rm CC}=7.07\times10^{-4} \text{eV}^2$) by a green solid line. The various neutrino mixing parameters used are taken from Table\,\ref{tab1}.}}
    \label{Fig7}
\end{figure}

In Figs.\,\ref{Fig5} and \ref{Fig6}, in the ultrarelativistic approximation ($t\approx L$), for the DUNE's baseline length ($L\approx 1300\,\rm km$) and energy ($E\approx 3.5\,\rm GeV$), the entanglement entropy $S_{EE}(\rho_\mu(t))$ and the capacity of entanglement $C_E(\rho_\mu(t))$, respectively, are plotted as a function of propagation length ($L$ in kilometers) for the initial muon flavor neutrino state $\ket{\nu_\mu}$, particularly in vacuum (red dashed line) and in the presence of a constant Earth-matter potential (green solid line). We observe that at $L=0$, the entanglement entropy [$S_{EE}(\rho_\mu(t))$] is zero, indicating that $\ket{\nu_\mu(t)}$ is in a separable state. Moreover, as $L$ increases ($L>0$), $S_{EE}(\rho_\mu(t))$ increases
across vacuum and matter backgrounds. Additionally, we observe a separation in entanglement entropy in the presence of matter compared to vacuum near $L\approx 1300\,\rm km$. In Fig.\,\ref{Fig6}, the capacity of entanglement $C_E$ (or the variance in the entanglement entropy) approaches zero when neutrino oscillates in vacuum (red dashed line) and matter (green solid line), in a separable state limit. However, $C_E$ attains a maximum in a partially entangled state. 
Hence, in the two-qubit system, at $L>0$, $S_{EE}(\rho_\mu(t))\neq0$ and $C_{E}(\rho_\mu(t))\neq0$ identify that the time-evolved muon flavor neutrino state $\ket{\nu_\mu(t)}$ is a bipartite pure entangled state, similar to the two-qubit entangled state used in quantum information. 

Moreover, in Fig.\,(\ref{Fig7}), using Eqs.\,(\ref{42}) and (\ref{43}) in Eq.\,(\ref{6}) with the $\Delta \mathcal{H}_{M}$ value, the QSL time for the entanglement entropy $\tau^{E}_{\rm QSL_{\mu}}$ of $\ket{\nu_\mu(t)}$ is estimated as a function of propagation length ($L\approx 1300\,\rm km$) across vacuum (red dashed line) and in a constant Earth-matter background (green solid line). Under the QSL time-bound condition for $S_{EE}$ i.e., $\tau^E_{\rm QSL_{\mu}}/L<1$, we observe a quick suppression of entanglement in vacuum (red dashed line) compared to matter (green solid line) at the DUNE's length scale and energy.

\section{Quantum speed limit time for evolution of the three-flavor neutrino state in matter}
\label{Sect5}
In the case of three-flavor neutrino oscillations in matter, we follow the formalism of the effective Hermitian Hamiltonian proposed in Ref.~\cite{Nguyen:2022snr}. The time-evolved muon-flavor neutrino state in the presence of matter can be expressed as \cite{Denton_2016,Barenboim_2019,Nguyen:2022snr}
\begin{equation}
     \ket{\nu_\mu (t)}=\tilde{\mathcal{A}}_{\mu e}(t)\ket{\nu_e}+\tilde{\mathcal{A}}_{\mu \mu}(t)\ket{\nu_\mu} + \tilde{\mathcal{A}}_{\mu\tau}(t)\ket{\nu_\tau},
     \label{45}
\end{equation} 
 where $|\tilde{\mathcal{A}}_{\mu \mu}(t)|^2+ |\tilde{\mathcal{A}}_{\mu e}(t)|^2+|\tilde{\mathcal{A}}_{\mu \tau}(t)|^2=1$. As a result, the survival and oscillation probabilities of the initial muon flavor neutrino state $\ket{\nu_\mu}$ in matter are determined as 
 \begin{equation}
P_{\mu\rightarrow\mu}=|\tilde{\mathcal{A}}_{\mu \mu}(t)|^2, \hspace{0.1cm} P_{\mu\rightarrow e}=|\tilde{\mathcal{A}}_{\mu e}(t)|^2,\hspace{0.1cm} \rm{and}\hspace{0.1cm} P_{\mu \rightarrow\tau}=|\tilde{\mathcal{A}}_{\mu \tau}(t)|^2,   
  \label{45a}
 \end{equation}
 respectively. The issue of neutrino mass ordering is closely linked to current experimental data on neutrino survival and oscillation probabilities, permitting two potential classes of solutions \cite{Giganti:2017fhf}. In the normal ordering (NO) scenario, the two lightest mass eigenstates demonstrate a marginal mass difference, while the third eigenstate is heavier such that  $m_3 > m_2 >m_1$. Conversely, in the inverted ordering (IO), the lightest mass eigenstate precedes a doublet of higher mass eigenstates. Within this doublet, there again exists a mass difference. Hence, $m_3 < m_1 < m_2$. The neutrino mixing parameters $\Delta m^2_{lk}$, $\theta_{ij}$, and $\delta_{\rm CP}$, along with their $1\sigma$ best-fit data for normal and inverted ordering, are presented in Tables\,\ref{tab2} and \ref{tab3}, respectively. Note that here $\Delta m^2_{3l} = \Delta m^2_{31} > 0$ for NO and $\Delta m^2_{3l} = \Delta m^2_{32} < 0$ for IO \cite{Esteban:2024eli,NuFIT}.

  \begin{table}[h!]
\centering
\begin{tabular}{ |p{0.3\linewidth}|p{0.3\linewidth}|}
 \hline
  Parameters & Best fit$\pm 1\sigma$  \\
\hline
$\theta_{12}/^{\circ}$ & $33.68^{+0.73}_{-0.70}$  \\
 \hline
 $\theta_{23}/^\circ$ &   $48.5^{+0.7}_{-0.9}$\\
 \hline
$\theta_{13}/^\circ$ & $8.52^{+0.11}_{-0.11}$ \\
 \hline
 $\frac{\Delta m^2_{21}}{10^{-5} eV^2}$ & $7.49^{+0.19}_{-0.19}$\\
 \hline
  $\frac{\Delta m^2_{3l}}{10^{-3} eV^2}$ & $+2.534^{+0.025}_{-0.023}$ \\
 \hline
 $\delta_{\rm CP}/^{\circ}$ & $177^{+19}_{-20}$\\
 \hline
\end{tabular}
\caption{\justifying{ In the three-flavor neutrino oscillation, the values of the neutrino mixing parameters for NO that we considered in our analysis are taken from Refs.\,\cite{Esteban:2024eli,NuFIT} along with their corresponding $1\sigma$ errors ($90\%$ CL)}.}
\label{tab2}
\end{table}
\begin{table}[h!]
\centering
\begin{tabular}{ |p{0.3\linewidth}|p{0.3\linewidth}|}
 \hline
  Parameters & Best fit$\pm 1\sigma$  \\
\hline 
$\theta_{12}/^{\circ}$ & $33.68^{+0.73}_{-0.70}$  \\
 \hline
 $\theta_{23}/^\circ$ &   $48.6^{+0.7}_{-0.9}$\\
 \hline
$\theta_{13}/^\circ$ & $8.58^{+0.11}_{-0.11}$ \\
 \hline
 $\frac{\Delta m^2_{21}}{10^{-5} eV^2}$ & $7.49^{+0.19}_{-0.19}$\\
 \hline
  $\frac{\Delta m^2_{3l}}{10^{-3} eV^2}$ & $-2.510^{+0.024}_{-0.025}$ \\
 \hline
  $\delta_{\rm CP}/^{\circ}$ & $285^{+25}_{-28}$\\
 \hline
\end{tabular}
\caption{\justifying{In the three-flavor oscillation, the values of the neutrino mixing parameters for IO that we considered in our analysis are taken from Refs.\,\cite{Esteban:2024eli,NuFIT} along with their corresponding $1\sigma$ errors ($90\%$ CL).}}
\label{tab3}
\end{table}

\begin{figure*}[!htbp]
  \centering
  \begin{subfigure}[b]{0.33\textwidth}
    \centering
    \includegraphics[width=\textwidth]{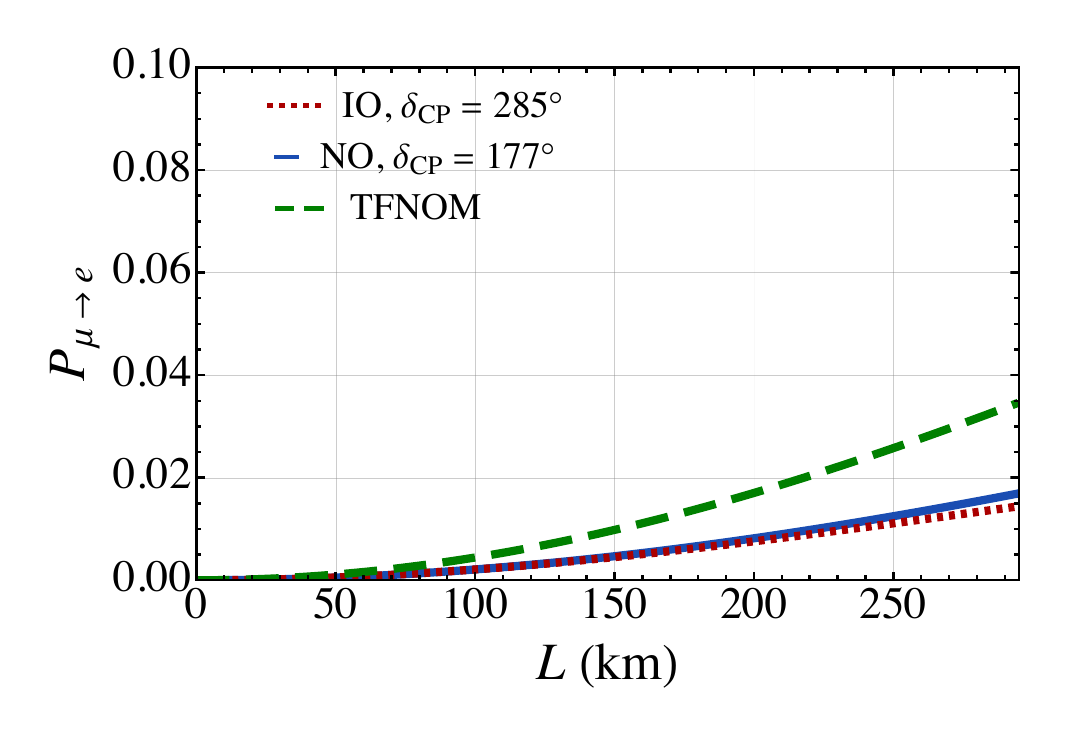}
    \caption{T2K, $\delta_{\rm CP}=177^\circ$, NO (blue solid line), $\delta_{\rm CP}=285^\circ$, IO (red dashed line), and TFNOM (green dashed line)}
    \label{8a_fig:sub1}
  \end{subfigure}
  \hfill
  \begin{subfigure}[b]{0.33\textwidth}
    \centering
    \includegraphics[width=\textwidth]{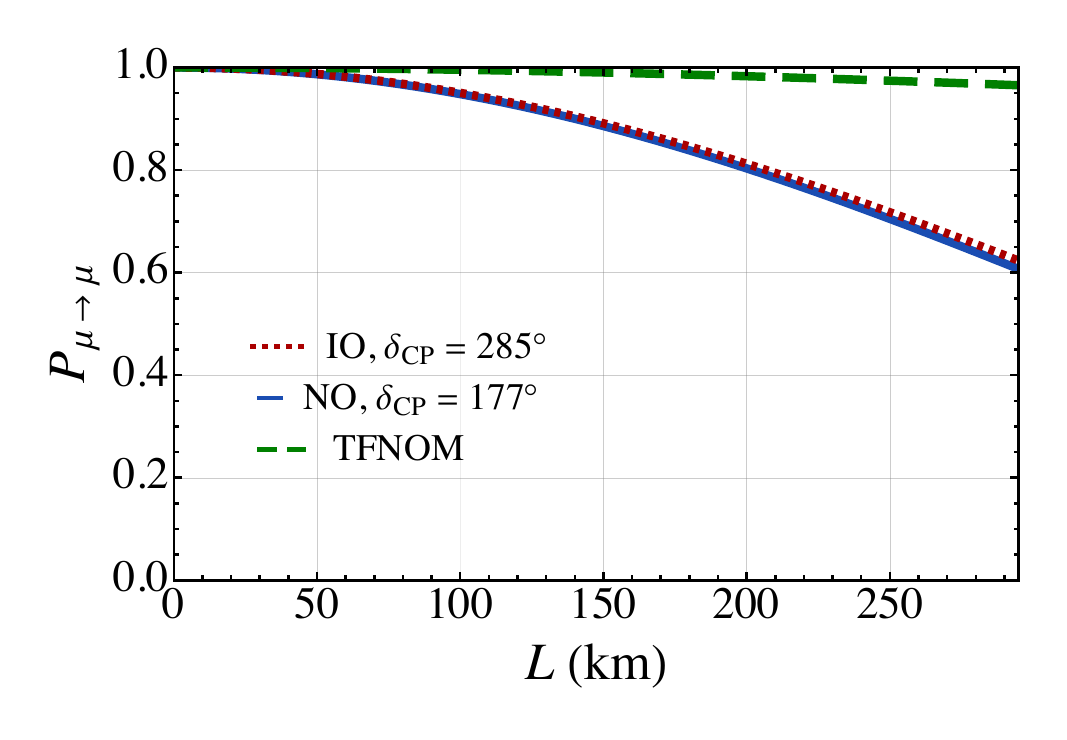}
    \caption{T2K, $\delta_{\rm CP}=177^\circ$, NO (blue solid line), $\delta_{\rm CP}=285^\circ$, IO (red dashed line), and TFNOM (green dashed line)}
    \label{8a_fig:sub2}
  \end{subfigure}
   \begin{subfigure}[b]{0.33\textwidth}
    \centering
    \includegraphics[width=\textwidth]{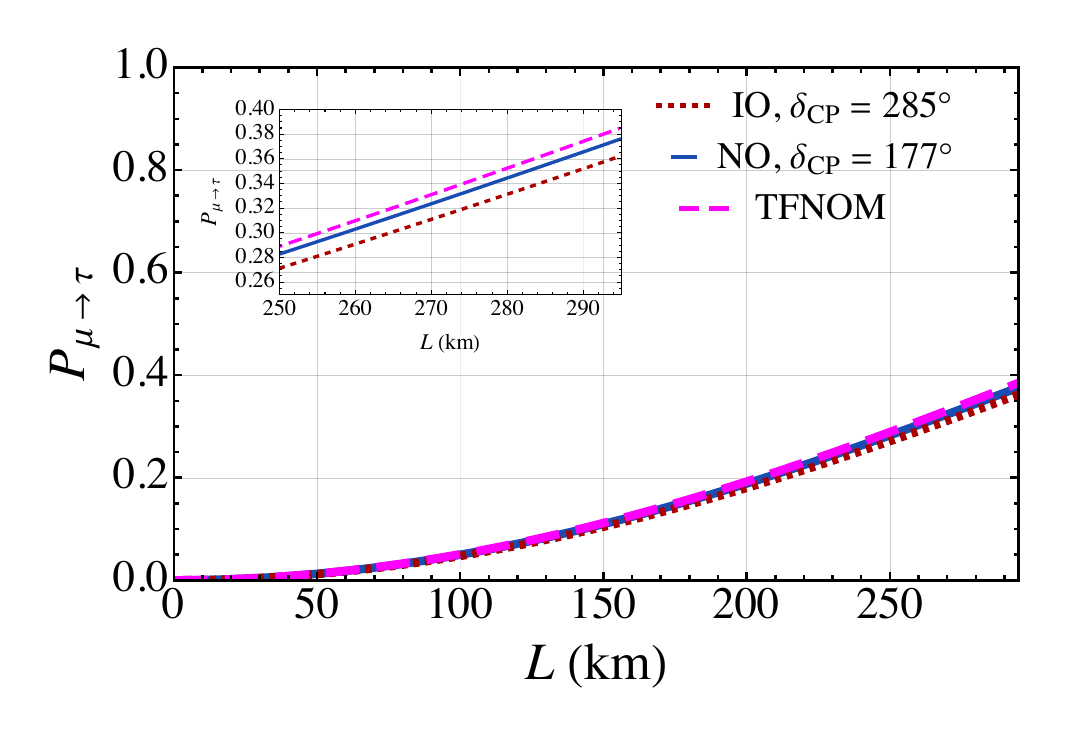}
    \caption{T2K, $\delta_{\rm CP}=177^\circ$, NO (blue solid line), $\delta_{\rm CP}=285^\circ$, IO (red dashed line), and TFNOM (magenta dashed line)}
    \label{8a_fig:sub3}
  \end{subfigure}
  \hfill
  \\
  \begin{subfigure}[b]{0.33\textwidth}
    \centering
    \includegraphics[width=\textwidth]{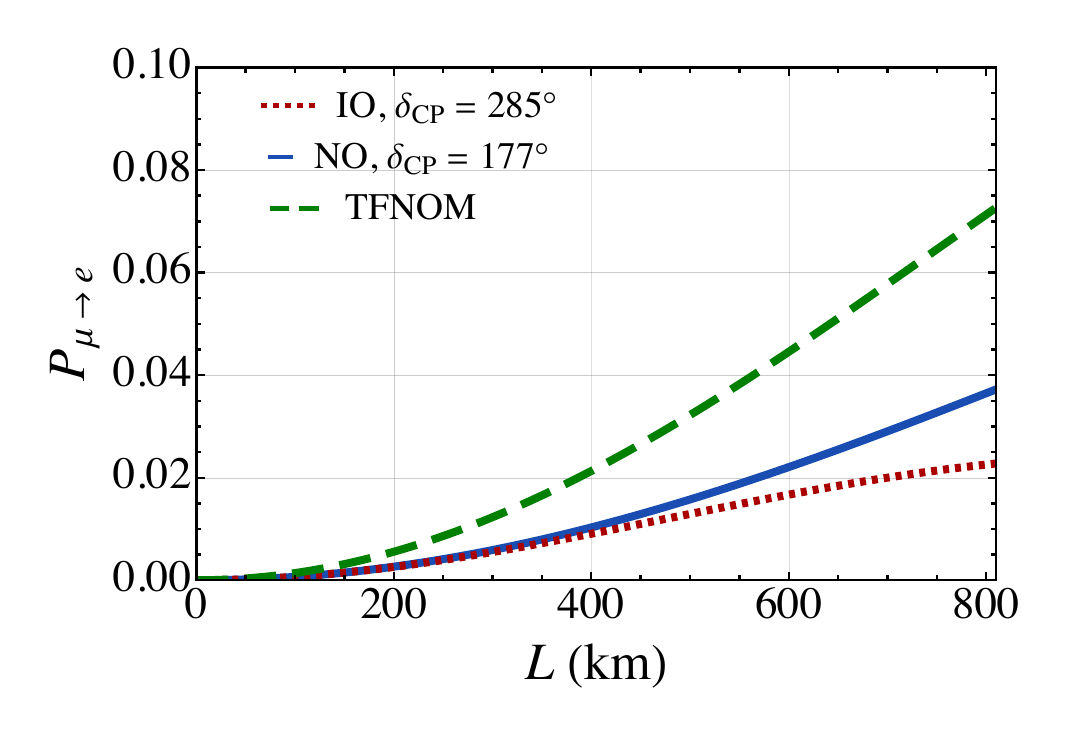}
    \caption{NOvA, $\delta_{\rm CP}=177^\circ$, NO (blue solid line), $\delta_{\rm CP}=285^\circ$, IO (red dashed line), and TFNOM (green dashed line)}
    \label{8a_fig:sub4}
  \end{subfigure}
  \hfill
  \begin{subfigure}[b]{0.33\textwidth}
    \centering
    \includegraphics[width=1.001\textwidth]{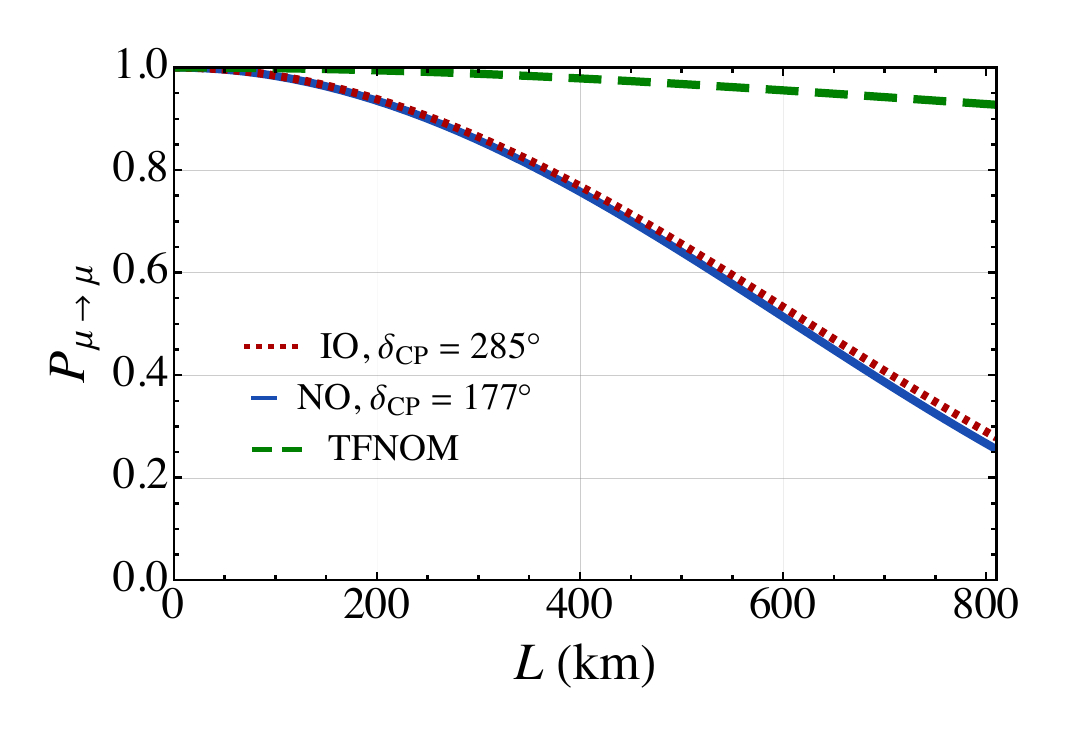}
    \caption{NOvA, $\delta_{\rm CP}=177^\circ$, NO (blue solid line), $\delta_{\rm CP}=285^\circ$, IO (red dashed line), and TFNOM (green dashed line)}
    \label{8a_fig:sub5}
  \end{subfigure}
  \hfill
  \begin{subfigure}[b]{0.33\textwidth}
    \centering
    \includegraphics[width=\textwidth]{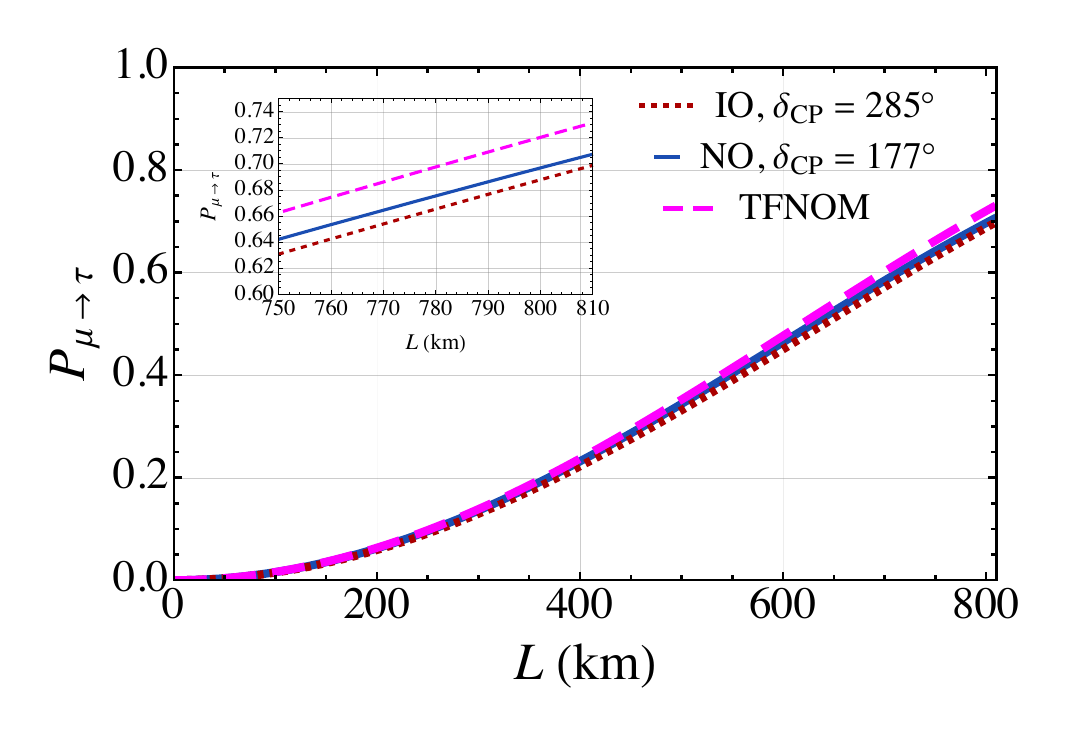}
    \caption{NOvA, $\delta_{\rm CP}=177^\circ$, NO (blue solid line), $\delta_{\rm CP}=285^\circ$, IO (red dashed line), and TFNOM (magenta dashed line)}
    \label{8a_fig:sub6}
  \end{subfigure}
\hfill
\\  
\begin{subfigure}[b]{0.33\textwidth}
    \centering
    \includegraphics[width=\textwidth]{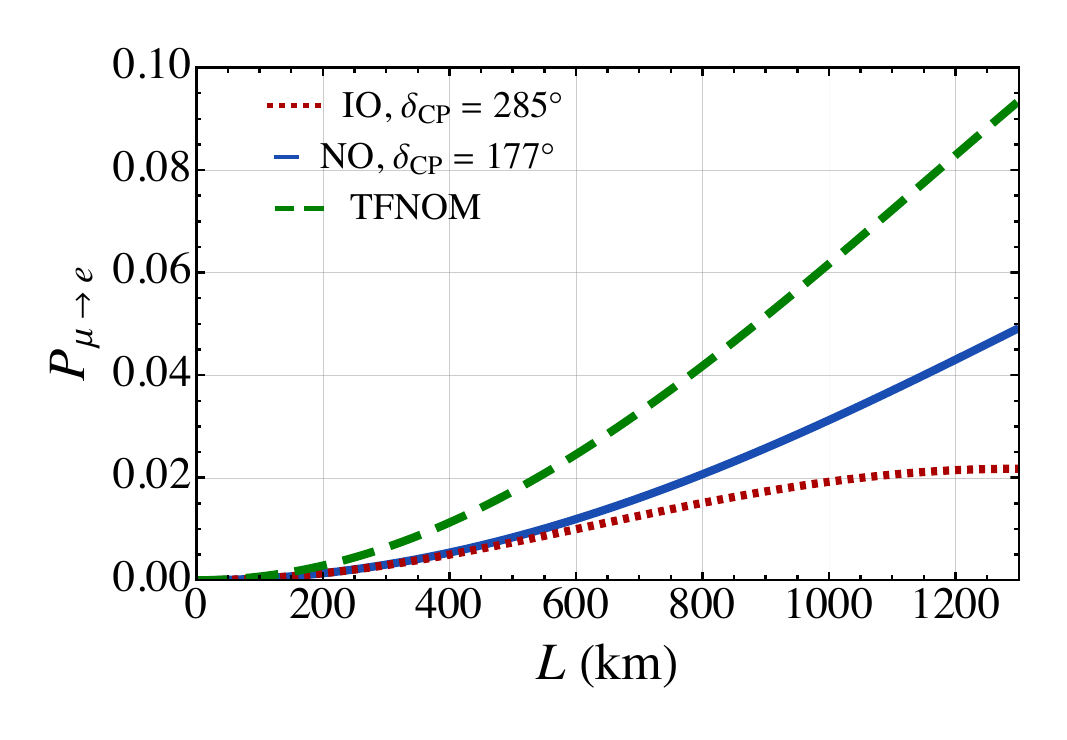}
    \caption{DUNE, $\delta_{\rm CP}=177^\circ$, NO (blue solid line), $\delta_{\rm CP}=285^\circ$, IO (red dashed line), and TFNOM (green dashed line)}
    \label{8a_fig:sub7}
  \end{subfigure}
  \hfill
  \begin{subfigure}[b]{0.33\textwidth}
    \centering
    \includegraphics[width=\textwidth]{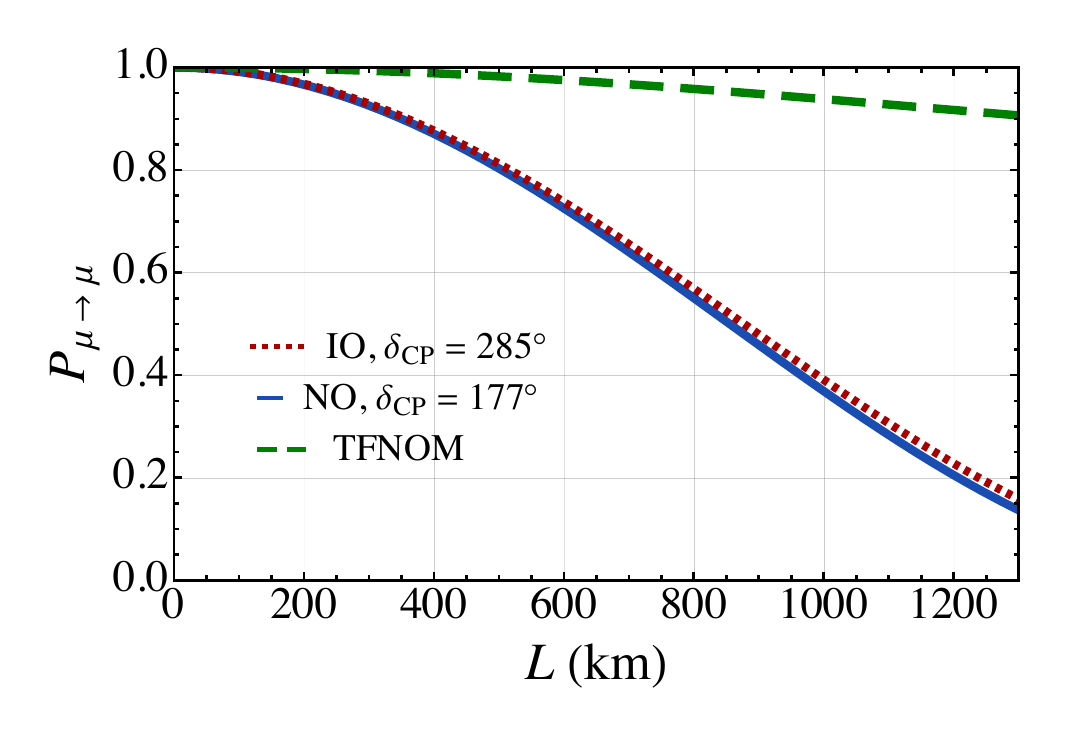}
    \caption{DUNE, $\delta_{\rm CP}=177^\circ$, NO (blue solid line), $\delta_{\rm CP}=285^\circ$, IO (red dashed line), and TFNOM (green dashed line)}
    \label{8a_fig:sub8}
  \end{subfigure}
  \hfill
  \begin{subfigure}[b]{0.33\textwidth}
    \centering
    \includegraphics[width=\textwidth]{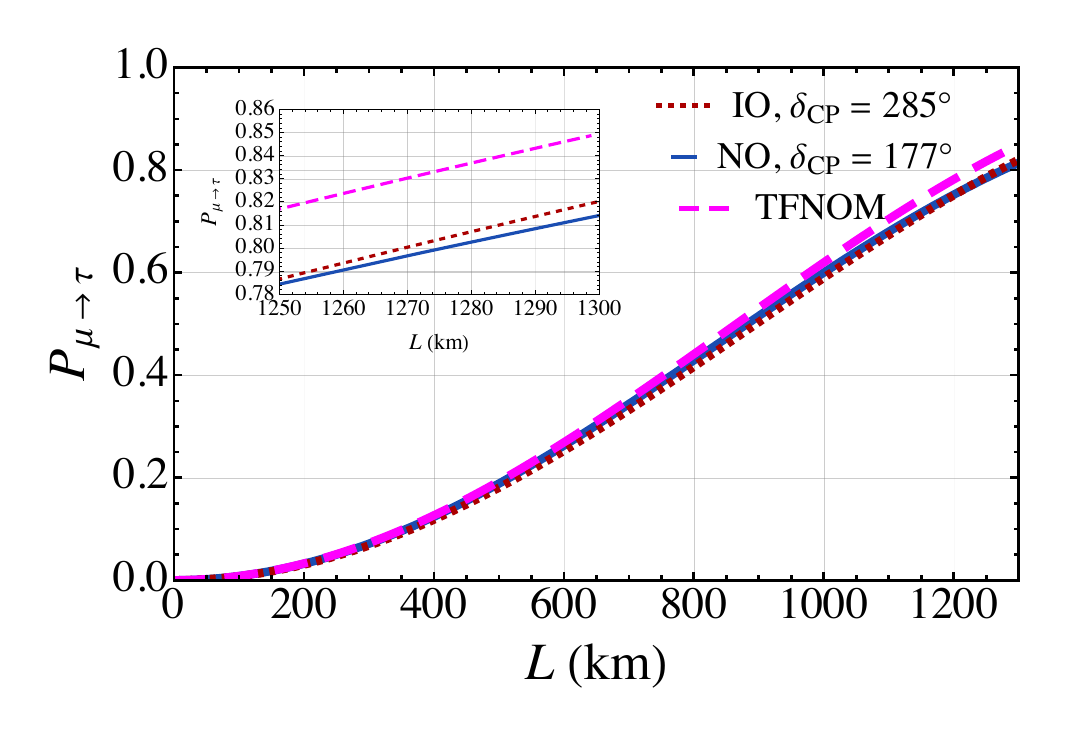}
    \caption{DUNE, $\delta_{\rm CP}=177^\circ$, NO (blue solid line), $\delta_{\rm CP}=285^\circ$, IO (red dashed line), and TFNOM (magenta dashed line)}
    \label{8a_fig:sub9}
  \end{subfigure}
  \caption{\justifying{In three-flavor neutrino oscillations, we depict the transition probabilities $P_{\mu\rightarrow e}$,  $P_{\mu\rightarrow \mu}$, and  $P_{\mu\rightarrow \tau}$ as functions of the propagation length $L\,\text{(km)}$ for the initial muon flavor neutrino state $\ket{\nu_\mu}$ and compare them between NO and IO. These comparisons are shown using the length scales and energies of T2K, NOvA and DUNE in Figs.\,\ref{8a_fig:sub1}, \ref{8a_fig:sub2}, \ref{8a_fig:sub3}, Figs.\,\ref{8a_fig:sub4}, \ref{8a_fig:sub5}, \ref{8a_fig:sub6}, and Figs.\,\ref{8a_fig:sub7}, \ref{8a_fig:sub8}, \ref{8a_fig:sub9}, respectively, under two different $CP$-violation phases and matter potentials. In the TFNOM, the transition probabilities $P_{\mu \rightarrow e}$, $P_{\mu \rightarrow \mu}$, and $P_{\mu \rightarrow \tau}$ as functions of the propagation length $L\,\text{(km)}$ for the initial muon flavor neutrino state $\ket{\nu_\mu}$ are also illustrated using the length scales and energies of T2K, NOvA and DUNE in Figs.\,\ref{8a_fig:sub1}, \ref{8a_fig:sub2}, \ref{8a_fig:sub3}, Figs.\,\ref{8a_fig:sub4}, \ref{8a_fig:sub5}, \ref{8a_fig:sub6}, and Figs.\,\ref{8a_fig:sub7}, \ref{8a_fig:sub8}, \ref{8a_fig:sub9}, respectively. The various neutrino mixing parameters used are taken from Tables\,\ref{tab2} for TFNOM and NO, and \ref{tab3} for IO along with their corresponding 1$\sigma$ errors ($90\%$ CL).}}
  \label{fig8a}
\end{figure*}

\begin{figure*}[!htbp]
  \centering
  \begin{subfigure}[b]{0.33\textwidth}
    \centering
    \includegraphics[width=\textwidth]{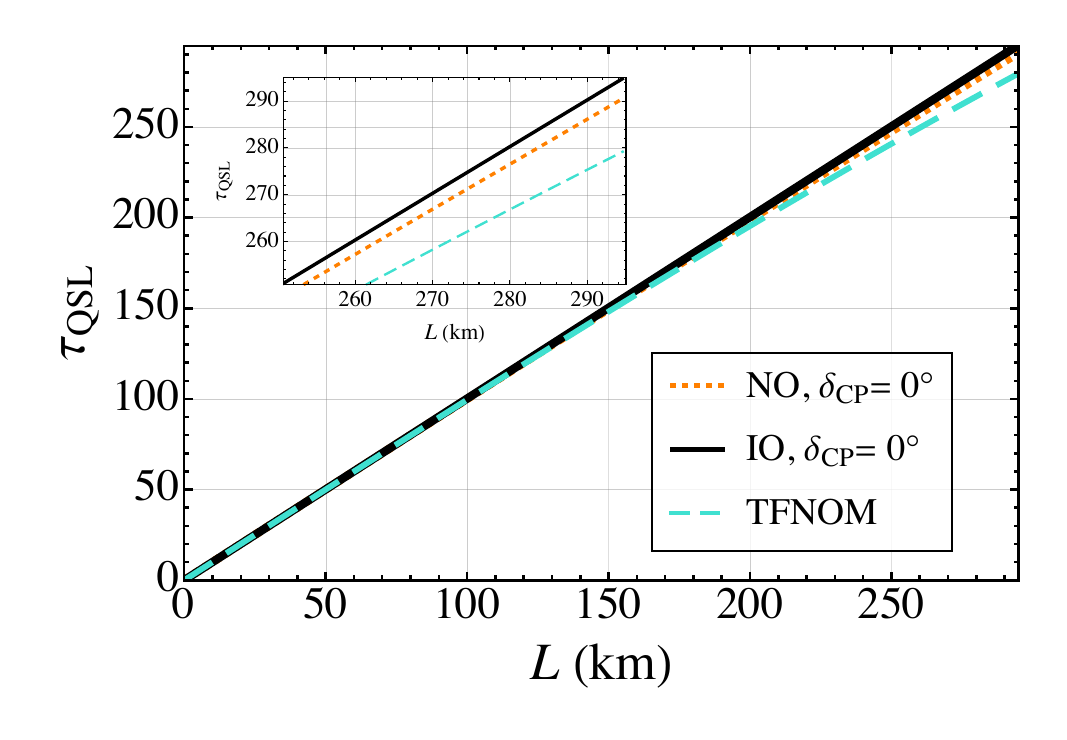}
    \caption{T2K, $\delta_{\rm CP}=0^\circ$, NO (orange dotted line), IO (black solid line), and TFNOM (cyan dashed line)}
    \label{8_fig:sub7}
  \end{subfigure}
  \hfill
  \begin{subfigure}[b]{0.33\textwidth}
    \centering
    \includegraphics[width=\textwidth]{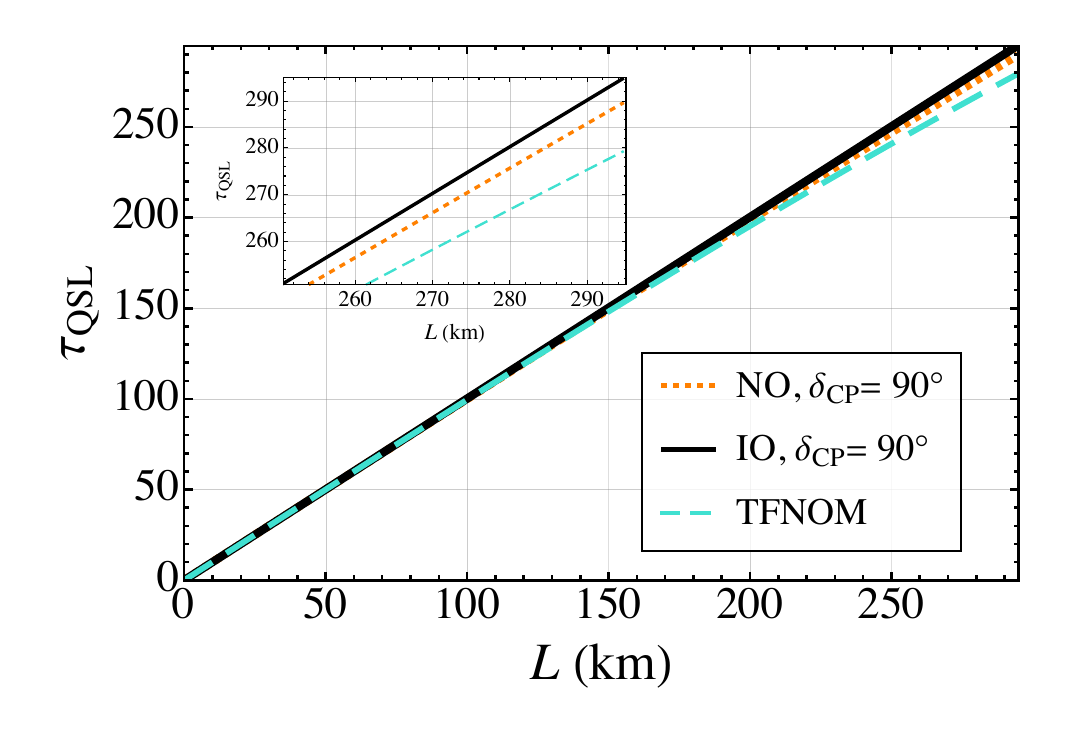}
    \caption{T2K, $\delta_{\rm CP}=90^\circ$, NO (orange dotted line), IO (black solid line), and TFNOM (cyan dashed line)}
    \label{8_fig:sub8}
  \end{subfigure}
  \hfill
  \begin{subfigure}[b]{0.33\textwidth}
    \centering
    \includegraphics[width=\textwidth]{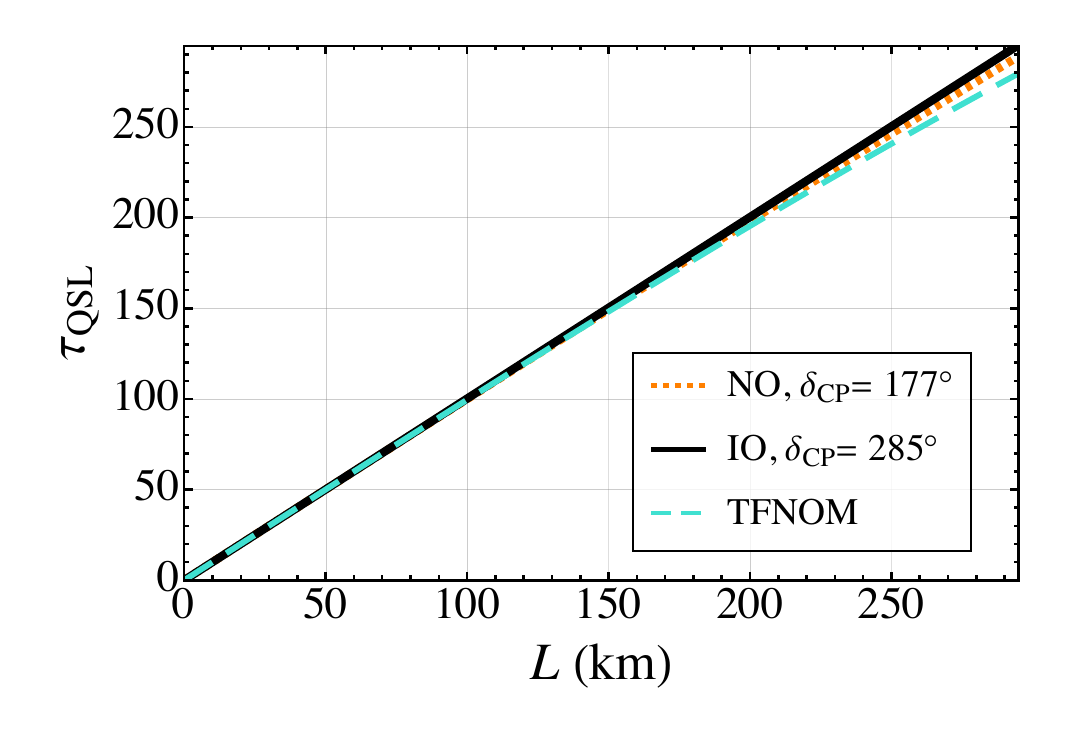}
    \caption{T2K, $\delta_{\rm CP}=177^\circ$, NO (orange dotted line), $\delta_{\rm CP}=285^\circ$, IO (black solid line), and TFNOM (cyan dashed line)}
    \label{8_fig:sub9}
  \end{subfigure}
  \hfill
  \\
  \begin{subfigure}[b]{0.33\textwidth}
    \centering
    \includegraphics[width=\textwidth]{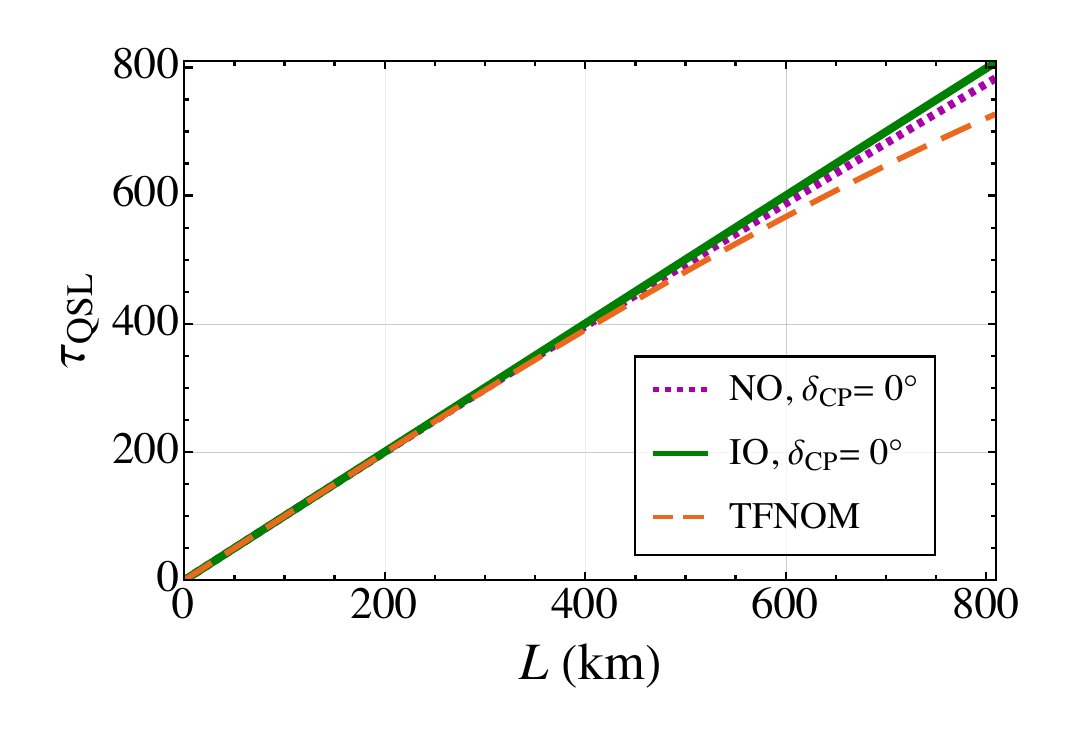}
    \caption{NOvA, $\delta_{\rm CP}=0^\circ$, NO (magenta dotted line), IO (green solid line), and TFNOM (orange dashed line)}
    \label{8_fig:sub4}
  \end{subfigure}
  \hfill
  \begin{subfigure}[b]{0.33\textwidth}
    \centering
    \includegraphics[width=1.001\textwidth]{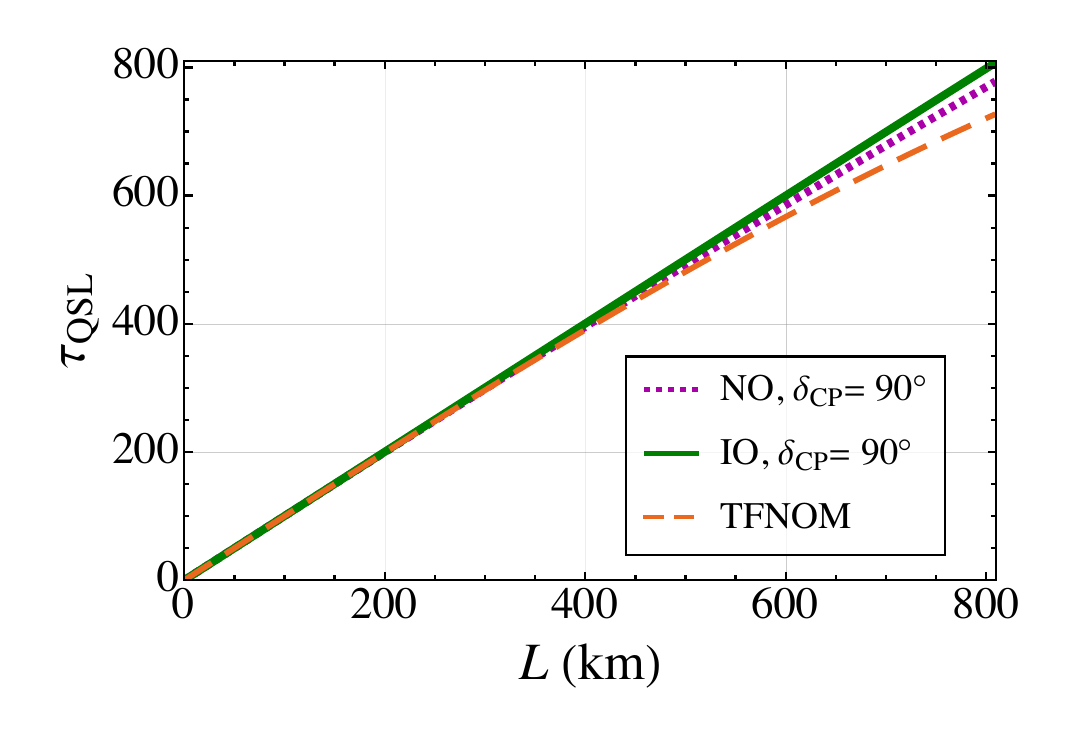}
    \caption{NOvA, $\delta_{\rm CP}=90^\circ$, NO (magenta dotted line), IO (green solid line), and TFNOM (orange dashed line)}
    \label{8_fig:sub5}
  \end{subfigure}
  \hfill
  \begin{subfigure}[b]{0.33\textwidth}
    \centering
    \includegraphics[width=\textwidth]{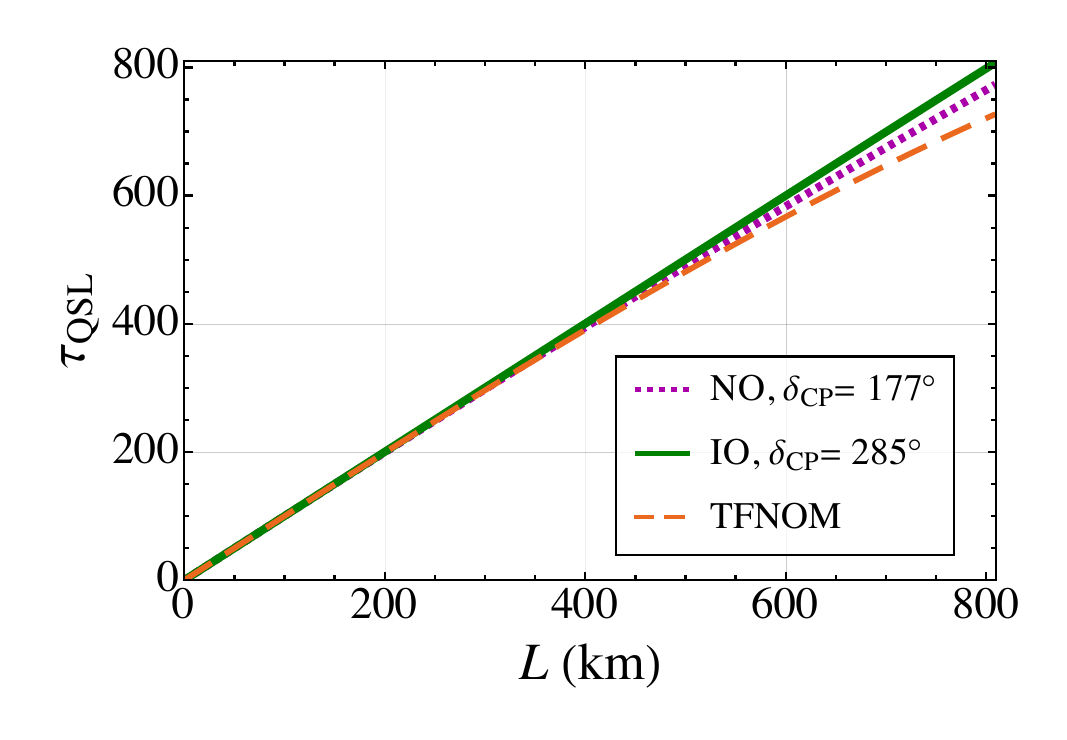}
    \caption{NOvA, $\delta_{\rm CP}=177^\circ$, NO (magenta dotted line), $\delta_{\rm CP}=285^\circ$, IO (green solid line), and TFNOM (orange dashed line)}
    \label{8_fig:sub6}
  \end{subfigure}
\hfill
\\  
 \begin{subfigure}[b]{0.33\textwidth}
    \centering
    \includegraphics[width=\textwidth]{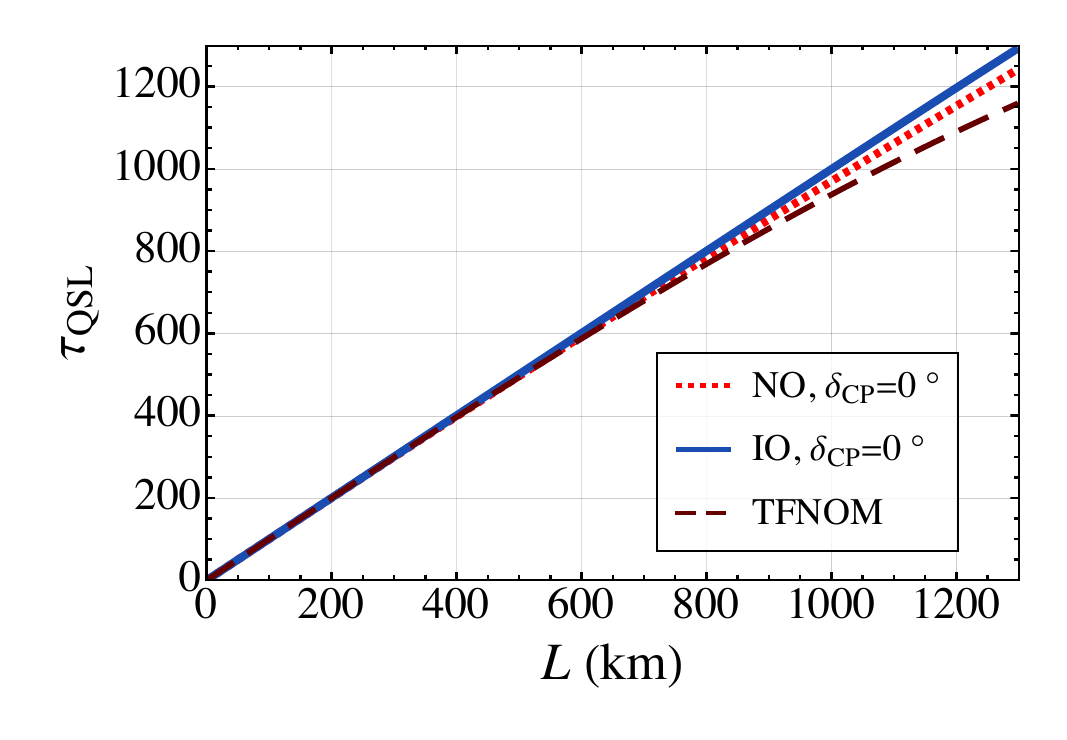}
    \caption{DUNE, $\delta_{\rm CP}=0^\circ$, NO (red dotted line), IO (blue solid line), and TFNOM (maroon dashed line)}
    \label{8_fig:sub1}
  \end{subfigure}
  \hfill
  \begin{subfigure}[b]{0.33\textwidth}
    \centering
    \includegraphics[width=\textwidth]{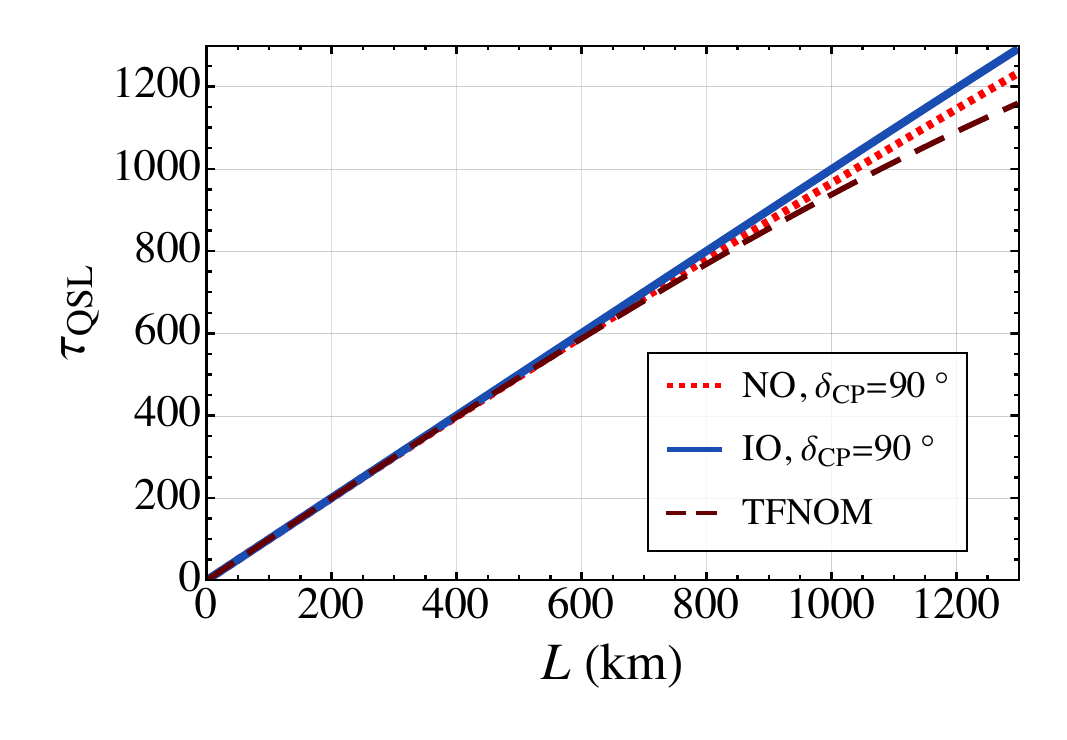}
    \caption{DUNE, $\delta_{\rm CP}=90^\circ$, NO (red dotted line), IO (blue solid line), and TFNOM (maroon dashed line)}
    \label{8_fig:sub2}
  \end{subfigure}
   \begin{subfigure}[b]{0.33\textwidth}
    \centering
    \includegraphics[width=\textwidth]{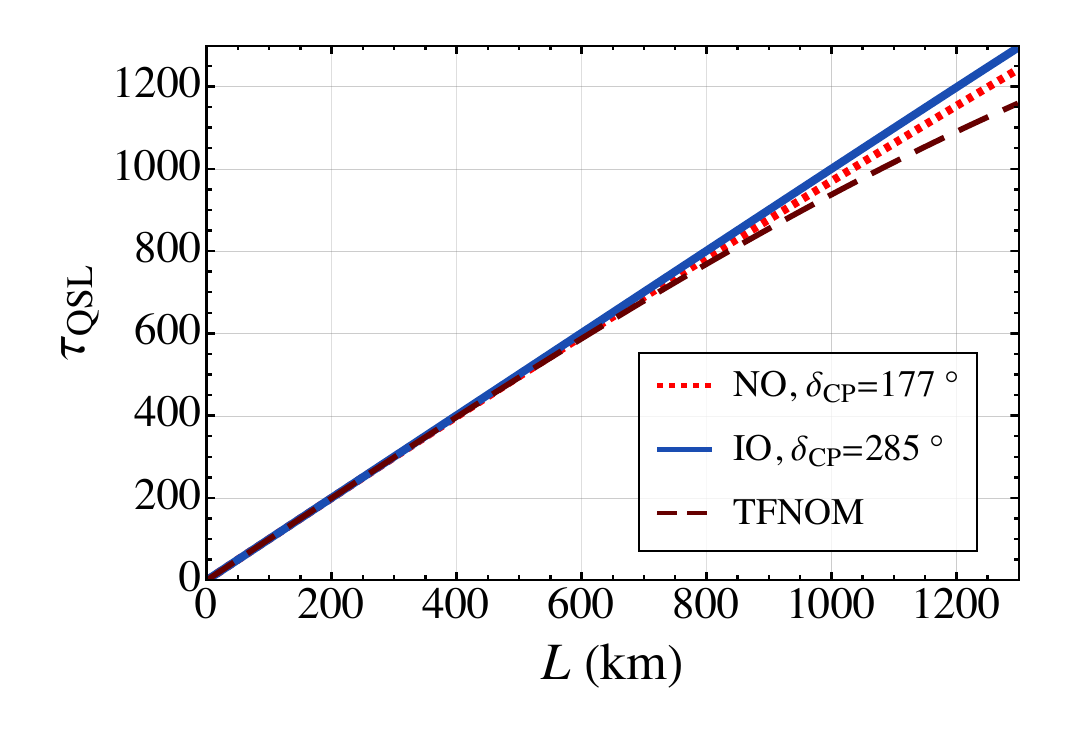}
    \caption{DUNE, $\delta_{\rm CP}=177^\circ$, NO (red dotted line), $\delta_{\rm CP}=285^\circ$, IO (blue solid line), and TFNOM (maroon dashed line)}
    \label{8_fig:sub3}
  \end{subfigure}
  \caption{\justifying{ In three-flavor neutrino oscillations, we compare the QSL time $\tau_{\rm QSL}$ versus $L\,\text{(km)}$ for the initial muon flavor neutrino state $\ket{\nu_\mu}$ in NO and IO. This comparison is illustrated using the length scales and energies of T2K, NOvA and DUNE  in Figs.\,\ref{8_fig:sub7}, \ref{8_fig:sub8}, \ref{8_fig:sub9}, Figs.\,\ref{8_fig:sub4}, \ref{8_fig:sub5}, \ref{8_fig:sub6}, and  Figs.\,\ref{8_fig:sub1}, \ref{8_fig:sub2}, \ref{8_fig:sub3}, respectively, under distinct $CP$-violation phases and matter potentials. In the TFNOM, the QSL time $\tau_{\rm QSL}$ versus $L\,\text{(km)}$ for the initial muon flavor neutrino state $\ket{\nu_\mu}$ is also illustrated using the length scales and energies of T2K, NOvA and DUNE in Figs.\,\ref{8_fig:sub7}, \ref{8_fig:sub8}, \ref{8_fig:sub9}, Figs.\,\ref{8_fig:sub4}, \ref{8_fig:sub5}, \ref{8_fig:sub6}, and Figs.\,\ref{8_fig:sub1}, \ref{8_fig:sub2}, \ref{8_fig:sub3}, respectively. The various neutrino mixing parameters used are taken from Tables\,\ref{tab2} for TFNOM and NO, and \ref{tab3} for IO along with their corresponding 1$\sigma$ errors ($90\%$ CL).}}
  \label{fig8}
\end{figure*}
\begin{figure*}[!htbp]
  \centering
  \begin{subfigure}[b]{0.33\textwidth}
    \centering
    \includegraphics[width=\textwidth]{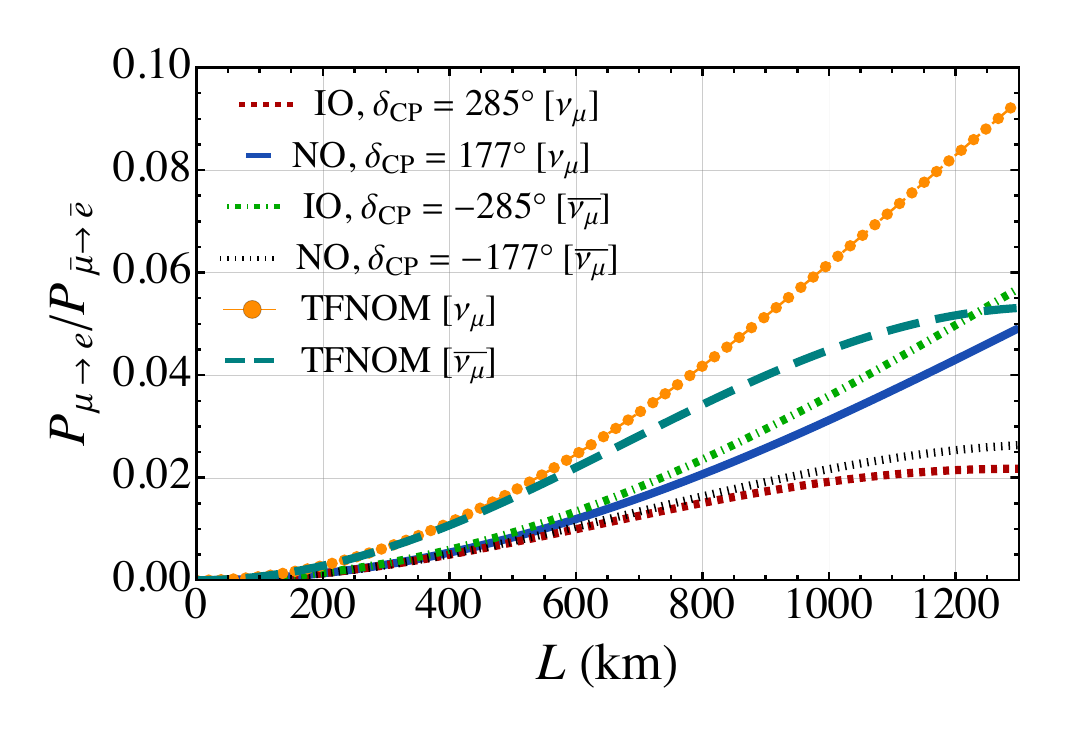}
    \caption{DUNE:\ NO (blue solid line), IO (red dashed line), and TFNOM (line with a small orange dot) for initial muon neutrino; NO (black dotted line), IO (green dot-dashed line), and TFNOM (cyan dashed line) for initial muon anti-neutrino.}
    \label{prob_anti_mu_e}
  \end{subfigure}
  \hfill
  \begin{subfigure}[b]{0.33\textwidth}
    \centering
    \includegraphics[width=\textwidth]{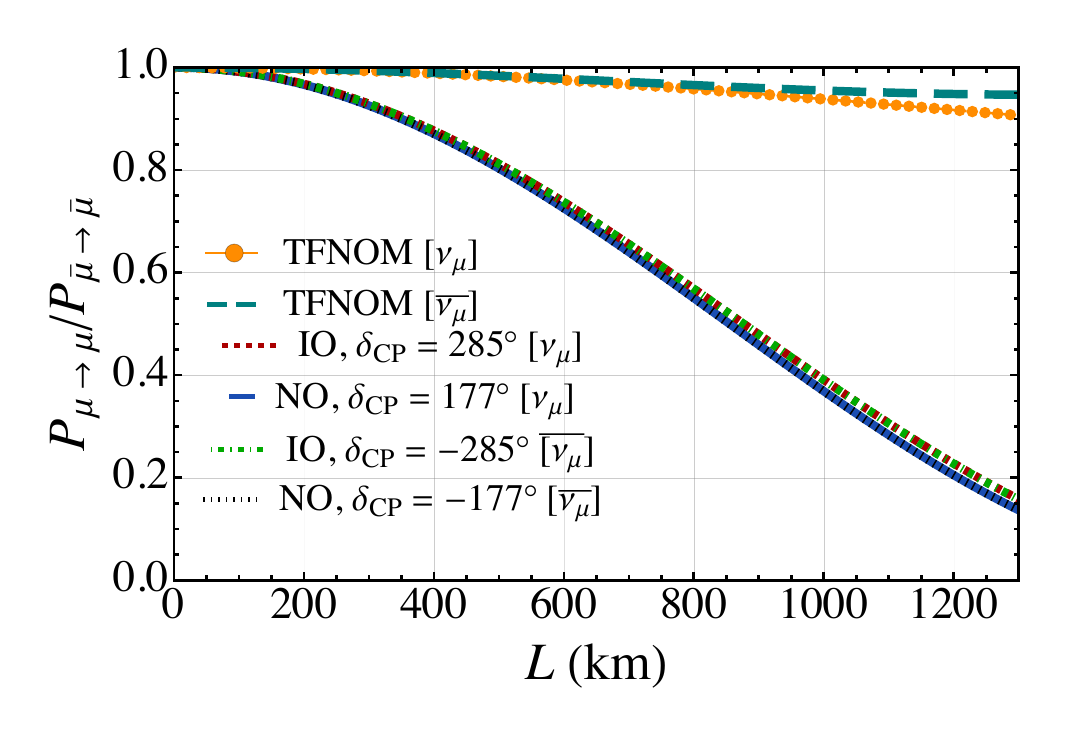}
    \caption{DUNE:\ NO (blue solid line), IO (red dashed line), and TFNOM (line with a small orange dot) for initial muon neutrino; NO (black dotted line), IO (green dot-dashed line), and TFNOM (cyan dashed line) for initial muon anti-neutrino.}
    \label{prob_anti_mu_mu}
  \end{subfigure}
   \begin{subfigure}[b]{0.33\textwidth}
    \centering
    \includegraphics[width=\textwidth]{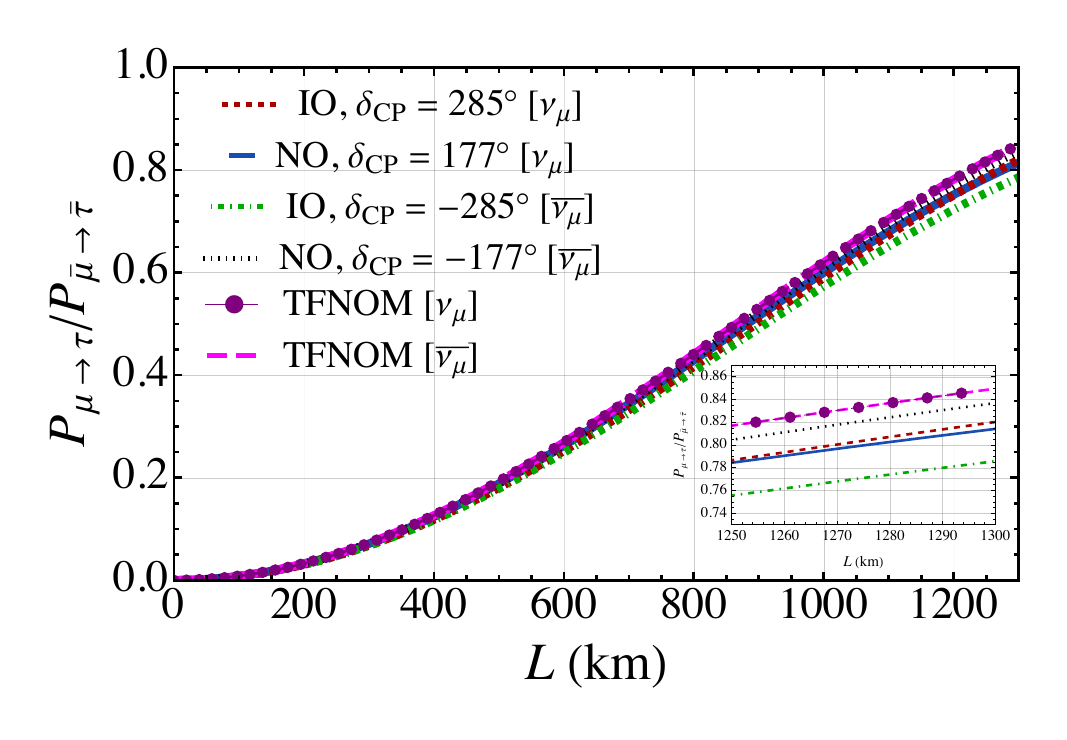}
    \caption{DUNE:\ NO (blue solid line), IO (red dashed line), and TFNOM (line with a small purple dot) for initial muon neutrino; NO (black dotted line), IO (green dot-dashed line),  and TFNOM (magenta dashed line) for initial muon anti-neutrino.}
    \label{prob_anti_mu_tau}
  \end{subfigure}
  \hfill
  \\
  \caption{\justifying{In the three-flavor neutrino oscillation, we depict the transition probabilities $P_{\mu \rightarrow e}$, $P_{\mu \rightarrow \mu}$, and $P_{\mu \rightarrow \tau}$ for the initial muon flavor neutrino state $\ket{\nu_\mu}$ (positive matter potential $+A_{\rm CC}$ and positive $CP$-violation phase $+\delta_{CP}$),
  and $P_{\overline{\mu} \rightarrow \overline{e}}$, $P_{\overline{\mu} \rightarrow \overline{\mu}}$, and $P_{\overline{\mu} \rightarrow \overline{\tau}}$ for the initial muon antineutrino flavor neutrino state ${\ket{\overline{\nu}_\mu}}$ (negative matter potential $-A_{\rm CC}$ and negative $CP$-violation phase $-\delta_{\rm CP}$), as functions of the propagation length $L\,\text{(km)}$. We compare their behaviors under the NO and IO scenarios. In the TFNOM, the transition probabilities $P_{\mu \rightarrow e}$ / $P_{\overline{\mu} \rightarrow \overline{e}}$, $P_{\mu \rightarrow \mu}$ / $P_{\overline{\mu} \rightarrow \overline{\mu}}$, and $P_{\mu \rightarrow \tau}$ / $P_{\overline{\mu} \rightarrow \overline{\tau}}$ as functions of the propagation length $L\,\text{(km)}$ for both the initial muon neutrino flavor state $\ket{\nu_\mu}$ (with positive matter potential $+A_{\rm CC}$) and the muon antineutrino flavor state $\ket{\overline{\nu}\mu}$ (with negative matter potential $-A_{\rm CC}$) are also shown in Figs.\,\ref{prob_anti_mu_e}, \ref{prob_anti_mu_mu}, and \ref{prob_anti_mu_tau}, respectively. This comparison is illustrated in DUNE setup using global fit data~\cite{Esteban:2024eli,NuFIT}.
}}
  \label{fig_prob_muon_antimuon}
\end{figure*}
\begin{figure*}
    \centering
    \includegraphics[scale=0.85]{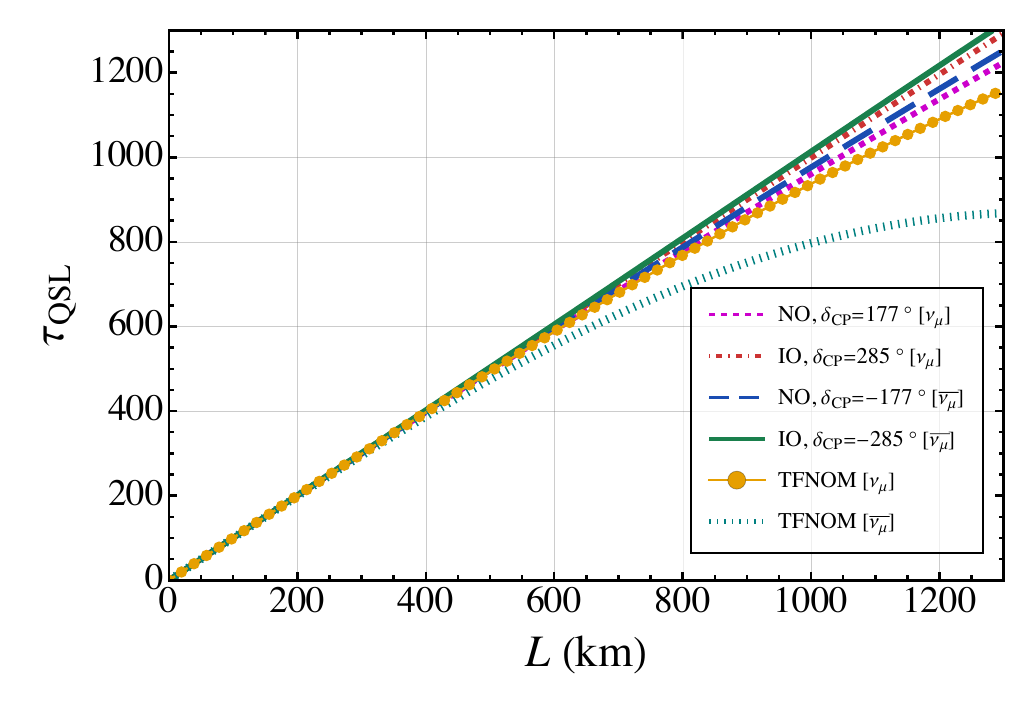}
    \caption{\justifying{In three-flavor neutrino oscillations, we compare the QSL time $\tau_{\rm QSL}$ versus $ L\,\text{(km)}$ for the initial muon flavor neutrino state $\ket{\nu_\mu}$ (positive matter potential $+A_{\rm CC}$ and positive $CP$-violation phase $+\delta_{\rm CP}$) and the initial antimuon flavor neutrino state $\ket{\overline{\nu}_\mu
}$ (negative matter potential $-A_{\rm CC}$ and negative $CP$-violation phase $-\delta_{\rm CP}$) in NO and IO. In the TFNOM, the QSL time $\tau_{\rm QSL}$ versus $ L\,\text{(km)}$ for both the initial muon neutrino flavor state $\ket{\nu_\mu}$ (positive matter potential $+A_{\rm CC}$) and the muon antineutrino flavor state $\ket{\overline{\nu}_\mu}$ (negative matter potential $-A_{\rm CC}$) are also depicted. This comparison is illustrated in the DUNE setup using global fit data~\cite{Esteban:2024eli,NuFIT}.}}
    \label{nu_annue_compare}
\end{figure*}

Moreover, including various mass-squared differences and mixing angles as outlined in Tables\,\ref{tab2} and \ref{tab3} for the NO and IO scenarios. The survival and oscillation probabilities of the state $\ket{\nu_{\mu}(t)}$ also rely on the ratio $L/E~\text{(km/GeV)}$, which can be fixed in various long-baseline accelerator neutrino source experiments setups like T2K, NOvA, and DUNE, presenting a more realistic scenario for neutrino oscillations.
\begin{itemize}
\item T2K (Tokai-to-Kamioka)  \cite{PhysRevLett.107.041801} operates as an off-axis experiment employing a $\nu_{\mu}$ neutrino beam originating at J-PARC (Japan Proton Accelerator Complex) with an energy range approximately between $100\,\mathrm{MeV}$ to $1\,\mathrm{GeV}$ and a baseline of $295 \mathrm{~km}$.
\item NOvA (NuMI Off-Axis $\nu_{e}$ Appearance) \cite{NOvA:2004blv}, a long-baseline experiment, utilizes neutrinos from the NuMI (Neutrinos at the Main Injector) beamline at Fermilab optimized to observe $\nu_{\mu} \rightarrow \nu_{e}$ oscillations. This experiment employs two detectors positioned $14~ \mathrm{mrad}$ off the axis of the NuMI beamline, with the near and far detectors located at $1$ and $810 \mathrm{~km}$ from the source, respectively. The flavor composition of the beam consists of $92.9 \%$ of $v_{\mu}$ and 5.8\% of $\bar{v}_{\mu}$ and $1.3 \%$ of $v_{e}$ and $\bar{v}_{e}$ with the energy of the neutrino beam varying from $1.5$ to $4 ~\mathrm{GeV}$.
\item DUNE (Deep Underground Neutrino Experiment) \cite{Kudryavtsev_2016,DUNE:2020ypp,LBNE:2013dhi} is an upcoming neutrino oscillation experiment. It is an experimental facility utilizing the NuMI neutrino beam with an energy range of $1-10\,\mathrm{GeV}$ from Fermilab and features a long baseline of $1300 \mathrm{~km}$. This configuration enables an $L/E~$ ratio of about $10^3 \mathrm{km/GeV}$, providing good sensitivity for the $CP$-violation ($\delta_{\rm CP}$) measurement and determination of mass ordering.  
\end{itemize}
The neutrino oscillation depends on a complex quantity in the PMNS mixing matrix related to a phase factor associated with $CP$ violation, $\delta_{\rm CP}$ \cite{Nguyen:2022snr}. However, the value of the $CP$-violating phase is still not well-constrained in neutrino experiments~\cite{T2K:2019bcf}. On the other hand, the $\delta_{\rm CP}$ is linked with matter-antimatter asymmetry (more baryons over antibaryons) in the observable Universe~\cite{Canetti:2012zc}. The hunt to find baryon number violation processes in nature has not been fruitful without caveats until now. However, if the number of leptons created in a reaction is not the same as the number destroyed, it could indirectly lead to a net change in the baryon number~\cite{Abada:2006ea,Burguet-Castell:2001ppm,Buchmuller:2003gz,Gribov:1968kq,Davidson:2008bu,Buchmuller:2004nz,Buchmuller:2005eh,Akhmedov:1998qx,Nardi:2006fx,Frampton:2002qc,Buchmuller:2002rq,Pascoli:2006ci,Branco:2011zb}. Neutrino, a lepton, can provide us with more insight into this. Comparing the difference between neutrino and antineutrino oscillations, ongoing long-baseline neutrino oscillation experiments, such as T2K  and NOvA~\cite{DeRujula:1998umv,T2K:2019bcf,Giganti:2017fhf,T2K:2019bcf,PhysRevLett.107.041801,ABE2011106,Wachala:2017pey,NOvA:2024lti} and also next-generation experiment, DUNE~\cite{DUNE:2020ypp,Brahma:2023pxj,LBNE:2013dhi}, have the potential to detect evidence for $CP$ violation. The global fit values~\cite{Esteban:2024eli,NuFIT} of $\delta_{\rm CP}$ in NO and IO are listed in Tables\,\ref{tab2} and \ref{tab3}, respectively. 

In our analysis, we consider the neutrino energies for T2K, NOvA, and DUNE to be $1.4$, $2.5$, and $3.5\,\rm GeV$, respectively\,\cite{PhysRevD.99.095001}. The corresponding matter parameter ($A_{\rm CC}=2EV_{\rm CC}$) calculated using Eq.\,({\ref{A14}}) is approximated as $2.83\times 10^{-4}\,\rm{eV^2}$, $5.05\times 10^{-4}\,\rm{eV^2}$ and $7.07\times 10^{-4}\,\rm{eV^2}$ for T2K, NOvA, and DUNE respectively. Substituting these $A_{\rm CC}$ values in Eq.\,({\ref{45a}}), in the ultrarelativistic regime $t\approx L$, Fig.\,\ref{fig8a} illustrates the variation of the survival probability ($P_{\mu\rightarrow \mu}$) and oscillation probabilities ($P_{\mu\rightarrow e}$ and $P_{\mu\rightarrow \tau}$) of the initial muon flavor neutrino state $\ket{\nu_\mu}$ in matter as functions of the long-baseline length $L\,(\rm km)$ for T2K ($L\approx 280\,\rm km$), NOvA ($L\approx 800\,\rm km$), and DUNE ($L\approx 1300\,\rm km$). These results are examined
for two distinct $\delta_{CP}$ values corresponding to NO ($\delta_{CP}=177^\circ$) and IO ($\delta_{CP}=285^\circ$), with comparative analyses presented for T2K, NOvA, and DUNE in Figs.\,(\ref{8a_fig:sub1}, \ref{8a_fig:sub2}, \ref{8a_fig:sub3}), Figs.\,\ref{8a_fig:sub4}, \ref{8a_fig:sub5}, \ref{8a_fig:sub6}, and  Figs.\,\ref{8a_fig:sub7}, \ref{8a_fig:sub8}, \ref{8a_fig:sub9}, respectively. 

In general, matter effects become significant only when one of the neutrino flavors is an electron neutrino ($\nu_e$). This is because $\nu_e$ interacts with electrons via charged current interactions, while $\nu_\mu$ and $\nu_\tau$ interact via neutral current interactions, which are flavor independent \cite{10.1093/acprof:oso/9780198508717.001.0001}. As a result, only the mixing angles $\theta_{12}$ and $\theta_{13}$ are affected by matter, while $\theta_{23}$ remains unchanged \cite{Nguyen:2022snr,Agarwalla:2013tza}. Therefore, for the case where only two neutrinos, $\nu_\mu$ and $\nu_\tau$, are involved,  $\nu_\mu\leftrightarrow \nu_\tau$ oscillations in matter are effectively the same as in vacuum \cite{Mocioiu:2000st}, whereas other transitions such as $\nu_\mu\leftrightarrow \nu_\mu$ and $\nu_\mu\leftrightarrow \nu_e$ are different in vacuum and matter, as shown in Figs.\,\ref{Fig1} and \ref{Fig2a}, respectively. For $\nu_\mu\leftrightarrow \nu_\tau$ oscillations in vacuum, 
one can reduce the three-flavor formalism \cite{Nguyen:2022snr} to an effective two-flavor framework by setting the parameters $\Delta m^2_{21}$ and $\theta_{13}$ to zero in the leading order \cite{Alok:2014gya}. Thus, in the two-flavor approximation, the oscillation probability for $\nu_\mu\leftrightarrow \nu_\tau$ is defined as \cite{Alok:2014gya,MINOS:2006foh,T2K:2013bzi} \begin{equation}
P(\nu_\mu\rightarrow\nu_\tau)\approx \sin^2(2\theta_{23})\sin^2(\Delta m^2_{32}L/4E),
\label{35new}
\end{equation} which is not the same as the one for $\nu_\mu\leftrightarrow \nu_e$ oscillations in
Eq.\,(\ref{36a}). Since the matter effect does not significantly alter $\nu_\mu\leftrightarrow\nu_\tau$ oscillations, Eq.\,(\ref{35new}) is valid in both vacuum and matter under the two-flavor approximation \cite{Mocioiu:2000st}. However, as shown in Fig.\,\ref{fig8a}, in the more realistic three-flavor framework, the presence of $\nu_e$ leads to matter-induced effects even on $\nu_\mu\leftrightarrow\nu_\tau$, as well as on other channels ($\nu_\mu\leftrightarrow \nu_e$ and $\nu_\mu\leftrightarrow \nu_\mu$).

Additionally, using the length scales and energies of T2K, NOvA and DUNE, the transition probabilities ($P_{\mu\rightarrow e}$, $P_{\mu\rightarrow\mu}$, and $P_{\mu\rightarrow \tau}$\footnote{Equation\,(\ref{35new}) remains applicable in both vacuum and matter within the two-flavor approximation \cite{Mocioiu:2000st}, and is therefore employed in the TFNOM framework to illustrate $P_{\mu\rightarrow\tau}$ (magenta dashed line) in Figs.\,\ref{8a_fig:sub3}, \ref{8a_fig:sub6}, and \ref{8a_fig:sub9}.}) in NO and IO, as shown in Figs.\,\ref{8a_fig:sub1}, \ref{8a_fig:sub2}, \ref{8a_fig:sub3}, Figs.\,\ref{8a_fig:sub4}, \ref{8a_fig:sub5}, \ref{8a_fig:sub6}, and Figs.\,\ref{8a_fig:sub7}, \ref{8a_fig:sub8}, \ref{8a_fig:sub9}, respectively, are also compared with results obtained from the two-flavor neutrino oscillation in matter (TFNOM), represented by a green dashed line for $P_{\mu\rightarrow e}$ and  $P_{\mu\rightarrow\mu}$, and a magenta dashed line for $P_{\mu\rightarrow\tau}$. Noticeable differences arise between the two- and three-flavor results in matter, primarily due to the additional mixing angles, squared-mass differences, and $CP$-violation phases accounted for in the three-flavor framework, which are absent in the two-flavor approximation. 

 Moreover, using Eq.\,(\ref{45}) and the matter-dependent effective Hermitian Hamiltonian \cite{Nguyen:2022snr} in Eq.\,(\ref{3}), the QSL time ($\tau_{\rm QSL}$) of the state $\ket{\nu_\mu(t)}$ in a constant Earth-matter background, taking into account $CP$-violation phases for both NO and IO scenarios can be studied. We explore these aspects across three long-baseline accelerator neutrino experiments: T2K, NOvA, and DUNE. In the ultrarelativistic regime, the top, middle, and bottom panels of Fig.\,\ref{fig8} illustrate $\tau_{\rm QSL}$ versus $L(\text{km})$ of the time-evolved muon flavor neutrino state $\ket{\nu_\mu(t)}$ in a constant Earth-matter background at distinct $CP$-violation phase values, under the assumptions of NO (Table \,\ref{tab2}) and IO (Table \,\ref{tab3}), and employing constraints on the ratio of $L/E~\text{(km/GeV)}$ based on long-baseline accelerator neutrino experiments such as T2K, NOvA and DUNE. 
 
 The top panel of Fig.\,\ref{fig8} displays $\tau_{\rm QSL}$ of the initial state $\ket{\nu_\mu}$ in constant matter as a function of propagation length $[L\,\text{(km)}]$ using the T2K length and energy scale. In Figs.\,\ref{8_fig:sub7},\,\ref{8_fig:sub8},\,\ref{8_fig:sub9}, near a length of 280 km, energy ($E\approx 1.4 \,\mathrm{GeV}$), and considering distinct values of $\delta_{\rm CP}$ for the NO ($\delta_{\rm CP}=0^\circ,90^\circ,177^\circ$) (orange dotted line) and for the IO ($\delta_{\rm CP}=0^\circ,\,90^\circ,\,285^\circ$) (black solid line), we observe that the separation between $\tau_{\rm QSL}$ in NO (orange dotted line) and IO (black solid line) increases as the $CP$-violation phase increases. Since $P_{\mu\rightarrow\mu}=1-P_{\mu\rightarrow e}-P_{\mu\rightarrow \tau}$, the $\tau_{\rm QSL}$ is connected to both the survival ($P_{\mu\rightarrow\mu}$) and oscillation probabilities ($P_{\mu\rightarrow e}$ and $P_{\mu\rightarrow \tau}$). In Figs.\,{\ref{8a_fig:sub1}, \ref{8a_fig:sub2}, \ref{8a_fig:sub3}}, it is shown that the difference between NO at $\delta_{CP}=177^\circ$ (blue solid line) and IO at $\delta_{CP}=285^\circ$ (red dashed line) for $P_{\mu\rightarrow\mu}$ and $P_{\mu\rightarrow\tau}$ is small, while $P_{\mu\rightarrow e}$ is large near $L\approx 280\,\rm km$. The difference between NO at $\delta_{CP}=177^\circ$ (orange dotted line) and IO at $\delta_{CP}=285^\circ$ (black solid line) for $\tau_{QSL}$ in Fig.\ref{8_fig:sub9} arises due to $P_{\mu\rightarrow e}$ near $L\approx 280\,\rm km$. Additionally, it is observed in Fig.\,{\ref{8_fig:sub9}} that under the time-bound condition $\rm \tau_{\rm QSL}/L<1$, the initial state $\ket{\nu_\mu}$ in the presence of a constant Earth-matter potential in NO (orange dotted line) undergoes more speedup compared to the IO (black solid line) case. 
 
Furthermore, within the middle panel of Fig.\,\ref{fig8}, we use the long-baseline length scale ($L\approx 800\text{km}$) with energy ($E\approx 2.5\,\text{GeV}$) provided by the NOvA experiment to examine the QSL time in NO and IO scenarios for the initial state $\ket{\nu_\mu}$ propagating in a constant Earth-matter background. Considering different values of $\delta_{CP}$, we compare $\tau_{\rm QSL}$ versus $L\text{(\rm km)}$ of the state $\ket{\nu_\mu(t)}$ in the NO (magenta dotted line) and IO (green solid line) scenarios in Figs.\,(\ref{8_fig:sub4}, \ref{8_fig:sub5}, \ref{8_fig:sub6}). Notably, at larger $CP$-violation phase values, $\tau_{\rm QSL}$ of the initial state  $\ket{\nu_\mu}$ in Fig.\,(\ref{8_fig:sub6}) demonstrates a more pronounced separation between the NO (magenta dotted line) and IO (green solid line) cases. This separation arises because the difference between NO (blue solid line) and IO (red dotted line) for $P_{\mu\rightarrow e}$ is more pronounced at the larger $CP$-violation near $L\approx 800\,\rm km$, as shown in Fig.\,\ref{8a_fig:sub4}. According to the time-bound condition $\rm \tau_{\rm QSL}/L<1$, the dynamic evolution of the initial state $\ket{\nu_\mu}$ in matter demonstrates faster evolution for the case of NO (magenta dotted line) as compared to the IO (green solid line), especially noticeable at substantial $CP$-violation phases, as can be seen in Fig.\,\ref{8_fig:sub6}.

The bottom panel of Fig.\,\ref{fig8} depicts $\tau_{\rm QSL}$ of the time-evolved muon flavor neutrino state $\ket{\nu_\mu(t)}$ using the DUNE's baseline length ($L\approx 1300\,\text{km}$) and energy ($E\approx 3.5\,\text{GeV}$). In Figs.\,\ref{8_fig:sub1}, \ref{8_fig:sub2}, \ref{8_fig:sub3}, we compute and compare $\tau_{\rm QSL}$ for the initial state $\ket{\nu_\mu}$ as a function of propagation length $L\,(\text{km})$ for NO (red dotted line) and IO (blue solid line), and for distinct $CP$-violation phase values. In all these figures, the results can distinguish between $\tau_{\rm QSL}$ for NO and IO. In Fig.\,{\ref{8a_fig:sub7}}, for larger $CP$-violation phase values, the separation between $P_{\mu\rightarrow e}$ in NO (blue solid line) and IO (red dotted line) is more pronounced at DUNE's length scale compared to T2K [see Fig.\,{\ref{8a_fig:sub1}}] and NOvA [see Fig.\,{\ref{8a_fig:sub4}}], and therefore in Fig.\,\ref{8_fig:sub3}, the behavior of $\tau_{\rm QSL}$ reflects a similar trend. Figure\,\ref{8_fig:sub3} highlights that under the constraint $\tau_{\rm QSL}/L<1$, the initial state $\ket{\nu_\mu}$ undergoes fast evolution at a large $CP$-violation phase in DUNE, particularly in the NO scenario. Hence, by using the length scales and energies of T2K, NOvA, and DUNE experiments, we observe that the faster evolution of neutrino muon flavor states in matter is sensitive to the significant $CP$-violation phase in the NO scenario. 

However, in all these experiments, including T2K, NOvA, and DUNE, as depicted in Fig.\,\ref{fig8}, under the time-bound condition $\tau_{\rm QSL}/L < 1$, the TFNOM exhibits a faster evolution of the neutrino muon flavor state compared to the three-flavor results in matter for both NO and IO (with zero and nonzero $CP$-violation phases).

So far, we have considered $\tau_{\rm QSL}$ for the initial muon flavor neutrino state $\ket{\nu_\mu}$. For completeness, we discuss $\tau_{\rm QSL}$ for the initial muon antineutrino flavor state $\ket{\overline{\nu}_\mu}$ relevant to accelerator experiments.

Using Eq.\,({\ref{45a}}), in Fig.\,{\ref{fig_prob_muon_antimuon}}, we show the variation of the survival probability ($P_{\mu\rightarrow \mu}$) and oscillation probabilities ($P_{\mu\rightarrow e}$ and $P_{\mu\rightarrow\tau}$) for the initial muon flavor neutrino state $\ket{\nu_\mu}$ [with $A_{CC}=+7.07\times10^{-4}\,\rm eV^2$, $\delta_{CP}=+177^\circ$ for NO (blue solid line), and $\delta_{CP}=+285^\circ$ for IO (red dashed line)], and, the survival probability ($P_{\overline{\mu}\rightarrow \overline{\mu}}$) and oscillation probabilities ($P_{\overline{\mu}\rightarrow \overline{e}}$ and $P_{\overline{\mu}\rightarrow\overline{\tau}}$) 
for the antimuon flavor neutrino state $\ket{\overline{\nu}_\mu}$ [with $A_{CC}=-7.07\times10^{-4}\,\rm eV^2$, $\delta_{CP}=-177^\circ$ for NO (black dotted line), and $\delta_{CP}=-285^\circ$ for IO (green dot-dashed line)], as functions of propagation length $L\,(\rm km)$. The comparison between the NO and IO scenarios is made using DUNE's baseline length and energy. It is observed that $P_{\mu\rightarrow e}$ / $P_{\overline{\mu}\rightarrow\overline{e}}$ [see Fig.\,\ref{prob_anti_mu_e}] exhibits more pronounced differences for the initial state $\ket{\nu_\mu}$ / $\ket{\overline{\nu}_\mu}$, between NO and IO, while $P_{\mu\rightarrow \mu}$ / $P_{\overline{\mu}\rightarrow \overline{\mu}}$ [see Fig.\,{\ref{prob_anti_mu_mu}}] and $P_{\mu\rightarrow\tau}$ / $P_{\overline{\mu}\rightarrow\overline{\tau}}$ [see Fig.\,{\ref{prob_anti_mu_tau}}] shows minimal differences near $L\approx 1300\,\rm km$. Additionally, Figs.\,\ref{prob_anti_mu_e}, \ref{prob_anti_mu_mu}, and \ref{prob_anti_mu_tau} depict the transition probabilities $P_{\mu \rightarrow e}$ / $P_{\overline{\mu} \rightarrow \overline{e}}$, $P_{\mu \rightarrow \mu}$ / $P_{\overline{\mu} \rightarrow \overline{\mu}}$, and $P_{\mu\rightarrow\tau}$\footnote{In the TFNOM, we use Eq.\,(\ref{35new}) to illustrate $P_{\mu\rightarrow\tau}$ (line with a small purple dot) and $P_{\overline{\mu}\rightarrow\overline{\tau}}$ (magenta dashed line) in Fig.\,\ref{prob_anti_mu_tau}. Since, the matter effect is negligible ($A_{CC}=0$) for $\nu_\mu\leftrightarrow\nu_\tau$ oscillations, the transition probabilities for the initial states $\ket{\nu_\mu}$ and $\ket{\overline{\nu}_\mu}$, are equal, i.e., $P_{\mu\rightarrow\tau}= P_{\overline{\mu}\rightarrow\overline{\tau}}$, and are effectively identical in vacuum and matter.} / $P_{\overline{\mu}\rightarrow\overline{\tau}}$, respectively, in the TFNOM for the initial states $\ket{\nu_\mu}$ (with $A_{\rm CC} = +7.07\times10^{-4}\,\rm eV^2$) and $\ket{\overline{\nu}_\mu}$ (with $A_{\rm CC} = -7.07\times10^{-4}\,\rm eV^2$), and compare them with the results obtained from three-flavor neutrino oscillations in matter (for both NO and IO with their respective $CP$-violation phases). Noticeable differences are observed between the two- and three-flavor results.

In Fig.\,\ref{nu_annue_compare}, the comparison illustrates $\tau_{\rm QSL}$ variation as a function of the propagation length $ L~(\text{km})$ for the initial states $\ket{\nu_\mu}$ and $\ket{\overline{\nu}_\mu}$ within both the NO and IO. Here, $\tau_{\rm QSL}$ for the initial muon neutrino state $\ket{\nu_\mu}$ is characterized by a positive matter potential ($+A_{\rm CC}$) and a positive $CP$-violation phase ($+\delta_{\rm CP}$) in NO (magenta dotted line) and IO (red dot-dashed line). Conversely, for the muon antineutrino state $\ket{\overline{\nu}_\mu}$, $\tau_{\rm QSL}$ is characterized by a negative matter potential ($-A_{\rm CC}$) and a negative $CP$-violation phase ($-\delta_{\rm CP}$) in NO (blue dashed line) and IO (green solid line). 
The significant difference between $P_{\mu\rightarrow e}$ / $P_{\overline{\mu}\rightarrow \overline{e}}$ in NO and IO for the initial states $\ket{\nu_\mu}$ / $\ket{\overline{\nu}_\mu}$, as observed above, explains the noticeable separation in $\tau_{\rm QSL}$ between NO and IO for the initial states $\ket{\nu_\mu}$ and $\ket{\overline{\nu}_\mu}$ near $L\approx 1300\,{\rm km}$, see Fig.\,{\ref{nu_annue_compare}}.
Moreover, under the constraint $\tau_{\rm QSL}/L<1$, in the NO scenario, a faster flavor evolution is estimated for both the initial states $\ket{{\nu}_\mu}$  and $\ket{\overline{\nu}_\mu}$, compared to the IO scenario. However, 
compared to the initial state $\ket{\overline{\nu}_\mu}$, significantly faster flavor evolution is observed for the initial state $\ket{{\nu}_\mu}$ near $L\approx 1300\,{\rm km}$, as shown in Fig.\,\ref{nu_annue_compare} by the magenta dotted line. 

Furthermore, in Fig.\,\ref{nu_annue_compare}, the QSL time $\tau_{\rm QSL}$ is also illustrated in the TFNOM for the initial states $\ket{\nu_\mu}$ and $\ket{\overline{\nu}_\mu}$. Under the constraint $\tau_{\rm QSL}/L < 1$, the results show a faster evolution for both the initial state $\ket{\nu_\mu}$~(line with a small orange dot) and $\ket{\overline{\nu}_\mu}$~(cyan dotted line) in the TFNOM compared to the three-flavor neutrino oscillations in matter (both for NO and IO with their respective $CP$-violation phases).

In the spirit of the two-flavor scenario, we now take up the case of entanglement in three-flavor neutrino oscillations in matter.

\section{Quantum Speed Limit time for entanglement in three-flavor neutrino
oscillations in matter}
\label{Sect6}

 In the three-neutrino system, we map the three-flavor neutrino state to three-qubit states as \cite{KumarJha:2020pke}
\begin{eqnarray}
{\ket{\nu_e}\equiv\ket{1}_{e}\otimes\ket{0}_{\mu}\otimes\ket{0}_\tau\equiv\ket{100};}
&\nonumber\\
{\ket{\nu_\mu}\equiv\ket{0}_{e}\otimes\ket{1}_{\mu}\otimes\ket{0}_\tau\equiv\ket{010};}&\nonumber\\
{\ket{\nu_\tau}\equiv\ket{0}_{e}\otimes\ket{0}_{\mu}\otimes\ket{1}_\tau\equiv\ket{001}.}&
\label{46}
\end{eqnarray}
 Following  Eq.\,(\ref{45}) and Eq.\,(\ref{46}), the time-evolved muon flavor neutrino states in matter can be represented in the three-qubit system as 
\begin{equation} 
    \ket{\nu_\mu (t)}=\tilde{\mathcal{A}}_{\mu e}(t)\ket{100}+\tilde{\mathcal{A}}_{\mu \mu}(t)\ket{010} + \tilde{\mathcal{A}}_{\mu\tau}(t)\ket{001}.
    \label{47}
    \end{equation}
 The corresponding density matrix $\rho(t)=\ket{\nu_\mu(t)}\bra{\nu_\mu(t)}$ in the standard basis $\ket{ijk}$, where each index takes the values 0 and 1 is given by $\rho(t)=$ 
 \begin{equation}
 \begin{pmatrix}  
 0 & 0 & 0 & 0 & 0 & 0 & 0 & 0\\ 
 0 & 0 & 0 & 0 & 0 & 0 & 0 & 0\\ 
0 & 0 & 0 & 0 & 0 & 0 & 0 & 0\\ 
0 & 0 & 0 & \vert{\tilde{\mathcal{A}}_{\mu e}(t)}\vert^2 & 0 & \tilde{\mathcal{A}}_{\mu e}(t) \tilde{\mathcal{A}}_{\mu \mu}^*(t) & \tilde{\mathcal{A}}_{\mu e}(t) \tilde{\mathcal{A}}_{\mu\tau}^*(t) & 0 \\
0 & 0 & 0 & 0 & 0 & 0 & 0 & 0\\ 
0 & 0 & 0 & \tilde{\mathcal{A}}_{\mu\mu}(t) \tilde{\mathcal{A}}_{\mu e}^*(t) & 0 & \vert{\tilde{\mathcal{A}}_{\mu\mu} (t)}\vert^2 & \tilde{\mathcal{A}}_{\mu\mu}(t)\tilde{\mathcal{A}}_{\mu\tau}^*(t) & 0 \\
0 & 0 & 0 & \tilde{\mathcal{A}}_{\mu\tau}(t) \tilde{\mathcal{A}}_{\mu e}^*(t) & 0 & \tilde{\mathcal{A}}_{\mu\tau}(t) \tilde{\mathcal{A}}_{\mu\mu}^* (t) & \vert{\tilde{\mathcal{A}}_{\mu\tau} (t)}\vert^2 & 0\\ 
0 & 0 & 0 & 0 & 0 & 0 & 0 & 0 
\end{pmatrix}.
\label{48}
\end{equation} 
\begin{figure*}[!htbp]
    \centering
    \begin{subfigure}{0.49\textwidth}
        \centering
        \includegraphics[width=\textwidth]{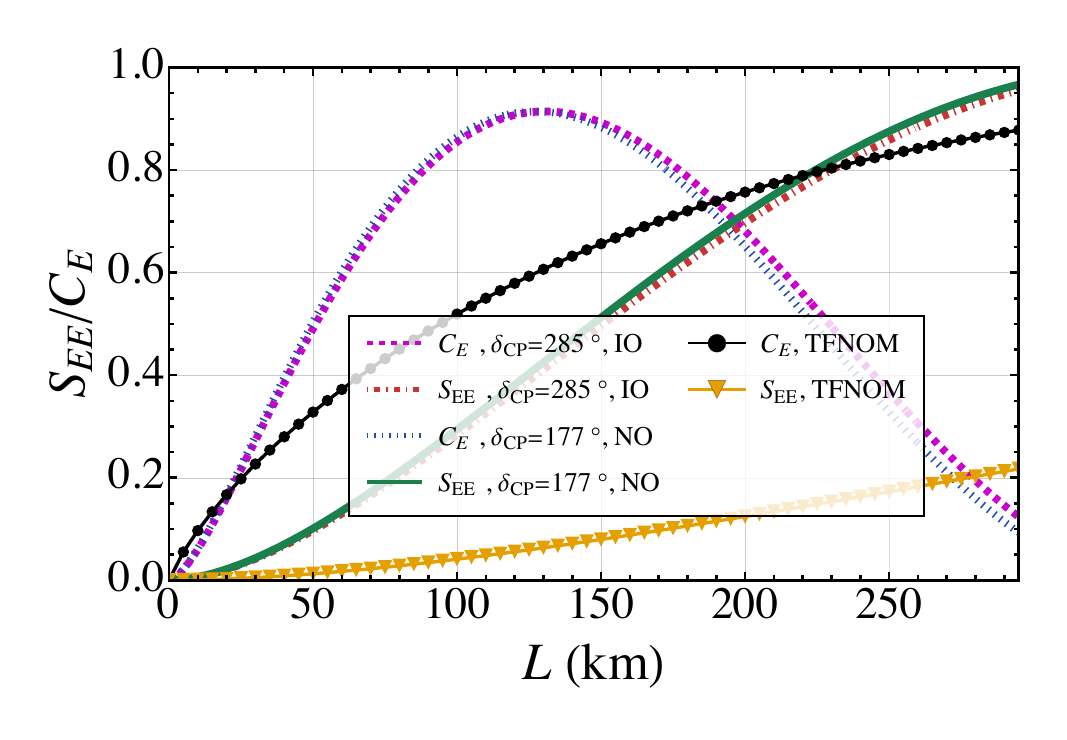}
        \caption{T2K, $\delta_{\rm CP}=177^\circ$ (NO), $\delta_{\rm CP}=285^\circ$ (IO)}
        \label{9_fig:sub1}
    \end{subfigure}
    \hfill
    \begin{subfigure}{0.49\textwidth}
        \centering
        \includegraphics[width=\textwidth]{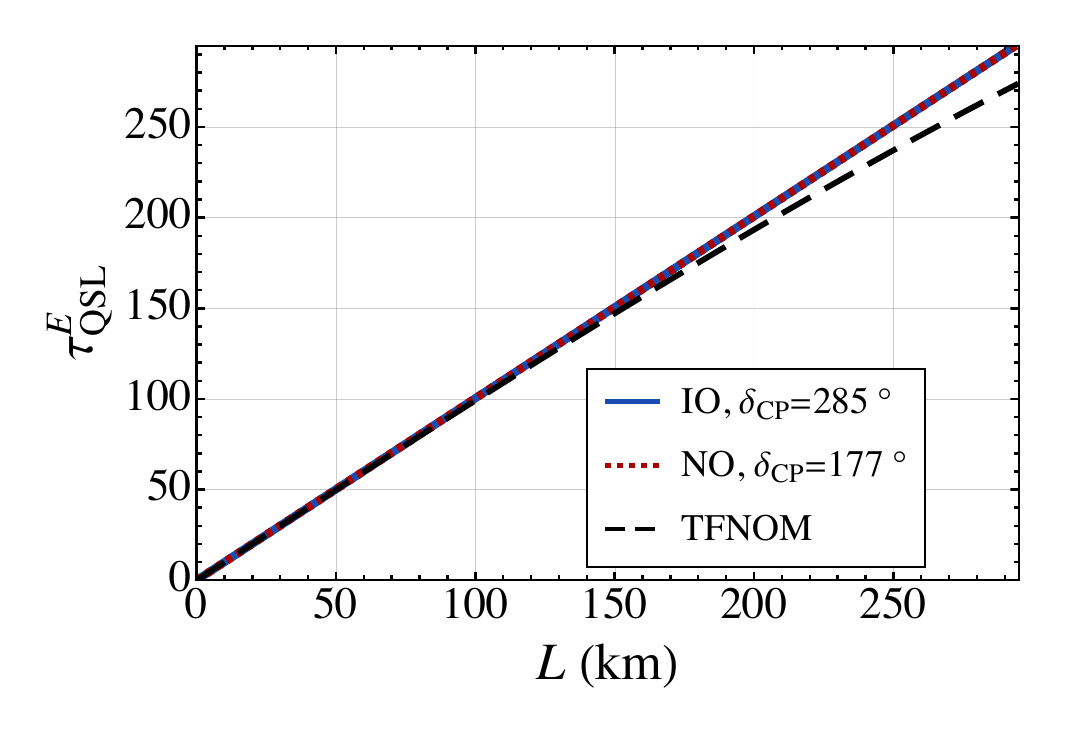}
        \caption{T2K, $\delta_{\rm CP}=177^\circ$ (NO), $\delta_{\rm CP}=285^\circ$ (IO)}
        \label{9_fig:sub2}
    \end{subfigure}
    
    \medskip
    
    \begin{subfigure}{0.49\textwidth}
        \centering
        \includegraphics[width=\textwidth]{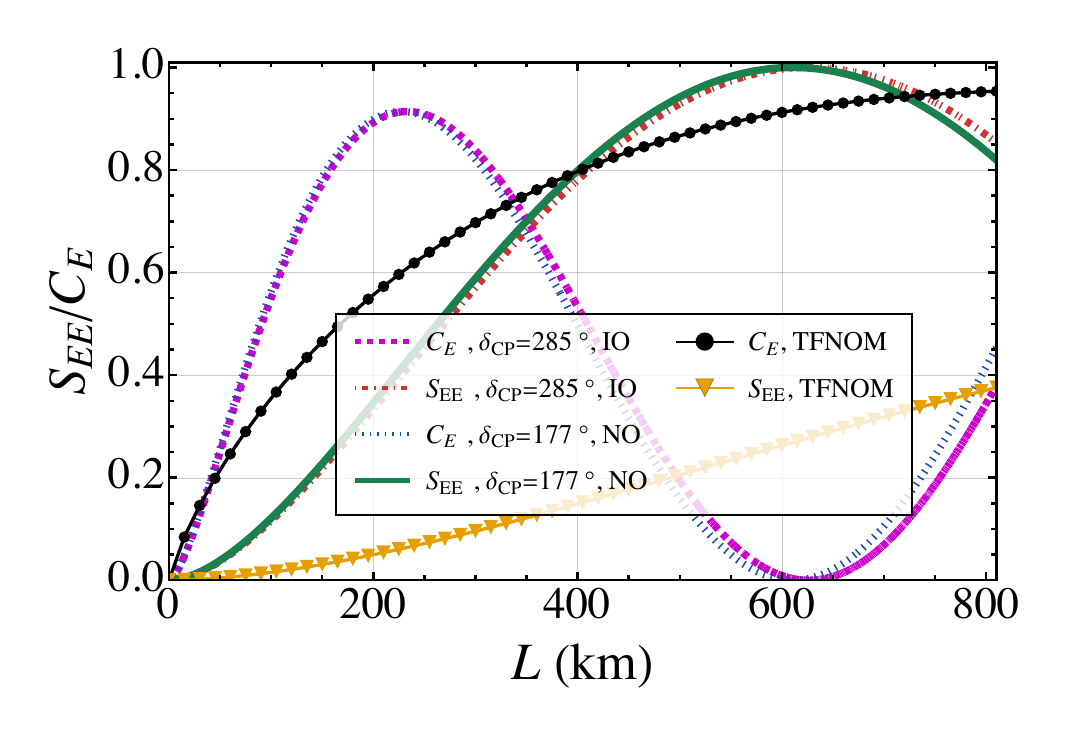}
        \caption{NOvA, $\delta_{\rm CP}=177^\circ$ (NO), $\delta_{\rm CP}=285^\circ$ (IO)}
        \label{9_fig:sub3}
    \end{subfigure}
    \hfill
    \begin{subfigure}{0.49\textwidth}
        \centering
        \includegraphics[width=\textwidth]{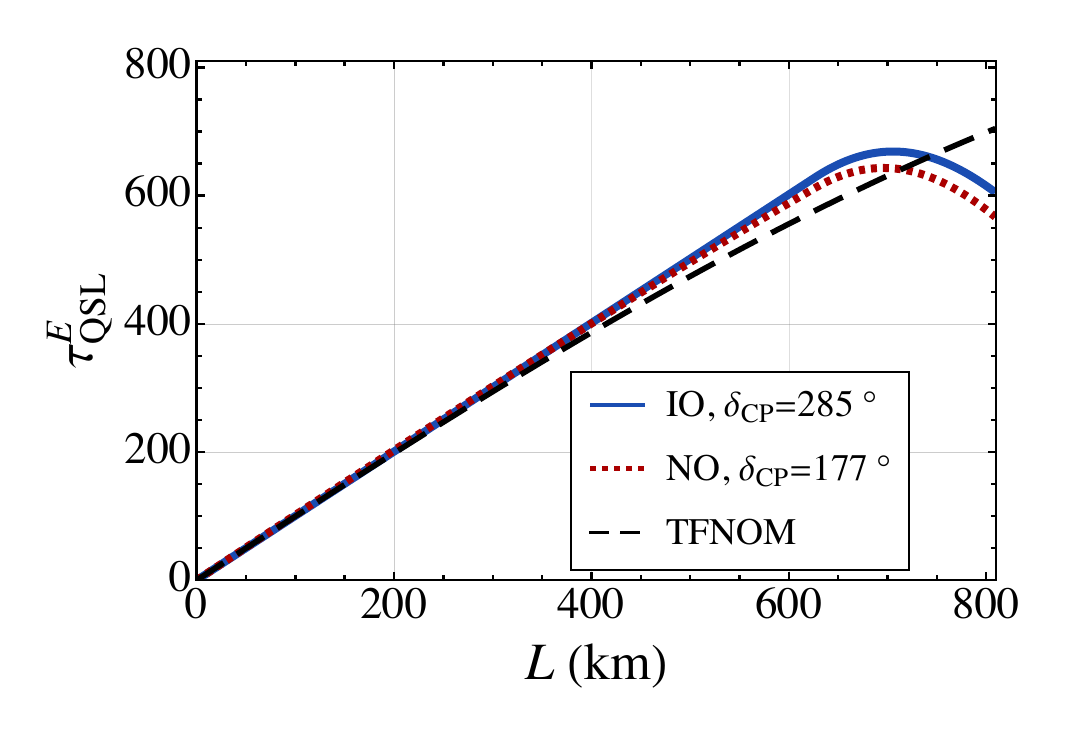}
        \caption{NOvA, $\delta_{\rm CP}=177^\circ$ (NO), $\delta_{\rm CP}=285^\circ$ (IO)}
        \label{9_fig:sub4}
    \end{subfigure}
    
    \medskip
    
    \begin{subfigure}{0.49\textwidth}
        \centering
        \includegraphics[width=\textwidth]{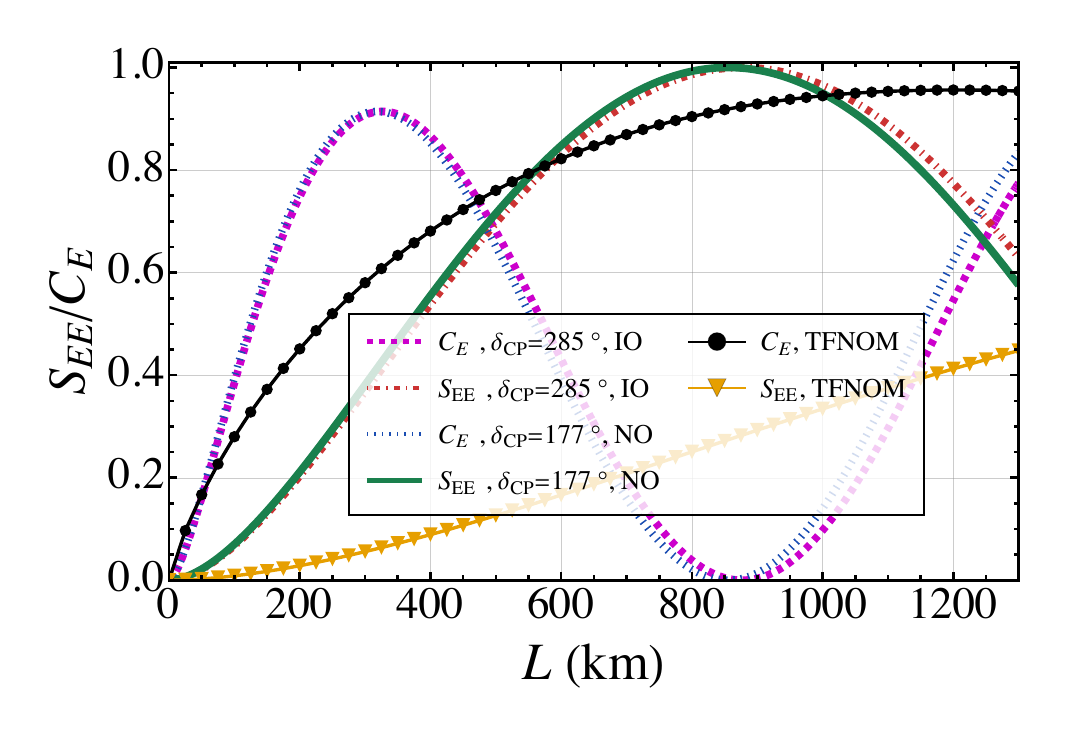}
        \caption{DUNE, $\delta_{\rm CP}=177^\circ$ (NO), $\delta_{\rm CP}=285^\circ$ (IO)}
        \label{9_fig:sub5}
    \end{subfigure}
    \hfill
    \begin{subfigure}{0.49\textwidth}
        \centering
        \includegraphics[width=\textwidth]{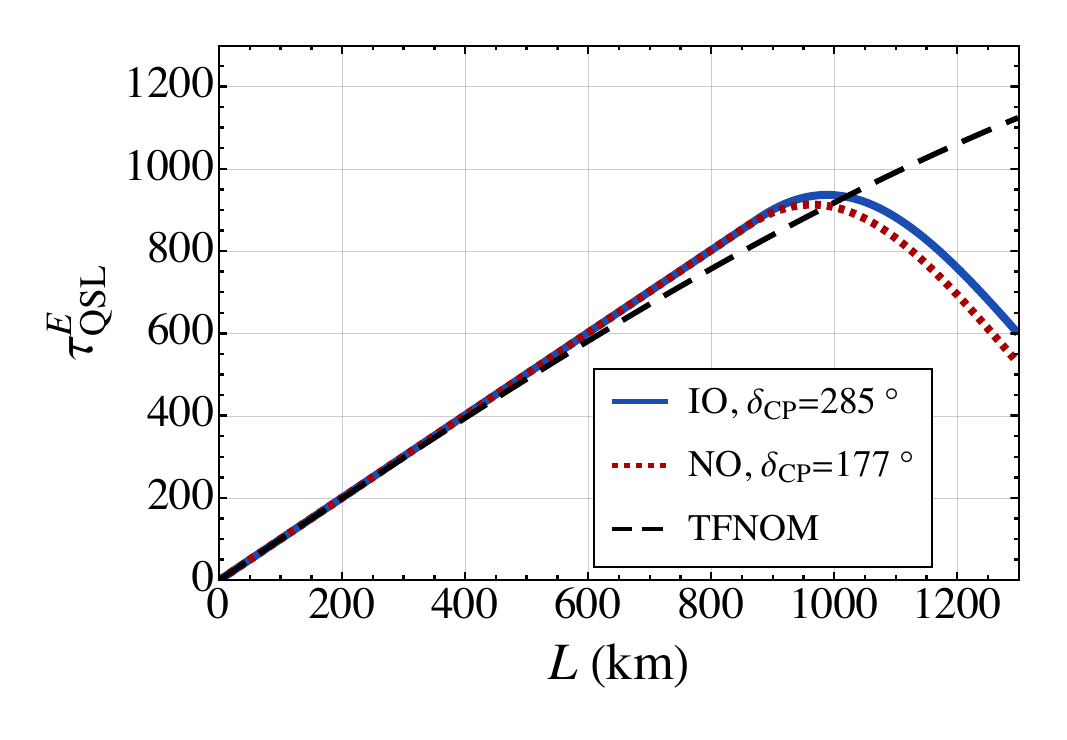}
        \caption{DUNE, $\delta_{\rm CP}=177^\circ$ (NO), $\delta_{\rm CP}=285^\circ$ (IO)}
        \label{9_fig:sub6}
    \end{subfigure}
    \caption{\justifying{For three-flavor neutrino oscillations, Figs.\,\ref{9_fig:sub1}, \ref{9_fig:sub3}, and \ref{9_fig:sub5} depict entanglement entropy [$S_{EE}(\rho_{\mu e}(t))$] and capacity of entanglement [$C_E(\rho_{\mu e}(t))$] for the initial state $\ket{\nu_\mu}$ in matter as functions of the propagation length $L\,\text{(km)}$ in both NO ($\delta_{\rm CP}=177^\circ$) and IO ($\delta_{\rm CP}=285^\circ$) using the length scales and energies corresponding to T2K, NOvA, and DUNE, respectively. Furthermore, we compare the QSL time $\tau^E_{\rm QSL}$ versus $L\,\text{(km)}$ for the initial state $\ket{\nu_\mu}$ in matter in NO ($\delta_{\rm CP}=177^\circ$) and IO ($\delta_{\rm CP}=285^\circ$). This comparison is illustrated in Figs.\,\ref{9_fig:sub2}, \ref{9_fig:sub4}, and \ref{9_fig:sub6} using the length scale and energy from T2K, NOvA, and DUNE, respectively. In the TFNOM, the entanglement measures $S_{EE}(\rho_\mu(t))$ and $C_E(\rho_\mu(t))$ in Figs.\,\ref{9_fig:sub1}, \ref{9_fig:sub3}, and \ref{9_fig:sub5}, and the QSL time $\tau^E_{\rm QSL}$ in Figs.\,\ref{9_fig:sub2}, \ref{9_fig:sub4}, and \ref{9_fig:sub6}, as functions of the propagation $L\,\text{(km)}$ for the initial muon-flavor neutrino state $\ket{\nu_\mu}$, are also illustrated using the length scales and energies corresponding to T2K, NOvA, and DUNE. The various neutrino mixing parameters used are taken from Tables\,\ref{tab2} for TFNOM and NO, and \ref{tab3} for IO along with their corresponding 1$\sigma$ errors ($90\%$ CL).}}
    \label{fig:subfigures}
\end{figure*}

Here $\rho=\rho^2$ and $ \rm Tr\,(\rho^2)=1$, i.e., a pure state. However, the reduced density matrix of the state $\rho$ [by tracing over the other qubit(s)] denoted by $\rho_{\mu e}(t)= \mathrm{Tr}_{(\tau)}(\rho(t))$ and $\rho_{\tau}(t)=\mathrm{Tr}_{(\mu e)}(\rho(t))$ represent  mixed states as $ \mathrm{Tr}\,(\rho^2_{\mu e}(t)<1$ and $ \mathrm{Tr}\,(\rho^2_{\tau}(t))<1$. Thus, the reduced states $\rho_{\mu e}(t)$ and $\rho_{\tau}(t)$ provide an example of two subsystems of a bipartite quantum system $\rho(t)$ in the framework of three-qubit mode (flavor) states. Treating one flavor mode as one part of the bipartite system $\rho(t)$ (e.g., $\tau$) and the other as the remaining part (e.g., $\mu e$), the eigenvalues of the reduced state $\rho_{\mu e}(t)$ are 
\begin{equation} \xi_1(t)=\xi_2(t)=0;\hspace{0.5em}\xi_3(t)=|\mathcal{A}_{\mu\mu}|^2; \hspace{0.5em} \text{and} \hspace{0.5em}\xi_4(t)=|\mathcal{A}_{\mu e}|^2+|\mathcal{A}_{\mu \tau}|^2;
    \label{49}
\end{equation}
  where $\xi_3(t)+\xi_4(t)=1$. Using the two nonzero eigenvalues, the bipartite entanglements measures such as entanglement entropy [Eq.\,(\ref{5})] and capacity of entanglement [Eq.\,(\ref{7})], can be calculated for the time-evolved muon flavor neutrino state $\ket{\nu_{\mu}(t)}$ in matter as
\begin{equation}
    S_{EE}(\rho_{\mu e}(t))= -\xi_3(t) \log_2\xi_3(t)-\xi_4(t)\log_2\xi_4(t),
    \label{50}
\end{equation}
and 
\begin{align}
     C_{E}(\rho_{\mu e}(t))=&-\xi_3(t) \log^2_2\xi_3(t)-\xi_4(t)\log^2_4\xi_4(t) \nonumber \\&
     -(-\xi_3(t) \log_2\xi_3(t)-\xi_4(t)\log_2\xi_4(t))^2,
     \label{51}
\end{align}
respectively. In a manner akin to the two-flavor neutrino oscillation in matter [as demonstrated in Eqs.\,(\ref{42}) and (\ref{43})], the two bipartite entanglement measures $S_{EE}(\rho_{\mu e}(t))$ and $C_{E}(\rho_{\mu e}(t))$ given in Eqs.\,(\ref{50}) and (\ref{51}) for the three-flavor neutrino oscillation in a constant matter background can likewise be elucidated in terms of neutrino transition probabilities.

In the ultra-relativistic limit, Figs.\,\ref{9_fig:sub1},\,\ref{9_fig:sub3}, and \ref{9_fig:sub5} illustrate $S_{EE}(\rho_{\mu e}(t))$ and $C_E(\rho_{\mu e}(t))$ for the initial state $\ket{\nu_\mu}$ in matter as a function of propagation length $L\text{(km)}$, for T2K, NOvA and DUNE, respectively. The length scales and energies used correspond to T2K ($L\approx 295\,\text{km}$, $E\approx 1.4\,\rm GeV$), NOvA ($L\approx 800\,\text{km}$, $E\approx 2.5\,\rm GeV$), and DUNE ($L\approx 1300\,\text{km}$, $E\approx 3.5\,\rm GeV$), respectively, considering two distinct $CP$-violation phases: $\delta_{\rm CP}=177^\circ$ in NO scenario and $\delta_{\rm CP}=285^\circ$ in the IO scenario. By using the mixing parameters and mass-squared differences mentioned in Tables\,\ref{tab2} and \ref{tab3}, it is observed that for $L>0$, entanglement measures $S_{EE}(\rho_{\mu e}(t))\neq0$ and $C_{ E}(\rho_{\mu e}(t))\neq0$. This suggests that the state $\ket{\nu_\mu(t)}$ can be considered to be a bipartite pure entangled state in both NO and IO, with $\delta_{\rm CP}$ taken into account.
In Fig.\,\ref{9_fig:sub1}, for the initial state $\ket{\nu_\mu}$, $S_{ EE}$ monotonically increases in both NO (green solid line) and IO (red dot-dashed line) and approximately coincide within the range of propagation length $L\approx 250\,\text{km}$, for the $\delta_{\rm CP}$ chosen. Consequently, $C_{ E}$ also approximately coincides in NO (blue dotted line) and IO (magenta dashed line) while taking into account $\delta_{\rm CP}$ for the state $\ket{\nu_\mu(t)}$. This analysis indicates that T2K is unable to discern between the entanglement entropy for the initial state $\ket{\nu_\mu}$ in NO and IO with $\delta_{\rm CP}$ under constant matter potential. However, in Fig.\,\ref{9_fig:sub3}, it is evident that NOvA's length scale and energy are able to capture the difference between the entanglement entropy $S_{EE}$ in NO (green solid line) and IO (red dot-dashed line) with the chosen $\delta_{\rm CP}$. Subsequently, a small difference between NO (blue dotted line) and IO (magenta dashed line) is observed for the capacity of entanglement $C_{ E}$. At $L\approx 800\,\text{km}$, a separation in entanglement entropy $S_{EE}$ is observed between the IO (red dot-dashed line) and the NO (green solid line), allowing them to be discerned. In Fig.\,\ref{9_fig:sub5}, the analysis using DUNE's length scale and energy reveals a significant split in $S_{EE}$ between NO (green solid line) and IO (red dot-dashed line). At $L\approx 1300\,\text{km}$, more suppression of entanglement is observed in the NO (green solid line). Consequently, at the same length, the capacity of entanglement $C_{ E}$ increases in the NO (blue dotted line) compared to the IO (magenta dashed line).

In addition, using the length scales and energies corresponding to the experiments T2K, NOvA, and DUNE, Figs.\,\ref{9_fig:sub1}, \ref{9_fig:sub3}, and \ref{9_fig:sub5}, respectively, also present the entanglement measures $S_{EE}(\rho_\mu(t))$ (line with an orange inverted triangle) and $C_E(\rho_\mu(t))$ (line with a small black dot) for the TFNOM as functions of the propagation length $L\,\text{(km)}$ for the initial muon-flavor neutrino state $\ket{\nu_\mu}$. The entanglement entropy $S_{EE}\rightarrow1$ for NOvA, DUNE, and very close to 1 for T2K. This indicates that the state $\ket{\nu_\mu(t)}$ achieves maximal bipartite pure entanglement for NOvA and DUNE in the three-flavor oscillations in matter for both NO and IO at their respective $CP$-violation phases, while $S_{EE}$ in the TFNOM (line with an orange inverted triangle) exhibits only partial bipartite pure entanglement across all three experiments.

Moreover, for the state $\ket{\nu_\mu(t)}$ in matter, using Eq.\,(\ref{45}) and the effective Hermitian Hamiltonian in matter \cite{Nguyen:2022snr} in Eq.\,({\ref{4}}), we determine the variance in the time-independent driving Hamiltonian ($\Delta \mathcal{H}_M$). Subsequently, applying Eqs.\,(\ref{50}) and (\ref{51}) with $\Delta\mathcal{H}_M$ value in Eq.\,(\ref{6}) enables us to numerically compute the QSL time for entanglement entropy $\tau^E_{\rm QSL}$ of the state $\ket{\nu_\mu(t)}$ in both NO and IO scenarios, with their respective $CP$-violation phases. In Fig.\,\ref{9_fig:sub2}, it is observed that $\tau^E_{\rm QSL}$ for the initial state $\ket{\nu_\mu}$ as a function of propagation length $ L\,(\rm km)$ in NO (red dotted line) and in IO (blue solid line) with their respective $CP$-violation phases coincide using the T2K length scale and energy. This leads to the time-bound condition $\tau^E_{\rm QSL}/L=1$. The result shows that the evolution speed of $\ket{\nu_\mu(t)}$ has already reached its maximum in both NO and IO at $ L\approx280\,\rm km$, and therefore T2K fails to capture any differences between the dynamic evolution of an entangled muon flavor neutrino state for speedup in NO and IO. In Fig.\,\ref{9_fig:sub4}, employing NOvA's length scale and energy, the plot of $\tau^E_{\rm QSL}$ versus $L\,(\rm km)$ demonstrates a slight disparity between the NO (red dotted line) and IO (blue solid line) scenarios, taking into account $\delta_{\rm CP}$. Under the constraint $\tau^E_{\rm QSL}/L<1$, for the initial state $\ket{\nu_\mu}$ in matter, a rapid suppression of entanglement is observed in the NO (red dotted line) compared to the IO (blue solid line) within the range of $\delta_{\rm CP}$. However, at DUNE's length scale and energy, Fig.\,\ref{9_fig:sub6} illustrates a significantly greater disparity between $\tau^E_{\rm QSL}$ versus $ L\,(\rm km)$ in the NO (red dotted line) and IO (blue solid line). It is evident that a much more rapid suppression of entanglement is observable in the NO (red dotted line) compared to the IO (blue solid line) within the range of $\delta_{\rm CP}$. Alternatively, one can say that within the framework of three-flavor neutrino oscillations in matter, for the initial state $\ket{\nu_\mu}$, faster growth of entanglement is observed in DUNE under the assumption of IO (blue solid line). Therefore, DUNE exhibits higher sensitivity in observing $\tau^E_{\rm QSL}$ compared to T2K and NOvA.

Further, using T2K length scale and energy in Fig.\,\ref{9_fig:sub2}, the time-bound condition $\tau^E_{\rm QSL}/L < 1$ indicates that the initial state $\ket{\nu_\mu}$ undergoes a more rapid suppression of entanglement in the TFNOM (black dashed line) compared to the three-flavor neutrino oscillation in matter, for both NO (red dotted line) and IO (blue solid line) with their respective best-fit $CP$-violation phases. However, for NOvA and DUNE in Figs.\,\ref{9_fig:sub4} and \ref{9_fig:sub6}, similar behavior is observed for most of the baselines, except towards the end of their respective baselines, where this behavior is reversed, i.e., faster suppression of entanglement is observed for three-flavor case instead of TFNOM.

\section{Discussion and Conclusion}
\label{Sect7}
We have studied $CP$-violation and the mass ordering problem of neutrino oscillations in matter, employing the QSL time as a key analytical tool. As a precursor to the three-flavor oscillations, the QSL time for two-flavor neutrino oscillations in matter has been initially analyzed. Our findings have revealed that rapid transitions of muon flavor neutrinos occur from initial to final states as the neutrino potential shifts from a constant Earth-matter potential to vacuum, based on the length scale and energy of a futuristic long-baseline accelerator neutrino experiment such as DUNE. Furthermore, we have explored two bipartite entanglement measures: entanglement entropy and capacity of entanglement—quantified in terms of two-flavor neutrino transition probabilities in matter, which are measurable quantities in neutrino experiments. The nonzero values of these entanglement measures indicate a bipartite pure entangled system during two-flavor neutrino oscillations in vacuum as well as in matter. For the initial muon flavor neutrino state evolution, the QSL time for entanglement entropy indicates rapid suppression of entanglement when neutrinos oscillate in vacuum instead of the constant Earth-matter background.

In order to probe the impact of $CP$-violation and the mass ordering problem of neutrino oscillations in matter, we have extended our analysis to three-flavor neutrino oscillations in the presence of a constant Earth-matter potential with $CP$-violation phases, considering two cases of neutrino mass ordering: NO and IO. We have focused on the initial muon flavor neutrino state evolution and used the length scales and energies of ongoing long-baseline neutrino experiments such as T2K, NOvA, and the upcoming DUNE experiment. Our results have revealed discrepancies in the QSL time, for the initial muon flavor neutrino state evolution, between NO and IO in the presence of their corresponding $CP$-violation phases across these experiments. This separation can be primarily ascribed to the oscillation probability $P_{\mu\rightarrow e}$. Notably, a faster evolution of the initial muon flavor neutrino state in matter is achievable in NO taking into account $CP$-violation phases in all three experiments. Moreover, at DUNE's length scale and energy, we have observed faster flavor state evolution for both the initial muon neutrino and muon antineutrino flavor states in the NO scenario. However, considering the baseline lengths and energies of current experiments T2K, NOvA, as well as the next-generation DUNE experiment, the results showed that both the initial muon neutrino and muon antineutrino flavor states evolve more rapidly in matter under the two-flavor approximation than in three-flavor cases for both NO and IO with zero and non-zero $CP$-violation phases.

In the three-flavor oscillations scenario, entanglement entropy and capacity of entanglement could be expressed in terms of three-flavor transition probabilities for the initial muon-flavor neutrino state. We have observed that, compared to the two-flavor neutrino oscillation results, the time-evolved muon-flavor neutrino state in the three-flavor framework exhibits a maximally bipartite pure entangled state. While there are no discrepancies in the entanglement entropy found for the best-fit $CP$-violation phase values in both NO and IO using the length scale and energy of T2K, discrepancies are observed when we use NOvA and DUNE.
Specifically, a greater decrease in entanglement entropy has been observed at DUNE's length scale and energy for the initial muon flavor neutrino state evolution in the presence of a constant Earth-matter background in NO with $CP$ violation.

Further, for the initial muon flavor neutrino state evolution, the QSL time for entanglement entropy has been numerically computed in NO and IO with $CP$-violation phases using T2K, NOvA, and DUNE length scales and energies. T2K fails to capture any differences in the QSL time for entanglement between NO and IO in the presence of $CP$-violation phase. However, NOvA and DUNE have shown sensitivity in capturing these differences. Faster suppression of entanglement, for the initial muon flavor neutrino state evolution in matter, has been observed in NO, in the presence of a significant $CP$-violation phase, at DUNE's length scale and energy. This behavior was compared with that of the two-flavor oscillation in matter across all the experiments (T2K, NOvA and DUNE). Except for T2K, our results showed that entanglement suppression occurred more rapidly near the ends of the NOvA and DUNE baselines in three-flavor neutrino oscillations in matter, compared to the two-flavor approximation.

Thus, by probing the transition probabilities and the QSL time for the evolution and entanglement of an initial muon neutrino state in three-flavor neutrino oscillations in matter, using baseline lengths and energies corresponding to the currently running T2K and NOvA experiments, and the upcoming DUNE experiment, we conclude that the results differ from those of the two-flavor approximation. These differences mainly stem from changes in the additional mixing angles and mass-squared differences due to the presence of matter, as well as from the $CP$-violation phases, which are included in the three-flavor neutrino oscillation framework in matter but excluded in the two-flavor approximation. These findings could help in ascertaining the role of $CP$ violation and determination of the mass ordering in neutrino oscillations.

\section*{Acknowledgments}
We thank Banibrata Mukhopadhyay, Ranjan Laha, Massimo Blasone, and Brij Mohan for their fruitful discussions and insightful input. Subhadip Bouri acknowledges the Council of Scientific and Industrial Research (CSIR), Government of India, for supporting his research under the CSIR Junior/Senior Research Fellowship program through Grant No. 09/0079(15488)/2022-EMR-I. A.K.J. and S.B. would like to acknowledge the project funded by SERB, India, with Ref. No. CRG/2022/003460, for supporting this research. We would like to thank the anonymous referees for their valuable comments, which helped us greatly in improving this work.

\bibliography{reference.bib}
\end{document}